\RequirePackage{fix-cm}
\documentclass[smallcondensed]{svjour3}       % onecolumn (second format)
\smartqed  % flush right qed marks, e.g. at end of proof
\usepackage{lscape}
\usepackage{graphicx}
\usepackage{cite}
\usepackage{listings}
\usepackage{amsmath,amssymb}
\usepackage{color}
\usepackage{stmaryrd}
\usepackage{xcolor}
\usepackage{multirow}
\usepackage{tikz}
\usepackage{verbatim}
\usepackage{array}

\journalname{Formal Methods in Systems Design}
\newcommand{\ldrf}{\ensuremath{\mathbb{L}}-DRF}
\newcommand{\ldrfa}{\ensuremath{\mathcal{A}^{\local}}} % analysis based on LDRF

\newcommand{\rdrf}{\ensuremath{\mathbb{L}}-RegDRF}
\newcommand{\Reg}{\textit{RegRel}}

\newcommand{\nat}{\mathbb{N}} % natural numbers

\newcommand{\A}[2][]{\left\langle #2 \right\rangle_{#1}}
\newcommand{\abracks}[1] {\ensuremath{\langle #1 \rangle}}
\newcommand\myeq{\mathrel{\stackrel{\makebox[0pt]{\mbox{\normalfont\tiny def}}}{=}}}
\newcommand{\eqdef}{\myeq}
\newcommand{\bb}[1]{\llbracket #1 \rrbracket}
\newcommand{\BB}[2][]{\llbracket #2 \rrbracket_{#1}}
\newcommand{\scfg}{sync-CFG}

\newcommand{\partialto}{\rightharpoonup}
\newcommand{\totalto}{\to}

\newcommand{\dom}{\ensuremath{\mathit{dom}}}
\newcommand{\domof}[1]{\dom(#1)}

\newcommand{\pset}[1]{\mathcal{P}(#1)}
\newcommand{\mtrue}{\mathit{true}}
\newcommand{\mfalse}{\mathit{false}}
\newcommand{\card}[1]{\left\vert{#1}\right\vert}

\newcommand{\ltrto}[1]{\rightarrow_{#1}} % general transition
\newcommand{\ltrtos}[1]{\Rightarrow^\std_{#1}} % standard interleaving transitions
\newcommand{\ltrtol}[1]{\Rightarrow^\local_{#1}} % LDRF transitions
\newcommand{\take}[1]{\textit{take}_{#1}}

\newcommand{\vars}{\mathcal{V}}
\newcommand{\threads}{\mathcal{T}}
\newcommand{\plocks}{\mathcal{M}}
\newcommand{\vals}{\mathbf{V}}

\newcommand{\cmd}{\mathit{cmd}} % set of program commands
\newcommand{\regions}{\mathit{R}}
\newcommand{\PC}{\mathit{PC}} % set of program counters
\newcommand{\LM}{\ensuremath{\mathit{LM}}} % set of lock maps
\newcommand{\Env}{\mathit{Env}} % set of environments
\newcommand{\RV}{\ensuremath{\mathit{Ver}}}
\newcommand{\VE}{\mathit{VE}} % set of versioned environments
\newcommand{\AdmStates}{\Sigma} % admissible states

\newcommand{\ent}{\mathit{ent}}
\newcommand{\inst}{\ensuremath{\iota}}
\newcommand{\instr}{\mathit{inst}}
\newcommand{\locs}{\mathcal{L}}	% program locations
\newcommand{\ploc}{\mathit{n}} % post-release points

\newcommand{\rellocs}[1][]{\locs^\mathit{rel}_{#1}}
\newcommand{\acqlocs}[1][]{\locs^\mathit{acq}_{#1}}

\newcommand{\tidof}[1]{\mathit{tid}(#1)} % thread identifier of a command
\newcommand{\cfg}{\ensuremath{G}}
\newcommand{\cmdof}[1]{\ensuremath{\mathit{cmd}(#1)}}
\newcommand{\instrof}[1]{\ensuremath{\mathit{instr}(#1)}}

\newcommand{\lock}{\mathtt{m}}
\newcommand{\acq}[1]{\mathtt{acquire(}#1\mathtt{)} }
\newcommand{\rel}[1]{\mathtt{release(}#1\mathtt{)} }

\newcommand{\assume}[1]{\mathtt{assume(}#1\mathtt{)} }
\newcommand{\assgn}[2]{#1\, \mathtt{:=} \, #2}

\newcommand{\cs}{\sigma}
\newcommand{\cS}{\Sigma}
\newcommand{\stds}{s}
\newcommand{\stdS}{\mathcal{S}}
\newcommand{\std}{\mathbb{S}}
\newcommand{\TR}{\mathit{TR}}
\newcommand{\reach}{\mathit{Reach}}
\newcommand{\lts}{\ensuremath{L}}
\newcommand{\ltsi}{\ensuremath{L^{\std}}}
\newcommand{\ltsl}{\ensuremath{L^{\local}}}

\newcommand{\pc}{\mathit{pc}}
\newcommand{\lmm}{\mu}
\newcommand{\tview}{\Theta}
\newcommand{\lview}{\Lambda}
\newcommand{\env}{\phi}
\newcommand{\rv}{\nu}
\newcommand{\ve}{\mathit{ve}}
\newcommand{\rfGsymb}[1]{\mathcal{G}_{#1}}
\newcommand{\Gsymb}{\mathcal{G}} % map relevant buffers
\newcommand{\f}{\chi}
\newcommand{\cstate}{\sigma} % concrete l-drf state
\newcommand{\rfpi}{\hat{\pi}} % l-drf execution
\newcommand{\csset}{C}

\newcommand{\reg}{\mathit{rg}}
\newcommand{\local}{\ensuremath{\mathbb{L}}}

\newcommand{\AbsState}[1][]{\mathcal{A}_{#1}}

\newcommand{\AbsLeq}[1][]{\sqsubseteq_{#1}}
\newcommand{\AbsJoin}[1][]{\sqcup_{#1}}
\newcommand{\Cart}{\times}

\newcommand{\mixfunc}{\mathit{mix}}
\newcommand{\absstate}[1][]{a_{#1}}

\newcommand{\adrf}{\textit{Rel}}

\newcommand{\ANa}{\mathbf{\overline{RT}}}
\newcommand{\ANb}{\mathbf{R\overline{T}}}
\newcommand{\ANc}{\mathbf{\overline{R}T}}
\newcommand{\ANd}{\mathbf{RT}}
\newcommand{\ANe}{\mathbf{VS}}

\newcommand{\aia}{\ensuremath{\mathcal{A}}}
\newcommand{\aic}{\ensuremath{\mathcal{C}}}
\newcommand{\join}{\ensuremath{\sqcup}}
\newcommand{\bigjoin}{\ensuremath{\bigsqcup}}
\newcommand{\lfp}{\ensuremath{\mathit{LFP}}}

\newcommand{\vrel}{\ensuremath{\mathit{VRel}}}
\newcommand{\rela}{\ensuremath{\mathit{Rel}}}
\newcommand{\vset}{\ensuremath{\mathit{ValSet}}}
\newcommand{\recent}[1]{\ensuremath{\mathit{recent(#1)}}}

\newcommand{\vrelseven}{\ensuremath{\{\A{x\!\mapsto\! 0^0,y\!\mapsto\! 0^0}\}}}
\newcommand{\vreleight}{\ensuremath{\{\A{x\!\mapsto\! 0^0,y\!\mapsto\! 0^0}\}}}
\newcommand{\vrelnine}{\ensuremath{\{\A{x\!\mapsto\! 1^1,y\!\mapsto\! 0^0}\}}}
\newcommand{\vrelten}{\ensuremath{\{\A{x\!\mapsto\! 1^1,y\!\mapsto\! 1^1}\}}}
\newcommand{\vreleleven}{\ensuremath{\{\A{x\!\mapsto\! 1^1,y\!\mapsto\! 1^1}\}}}
\newcommand{\vrelone}{\ensuremath{\{\A{x\!\mapsto\! 0^0,y\!\mapsto\! 0^0}\}}}
\newcommand{\vreltwo}{\ensuremath{\{\A{x\!\mapsto\! 1^1,y\!\mapsto\! 1^1}\}}}
\newcommand{\vrelthree}{\ensuremath{\{\A{x\!\mapsto\! 1^2,y\!\mapsto\! 1^1}\}}}
\newcommand{\vrelfour}{\ensuremath{\{\A{x\!\mapsto\! 2^3,y\!\mapsto\! 1^1}\}}}
\newcommand{\relseven}{\ensuremath{\{\A{x\!\mapsto\! 0,y\!\mapsto\! 0}\}}}
\newcommand{\releight}{\ensuremath{\{\A{x\!\mapsto\! 0,y\!\mapsto\! 0}\}}}
\newcommand{\relnine}{\ensuremath{\{\A{x\!\mapsto\! 1,y\!\mapsto\! 0}\}}}
\newcommand{\relten}{\ensuremath{\{\A{x\!\mapsto\! 1,y\!\mapsto\! 1}\}}}
\newcommand{\releleven}{\ensuremath{\{\A{x\!\mapsto\! 1,y\!\mapsto\! 1}\}}}
\newcommand{\relone}{\ensuremath{\{\A{x\!\mapsto\! 0,y\!\mapsto\! 0}\}}}
\newcommand{\reltwoa}{\ensuremath{\{\A{x\!\mapsto\! 0,y\!\mapsto\!0}\!,}}
\newcommand{\reltwob}{\ensuremath{\ \A{x\!\mapsto\! 1,y\!\mapsto\!0}\!,}}
\newcommand{\reltwoc}{\ensuremath{\ \A{x\!\mapsto\! 0,y\!\mapsto\!1}\!,}}
\newcommand{\reltwod}{\ensuremath{\ \A{x\!\mapsto\! 1,y\!\mapsto\!1}\}}}
\newcommand{\relthreea}{\ensuremath{\{\A{x\!\mapsto\! 0,y\!\mapsto\!0}\!,}}
\newcommand{\relthreeb}{\ensuremath{\ \A{x\!\mapsto\! 1,y\!\mapsto\!1}\}}}
\newcommand{\relfoura}{\ensuremath{\{\A{x\!\mapsto\! 1,y\!\mapsto\!0}\!,}}
\newcommand{\relfourb}{\ensuremath{\ \A{x\!\mapsto\! 2,y\!\mapsto\!1}\}}}
\newcommand{\vsseven}{\ensuremath{x\!\mapsto\!\{0\},y\!\mapsto\!\{0\}}}
\newcommand{\vseight}{\ensuremath{x\!\mapsto\!\{0\},y\!\mapsto\!\{0\}}}
\newcommand{\vsnine}{\ensuremath{x\!\mapsto\!\{1\},y\!\mapsto\!\{0\}}}
\newcommand{\vsten}{\ensuremath{x\!\mapsto\!\{1\},y\!\mapsto\!\{1\}}}
\newcommand{\vseleven}{\ensuremath{x\!\mapsto\!\{1\},y\!\mapsto\!\{1\}}}
\newcommand{\vsone}{\ensuremath{x\!\mapsto\!\{0\},y\!\mapsto\!\{0\}}}
\newcommand{\vstwo}{\ensuremath{x\!\mapsto\!\{0,1\},y\!\mapsto\!\{0,1\}}}
\newcommand{\vsthree}{\ensuremath{x\!\mapsto\!\{0,1\},y\!\mapsto\!\{0,1\}}}
\newcommand{\vsfour}{\ensuremath{x\!\mapsto\!\{1,2\},y\!\mapsto\!\{0,1\}}}
\newcommand{\ldrfsevena}{\ensuremath{\A{1,7}}}
\newcommand{\ldrfsevenb}{\ensuremath{\A{x\!\mapsto\!
      0^0,y\!\mapsto\! 0^0}}}
\newcommand{\ldrfsevenc}{\ensuremath{\A{x\!\mapsto\! 0^0,y\!\mapsto\! 0^0}}}
\newcommand{\ldrfsevend}{\ensuremath{\A{}}}
\newcommand{\ldrfeighta}{\ensuremath{\A{1,8}}}
\newcommand{\ldrfeightb}{\ensuremath{\A{x\!\mapsto\!
      0^0,y\!\mapsto\! 0^0}}}
\newcommand{\ldrfeightc}{\ensuremath{\A{x\!\mapsto\! 0^0,y\!\mapsto\! 0^0}}}
\newcommand{\ldrfeightd}{\ensuremath{\A{}}}
\newcommand{\ldrfninea}{\ensuremath{\A{1,9}}}
\newcommand{\ldrfnineb}{\ensuremath{\A{x\!\mapsto\!
      0^0,y\!\mapsto\! 0^0}}}
\newcommand{\ldrfninec}{\ensuremath{\A{x\!\mapsto\! 1^1,y\!\mapsto\! 0^0}}}
\newcommand{\ldrfnined}{\ensuremath{\A{}}}
\newcommand{\ldrftena}{\ensuremath{\A{1,10}}}
\newcommand{\ldrftenb}{\ensuremath{\A{x\!\mapsto\!
      0^0,y\!\mapsto\! 0^0}}}
\newcommand{\ldrftenc}{\ensuremath{\A{x\!\mapsto\! 1^1,y\!\mapsto\! 1^1}}}
\newcommand{\ldrftend}{\ensuremath{\A{}}}
\newcommand{\ldrfelevena}{\ensuremath{\A{1,11}}}
\newcommand{\ldrfelevenb}{\ensuremath{\A{x\!\mapsto\!
      0^0,y\!\mapsto\! 0^0}}}
\newcommand{\ldrfelevenc}{\ensuremath{\A{x\!\mapsto\! 1^1,y\!\mapsto\! 1^1}}}
\newcommand{\ldrfelevend}{\ensuremath{11\!\mapsto\!\A{x\!\mapsto\!1^1\!,\!y\!\mapsto\! 1^1}}}
\newcommand{\ldrfonea}{\ensuremath{\A{1,11}}}
\newcommand{\ldrfoneb}{\ensuremath{\A{x\!\mapsto\!
      0^0,y\!\mapsto\! 0^0}}}
\newcommand{\ldrfonec}{\ensuremath{\A{x\!\mapsto\! 1^1,y\!\mapsto\! 1^1}}}
\newcommand{\ldrfoned}{\ensuremath{11\!\mapsto\!\A{x\!\mapsto\! 1^1\!,\!y\!\mapsto\! 1^1}}}
\newcommand{\ldrftwoa}{\ensuremath{\A{2,11}}}
\newcommand{\ldrftwob}{\ensuremath{\A{x\!\mapsto\!
      1^1,y\!\mapsto\! 1^1}}}
\newcommand{\ldrftwoc}{\ensuremath{\A{x\!\mapsto\! 1^1,y\!\mapsto\! 1^1}}}
\newcommand{\ldrftwod}{\ensuremath{11\!\mapsto\!\A{x\!\mapsto\! 1^1\!,\!y\!\mapsto\! 1^1}}}
\newcommand{\ldrfthreea}{\ensuremath{\A{3,11}}}
\newcommand{\ldrfthreeb}{\ensuremath{\A{x\!\mapsto\!
      1^2,y\!\mapsto\! 1^1}}}
\newcommand{\ldrfthreec}{\ensuremath{\A{x\!\mapsto\! 1^1,y\!\mapsto\! 1^1}}}
\newcommand{\ldrfthreed}{\ensuremath{11\!\mapsto\!\A{x\!\mapsto\! 1^1\!,\!y\!\mapsto\! 1^1}}}
\newcommand{\ldrffoura}{\ensuremath{\A{4,11}}}
\newcommand{\ldrffourb}{\ensuremath{\A{x\!\mapsto\!
      2^3,y\!\mapsto\! 1^1}}}
\newcommand{\ldrffourc}{\ensuremath{\A{x\!\mapsto\! 1^1,y\!\mapsto\! 1^1}}}
\newcommand{\ldrffourd}{\ensuremath{11\!\mapsto\!\A{x\!\mapsto\! 1^1\!,\!y\!\mapsto\! 1^1}}}

\begin{document}

\title{A Thread-Local Semantics and Efficient Static Analyses for Race Free Programs%\thanks{Grants or other notes
}

\titlerunning{Thread-Local Analyses}        % if too long for running head

\author{Suvam Mukherjee \and
        Oded Padon \and
        Sharon Shoham \and
        Deepak D'Souza \and
        Noam Rinetzky
}

\authorrunning{S.\@ Mukherjee et al} % if too long for running head

\institute{Suvam Mukherjee \at
              Microsoft Research, \\
              India \\
              \email{suvamm@outlook.com}             \\
          \emph{Work done while the author was at the Indian Institute of Science, Bangalore, India}
           \and
           Oded Padon \at
              Stanford University, \\
              USA \\
              \email{padon@cs.stanford.edu}
          \and
          Sharon Shoham \at
              Tel Aviv University, \\
              Israel \\
              \email{sharon.shoham@gmail.com}
          \and
          Deepak D'Souza \at
              Indian Institute of Science, \\
              India \\
              \email{deepakd@iisc.ac.in}
          \and
          Noam Rinetzky \at
              Tel Aviv University, \\
              Israel \\
              \email{maon@cs.tau.ac.il}
}


\maketitle

\begin{abstract}
Data race free (DRF) programs constitute an important class of
concurrent programs.
In this paper we provide a framework for designing and proving the correctness
of data flow analyses that target this class of
programs. These analyses are in the same spirit as the ``sync-CFG''
analysis proposed in earlier literature.
To achieve this, we first propose a novel concrete
semantics for DRF programs, called \ldrf\, that is
\emph{thread-local} in nature -- each thread operates on its
own copy of the data state.
We show that abstractions of our semantics
allow us to reduce the analysis of DRF programs to a
\emph{sequential} analysis.
This aids in rapidly porting existing sequential analyses to
sound and scalable analyses for DRF programs.
Next, we parameterize \ldrf\ with
a partitioning of the program variables into ``regions'' which are
accessed atomically.
Abstractions of the region-parameterized semantics
yield more precise analyses for \emph{region-race} free
concurrent programs.
We instantiate these abstractions to devise efficient relational analyses for race free
programs, which we have implemented in a prototype tool called
RATCOP. On the benchmarks, RATCOP was able to prove upto 65\%
of the assertions, in comparison to 25\% proved by our baseline. Moreover, in a comparative study with a recent concurrent static analyzer, RATCOP was up to 5 orders of magnitude faster.
\end{abstract}

%!TEX root=./sas.tex
\section{Introduction}\label{Se:intro}

Our aim in this work is to provide a framework for developing
data-flow analyses which specifically target the class of data race
free (DRF) concurrent programs. DRF programs constitute an important
class of concurrent programs, as most programmers strive to write
race free code.
There are a couple of reasons why programmers do so.
Firstly, even assuming sequential consistency (SC) semantics, a racy program
often leads to undesirable effects like atomicity violations.
%% since data races are the sources of many
%% concurrency defects. All data races, ``benign" or otherwise, should be
%% considered as errors \cite{boehm2012, adve2010memory,
%%   boehm2011miscompile, andersonFunWithConcProbs, vyukov}.
Secondly, under the prevalent ``SC-for-DRF'' policy
only DRF programs are guaranteed to have
sequentially consistent execution behaviors in many weak memory
models \cite{adve1990weak, BoehmA08, jmm-manson}.
Non-DRF programs do not have this guarantee: for
example the Java Memory Model \cite{jmm-manson} gives some weak
guarantees, while the C++ semantics \cite{BoehmA08} gives
essentially \emph{no} 
guarantees, for the execution semantics of racy programs.
Thus ensuring that a racy program does something useful is a
difficult job for a programmer.
For these and other reasons, programmers tend to write race free programs.
There is thus is a large code base of DRF programs that can
benefit from data-flow 
analysis techniques that leverage the property of race-freedom to
provide analyses that run efficiently.
% (leading to insidious defects).
% As such, programmers are expected to write DRF programs
% \cite{boehm2012}.

The starting point of this work is the ``sync-CFG'' style of statically
analyzing DRF programs, proposed in \cite{de2011dataflow}.
The analysis here essentially runs a sequential analysis on each
thread, communicating data-flow facts between threads only via
``synchronization edges'', that go from a release statement in one
thread to the corresponding acquire statement in another thread.
The analysis thus runs on the
control-flow graphs (CFGs) of the threads, augmented with
synchronization edges, as shown in the center of Fig.~\ref{fi:motEx},
which explains the name for this style of analysis.
The analysis computes data flow facts about the value of
a variable that are sound
\emph{only} at points where that variable is \emph{relevant}, in that
it is read or written to at that location.
The analysis thus trades unsoundness of facts at irrelevant points for
the efficiency gained by restricting interference between threads to points of synchronization alone.

However, the analysis in \cite{de2011dataflow} suffers from
some drawbacks.
Firstly, the analysis is intrinsically a
``value-set'' analysis, which can only keep track of the set of values
each variable can assume, and not the relationships \emph{between}
variables.
Any naive attempt to extend the analysis to a more precise relational
one quickly leads to unsoundness.
The second issue is to do with the technique for establishing
soundness.
A convenient way to prove soundness of an analysis is to show that it
is a consistent abstraction \cite{cousot1977abstract} of a canonical
analysis, like the collecting semantics for sequential programs \cite{cousot1977abstract} or the
interleaving semantics for concurrent programs
\cite{lamport1997make}.
For this one typically makes use of the ``local'' sufficient
conditions for consistent abstration given in \cite{cousot1977abstract}.
However, for a \scfg-based analysis, it appears difficult to use
this route to show it to be a consistent abstraction of the
standard interleaving semantics.
This is largely due to the thread-local nature of the states and
the unsoundness at irrelevant points, which makes it difficult to
come up with natural abstraction and concretization functions that
form a Galois connection.
Instead, one needs to resort to an intricate argument, as done
in \cite{de2011dataflow},
which essentially
shows that in the least fixed point of the analysis, every write to a
variable will flow to a read of that variable via a happens-before
path (that is guaranteed to exist by the property of race-freedom).
Thus, while one can argue soundness of \emph{abstractions} of the
value-set analysis by demonstrating a consistent abstraction with
the latter,
%of an analysis that abstracts
%the value-set analysis by showing it to be a consistent abstraction
%of the value set analysis,
to argue soundness of any \emph{other} proposed \scfg\ style
analysis (in particular one that uses a more precise domain than
value-sets), one would
have to work out a similar involved proof as in
\cite{de2011dataflow}.

Towards addressing these issues, we propose a framework that
facilitates the design of different \scfg\ analyses with varying
degrees of precision and efficiency.
The foundation of this framework is a novel \emph{thread-local} semantics for DRF
programs, which can play the role of a ``most precise'' analysis which
other \scfg\ analyses can be shown to be consistent abstractions of.
This semantics, which we call \ldrf \cite{sas17-ldrf}, is similar to the interleaving
semantics of concurrent programs, but keeps thread-local (or
per-thread) copies of the shared state.
Intuitively, our semantics works as follows.
Apart from its local copy of the shared data state,
each thread $t$ also maintains a per-variable
version count, which is incremented whenever $t$ writes to the variable.
The exchange of information between threads is
via buffers, associated with release program points in the program.
When a thread releases a lock $\lock$, it stores its local data state to the
corresponding buffer, along with the version counts of the variables. As
a result, the buffer of a release point records both the local data
state and the variable %region
versions, as they were, when the release was last
executed.
When some thread $t'$ subsequently acquires $\lock$,
it compares its
per-variable version count with those in the
buffers pertaining to release points associated with $\lock$. The thread $t'$ then copies over the valuation (and the version) of a %region
variable to its local state,
if it is newer in some buffer (as indicated by a higher version count).
%Similar to a \scfg\ analysis,
The value of a shared variable in the local
state of a thread may be ``\emph{stale}'', in that the variable
has subsequently been updated by another thread but has not yet been
reflected here.
The \ldrf\ semantics leverages the race
freedom property to ensure that the value of a variable is correct in
the local state at program points where it is \emph{relevant}
(read or written to).
% We prove that
% our new semantics is equivalent to the standard SC interleaving
% semantics for DRF programs.
%
It thus captures the essence of a \scfg\ analysis.
The \ldrf\ semantics is also of independent interest, since it can be
viewed as an alternative characterization of the behavior of data race
free programs.

The analysis induced by the \ldrf\ semantics is shown to be sound
for DRF programs.
In addition, the analysis is, in some sense, the most precise \scfg\
analysis one can hope for: at every point in a thread, the
relevant part of the thread-local copy of the shared state is
\emph{guaranteed} to arise in some execution of the program.

Using the \ldrf\ semantics as a basis, we now propose several precise
and efficient \emph{relational} \scfg\ analyses.
The soundness of these analyses all follow immediately, since they can
easily be shown to be consistent abstractions of \ldrf.
The key idea behind obtaining a sound relational analysis is suggested by
the \ldrf\ analysis: we preserve variable correlations within a thread, whereas
at each $\mathtt{acquire}$ point,
we apply a \emph{mix} operator on the abstract values. The \emph{mix} operation
essentially amounts to forgetting all correlations between the
variables.

While these analyses allow maintaining fully-relational
properties within thread-local states,
communicating information over cross-thread edges loses all
correlations due to the \emph{mix} operation.
To improve precision further,
we refine the \ldrf\ semantics to take into account \emph{data regions}.
Technically, we introduce the notion of \emph{region race freedom} and
develop the \rdrf\ semantics \cite{sas17-ldrf}:
the programmer can partition  the program variables into ``regions''
that should be accessed \emph{atomically}.
A program is \emph{region race free} if it does not contain
conflicting accesses to variables in the same region, that are
unordered by the happens-before relation \cite{Lamport78}.
The classical notion of data race freedom is a special case of region
race freedom, where each region consists of a single variable. Techniques to determine whether a program is race free can be naturally
extended to determine region race freedom (see Sec.~\ref{ch:rdrf}).
\rdrf\ refines \ldrf\ by
taking into account the atomic nature of accesses that the program
makes to variables in the same region. For programs which are free from region-races, \rdrf\ produces executions which are indistinguishable, with respect to reads
of the regions, from the ones produced by  \ldrf.
By leveraging the %more precise
\rdrf\ semantics as a starting point, we
obtain more precise sequential analyses
%than~\cite{de2011dataflow}
that track relational properties \emph{within regions} and \emph{across} threads.
This is obtained
by refining the granularity of the \emph{mix} operator from single
variables to regions.

We have implemented the new relational analyses (based on \ldrf\ and \rdrf) in a prototype analyzer called
RATCOP \cite{ratcop}, and provide a thorough empirical evaluation in
Sec.~\ref{ch:exp-ratcop}. We show that RATCOP attains a precision of up to
$65\%$ on a subset of race-free programs from the SV-COMP15 suite. This subset contains programs which have interesting relational invariants.
In contrast, an interval based value-set analysis derived from
\cite{de2011dataflow} (which we use as our baseline) was able to prove only $25\%$ of the assertions.
On a
separate set of experiments, RATCOP turns out to be nearly $5$ orders
of magnitude faster than an existing state-of-the-art abstract
interpretation based tool \cite{monat2017precise}.

The rest of this paper is organized as follows.
In the next section we give an overview of our thread-local semantics
and the associated analyses.
In Sec.~\ref{se:prelims} we define our programming language and its standard interleaving semantics.
Sec.~\ref{sec:ldrf} contains the \ldrf\ semantics and the proof of its
soundness and completeness vis-a-vis the standard semantics. We then
introduce some analyses inspired by the \ldrf\ semantics, and formally
show how we can prove their soundness by showing them to be a
consistent abstraction of the \ldrf\ semantics.
In Sec.~\ref{ch:rdrf} we introduce our region-based analysis.
In Sec.~\ref{ch:exp-ratcop} we describe the implementation of our
analyses, and experimental evaluation.
We conclude in Sec.~\ref{Se:related} with related work and discussion.

\section{Overview}
\label{sec:overview}

We illustrate the \ldrf\ semantics, and its sequential abstractions, on the simple program in Fig.~\ref{fi:motExProg}. We assume that all variables are shared and are initialized to $0$. The threads access $x$ and $y$ only after acquiring lock $m$. The program is free from data races.

\begin{figure}[!htb]
\centering
\begin{minipage}{0.5\textwidth}
\begin{lstlisting}
Thread t$_1$() {
1:  acquire(m);
2:  x := y;
3:  x++;
4:  y++;
5:  assert(x=y);
6:  release(m);
7:
}
\end{lstlisting}
\end{minipage}%
\begin{minipage}{0.5\textwidth}
\begin{lstlisting}
Thread t$_2$() {
 8:  z++;
 9:  assert(z=1);
10:  acquire(m);
11:  assert(x=y);
12:  release(m);
13:
}
\end{lstlisting}
\end{minipage}
\caption{\label{fi:motExProg}A simple race free multi-threaded program. The variables $\mathtt{x}$, $\mathtt{y}$ and $\mathtt{z}$ are shared and initialized to $0$.}
\end{figure}

\begin{figure}[!ht]
\centering
\includegraphics[width=\textwidth]{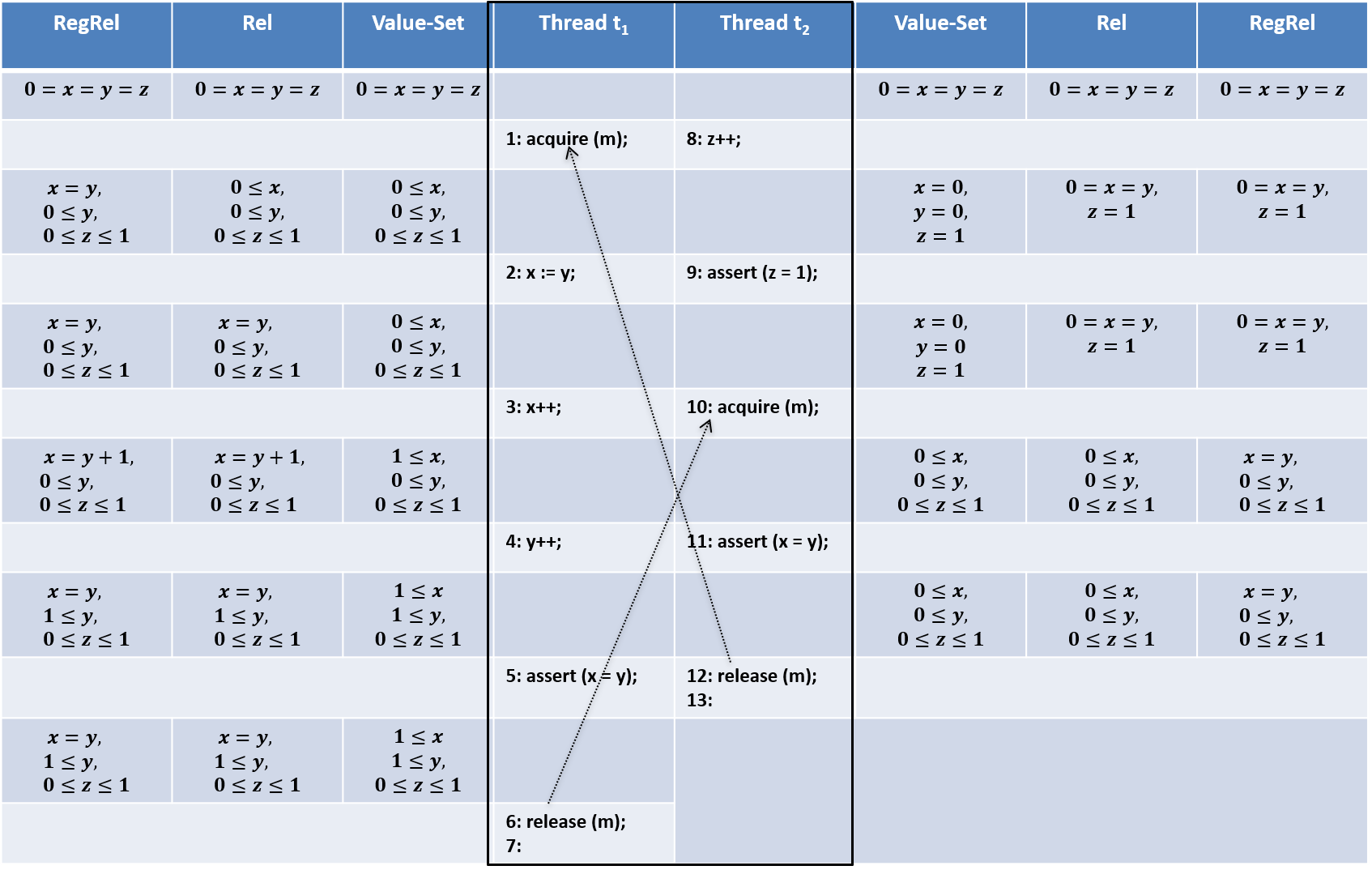}
\caption{\label{fi:motEx}\small{The \scfg\ representation of the
    program of Fig.~\ref{fi:motExProg} (center), with the facts
    computed by three analyses based on the \ldrf\ semantics shown
    in the three columns on the sides.
    In the  \scfg\ the intra-thread control flow
    edges are omitted for clarity, and only the synchronization edges are
    shown. The columns \adrf\ and \Reg\ show the facts computed by
    polyhedral-based relational abstractions of the \ldrf\ 
    semantics and its region-parameterized version, respectively. The
    Value-Set column
    shows the facts computed by interval abstractions of the Value-Set
    analysis of \cite{de2011dataflow}.
%All the three abstract analyses treat the program as a non-deterministic \emph{seqential} program, with inter-thread edges added between lock release-acquire points, as shown in the figure.
The \Reg\ analysis is able to prove all $3$ assertions, while
\adrf\ fails to prove the assertion at line~$11$. Value-Set manages to
prove only the assertion at line $9$.
%% Some of the facts have been
%% simplified for clarity: for example, we omit the fact $x \ge 0$ before
%% program point $2$ in the \Reg\ semantics, as it is implied. Also note
%% that we cannot establish an upper bound for both $x$ and $y$ because
%% of the ``spurious" cycle created due to the synchronization edges.
}}
% \vspace{-37.5pt}
\end{figure}

Fig.~\ref{fi:motEx} shows the \scfg\ representation of the program
(the control-flow graphs of the threads have been made implicit to
improve clarity) in the center. The columns to the left and right show
data flow facts obtained using three different analyses based on the 
\ldrf\ semantics, which we will describe later.

%\vspace{-10pt}

A state in the \ldrf\ semantics keeps track of the following
components: a location map  $\pc$ mapping each thread to the location of the next command to be executed, a lock map $\lmm$ which maps each lock to the thread holding it, a local environment (variable to value map)
$\tview$ for each thread, and a function $\lview$
which maps each buffer (associated with each program location following a release command) to an environment. Every release point of each lock $\lock$ has an associated buffer, where a thread stores a copy of its local environment when it executes the corresponding release instruction.
In the environments, each variable $x$ has a
version count associated with it which, along any execution $\pi$, essentially associates this valuation of $x$ with a unique prior write to it in $\pi$.
As an example, the ``versioned" environment $\langle x\mapsto 1^2, y
\mapsto 1^1, z \mapsto 0^0 \rangle$, obtained at some point in an
execution $\pi$, says that $x$ has the value $1$ by the second
write to $x$,  $y$ has the value 1 by the first write to $y$ in
$\pi$, and $z$ has not been
written to. An execution is an interleaving of commands from different
threads. Consider an execution of the program in Fig.~\ref{fi:motExProg}
where, after a certain number of interleaved steps, we have the state

\begin{align*}
\pc: t_1 \mapsto 6, t_2 \mapsto 10  \nonumber \\
\tview(t1): x \mapsto 1^2, y \mapsto 1^1, z \mapsto 0^0 \nonumber \\
\tview(t2): x \mapsto 0^0, y \mapsto 0^0, z \mapsto 1^1 \nonumber \\
\lmm: m \mapsto t_1 \nonumber \\
\lview: 7 \mapsto \bot, 13 \mapsto \bot \nonumber \\
\end{align*}

The release buffers are all empty as no thread has executed a
$\mathtt{release}$ yet. Note that the values (and versions) of $x$ and
$y$ in $t_2$ (similarly for $z$ in $t_1$) are
\emph{stale}, as they do not have the latest value of these variables
which were updated by another thread.
% since it was $t_1$ which last modified them (similarly for $z$ in
% $\tview(t_1)$). 
Next, $t_1$ can
execute the $\mathtt{release}$ at line~$6$, thereby setting $\lmm(m) =
\bot$ and storing its current local versioned environment to
$\lview(7)$. 
Now $t_2$ can execute the $\mathtt{acquire}$ at line~$10$.
In doing so, the following state changes take place.
As usual, the $\pc$ is updated to say that $t_2$ is now at line 11,
and the lock map is updated to say that $t_2$ now holds lock $m$.
Additionally $t_2$ 
``imports" the most up-to-date values (and versions) of %variables
$x$ and $y$ from the release buffer $\lview(7)$. 
We call this inter-thread join operation a \emph{mix}. This results in
its local state becoming $\langle x \mapsto 1^2, y \mapsto 1^1, z
\mapsto 1^1\rangle$ (the valuations of $x$ and $y$ are pulled in from
the buffer, while the valuation of $z$ in $t_2$'s local state
persists).
The state thus becomes
\begin{align*}
\pc: t_1 \mapsto 7, t_2 \mapsto 11  \nonumber \\
\tview(t1): x \mapsto 1^2, y \mapsto 1^1, z \mapsto 0^0 \nonumber \\
\tview(t2): x \mapsto 1^2, y \mapsto 1^1, z \mapsto 1^1 \nonumber \\
\lmm: m \mapsto t_2 \nonumber \\
\lview(7): x \mapsto 1^2, y \mapsto 1^1, z \mapsto 0^0 \nonumber \\
\lview(13): \bot \nonumber
\end{align*}
We note that the values of $x$ and $y$ in $\tview(t_2)$ are no longer
stale: the \ldrf\ semantics leverages race freedom to ensure that the values of
$x$ and $y$ are correct when they are read at line~$11$.

Roughly, we obtain \emph{sequential} data-flow abstractions of the
\ldrf\ semantics via the
following steps:
\begin{itemize}
\item Provide a data abstraction of sets of environments.
\item Define the state to be a map from locations to these abstract data values.
\item Compute the \scfg\ representation of the program by drawing inter-thread edges which connect releases and acquires of the same lock (as shown in the center of Fig.~\ref{fi:motEx}).
\item Define an abstract \emph{mix} operation which soundly approximates the ``import" step outlined earlier.
\item Analyze the program as if it was a sequential program, with \emph{inter}-thread join points (the $\mathtt{acquire}$'s) using the \emph{mix} operator.
\end{itemize}

The analysis in \cite{de2011dataflow} is precisely such a sequential
abstraction, where the abstract data values are abstractions of
\emph{value-sets} (variables mapped to sets of values). Value sets do
not track correlations between variables, and only allow coarse
abstractions like Intervals \cite{cousot1976static}. The \emph{mix}
operator, in this case, turns out to be the standard join (union of
value-sets). For the program of Fig.~\ref{fi:motExProg}, the interval
based value-set 
analysis, shown in the column ``Value-Set'' in Fig.~\ref{fi:motEx},
only manages to prove the assertion at line~$9$.

A more precise relational abstraction of \ldrf\, which we call \adrf\, 
can be obtained by keeping track of a set of environments at each
point. Fig.~\ref{fi:motEx} shows (in the column ``Rel'') the results of
such an analysis implemented using convex polyhedra
\cite{cousot1978polyhedra}.
The resulting analysis is more precise than the interval analysis, being
able to prove the assertions at lines $5$ and $9$. However, in this
case, the \emph{mix} must forget the correlations among variables in
the incoming states: it essentially treats them as value sets. This is
essential for soundness. Thus, even though the $\mathtt{acquire}$ at
line $10$ obtains the fact that $x = y$ from the buffer at $7$, and
the incoming fact from $9$ also has $x = y$, it fails to maintain this
correlation \emph{after} the \emph{mix}. Consequently, it fails to
prove the assertion at line $11$.

Finally, one can exploit the fact that $x$ and $y$ form a data
``region'' in that they are protected by the same lock. 
% are always accessed atomically by the two threads. 
The variable $z$ constitutes a region by itself. As we show in later in
Sec.~\ref{ch:rdrf}, the program is \emph{region race free} for this
particular region definition. One can parameterize the
\ldrf\ semantics with this region definition, to yield the
\rdrf\ semantics. The resulting analysis called \Reg\ maintains
relational information as in the \adrf\ analysis, 
but has a more precise \emph{mix} operator which preserves
relational facts that hold \emph{within} a region. Since both the
incoming facts at line $10$ satisfy $x = y$, the \emph{mix} preserves
this fact, and the analysis is able to prove the assertion at line
$11$.

Note that in all the three analyses, we are guaranteed to compute
sound facts for variables \emph{only} at points where they are
accessed. For example, all three analyses claim that $x$ and $y$ are
both $0$ at line~$9$, which is clearly wrong. However, we note that
$x$ and $y$ are not \emph{accessed} at this point.
This loss of soundness at ``irrelevant'' points helps us gain
efficiency in the analysis by not having to propagate all
interferences from one thread to \emph{all} points of another thread.
%% We make this trade-off for the soundness
%% guarantee in order to achieve a more efficient analysis. 
We also point out 
that in Fig. \ref{fi:motEx}, the inter-thread edges add a spurious
loop in the \scfg\ (and, therefore, in the analysis of the
program), which prevents us from computing an upper bound for the
values of $x$ and $y$. We show in Sec. \ref{se:ldrf-more-abs} how we can
appropriately abstract the versions to avoid some of these spurious
loops.

\section{Programming Language and Semantics}
\label{se:prelims}
In this section we introduce the programming language
we use to describe multi-threaded programs, and describe the standard
interleaving semantics for programs in this language.

\subsection{Preliminaries}

We begin by introducing some of the mathematical notation we will use
in this paper.
We denote the set of natural numbers $\{0, 1, \ldots,
\}$ by $\nat$.
We use $\totalto$ and $\partialto$ to denote total and partial
functions, respectively.
%and $\undefunc$ to denote a function which is not defined anywhere.
We use ``$\bot$" to denote an undefined value, which we assume is
included in every domain under consideration. 
%% We write $\bar{S}$ to
%% denote a (possibly empty) finite sequence of elements coming from a
%% set $S$. 
We denote the \emph{length} of a finite sequence of elements $\pi$ by $|\pi|$,
and the $i$-th element of $\pi$, for $0 \leq i < |\pi|$, by $\pi_i$.
For a function $f: A \rightarrow B$, we denote by $\domof{f}$ its
domain $A$, and for $a \in A$ and $b \in B$, we write $f[a\mapsto b]$
to denote the function $f':A \rightarrow B$ such that $f'(x) = b$ if
$x = a$, and $f(x)$ otherwise.
For a pair of elements $\ve = \langle\env,\rv\rangle$, we write
$\ve.1$ to denote the first component
$\env$, and $\ve.2$ to denote the second
component $\rv$, of the pair $\ve$.

We will make use of the standard notion of labelled transition systems
to describe the semantics we will give to our programs.
A \emph{Labelled Transition System} (LTS) is a structure
$\lts = (S, \Gamma, s_{0}, \rightarrow)$,
where $S$ is a set of \emph{states}, $\Gamma$ is a set of \emph{transition
labels}, $s_0 \in S$ is the \emph{initial state},
and $\rightarrow \subseteq S \times \Gamma \times S$ is the (labelled)
transition relation.
We sometimes write a transition $t = \abracks{s,l,s'}$ as
$s \rightarrow_l s'$.

An \emph{execution} of an LTS $\lts = (S, \Gamma,
s_{0}, \rightarrow)$, is
a finite sequence of transitions $\pi = t_1, t_2, \ldots,t_n$ ($n \geq
0$) from $\rightarrow$, such that there exists a sequence of states
$q_0, q_1, \ldots, q_{n}$ from $S$, with $q_0 = s_0$ and $t_i =
(q_{i-1},l_i,q_{i})$ for each $1 \leq i \leq n$.
Wherever convenient we will also represent an execution like $\pi$ above
as an interleaved sequence of the form
\[
q_0 \ltrto{l_1} q_1 \ltrto{l_2} \cdots \ltrto{l_n}
q_{n}.
\]

We also define $\reach(\lts)$ to be the set of states reachable by an
execution of $\lts$. Thus
\[
\reach(\lts) = \{ s \in S \ | \ \exists \mbox{ an execution } q_0
\ltrto{l_1} \cdots \ltrto{l_n} q_{n} \mbox{ with }
s = q_{n} \}.
\]

\subsection{Programming Language}
\label{Se:PL}

We consider a simple multi-threaded programming language where each
program has a fixed number of static threads. There is no dynamic
memory allocation, no dynamic creation of threads and no procedure calls.
A program has a finite number of variables $\vars$ and locks $\plocks$
which are shared by the threads of the program.
We denote by $\vals$ the set of values that the program
variables can assume. In this
work we will take $\vals$ to be simply the set of integers.

Each thread in the program is a control-flow graph in which each edge
is labelled by a basic statement (or command) over the set of
variables $\vars$ and locks $\plocks$.
We allow a small set of basic commands over $\vars$ and $\plocks$,
which we denote by $\cmd_{\vars, \plocks}$, as shown in
Tab.~\ref{tab:progCmd}.
For generality, we refrain from defining the syntax of the expressions
$e$ and boolean conditions $b$.

\begin{table}[!htb]
\begin{center}
\begin{tabular}{| l | l | l |}
\hline
\textbf{Type} & \textbf{Syntax} & \textbf{Description} \\ \hline
Assignment & $\assgn{x}{e}$ & Assigns the value of expression $e$ to variable $x \in \vars$\\ \hline
Assume & $\assume{b}$ & Blocks execution if condition $b$ does not hold
%Enabled only if boolean condition $b$ holds (only updates $\pc$)
\\ \hline
Acquire & $\acq{\lock}$  & Acquires lock $\lock \in \plocks$, provided $\lock$ is not held by any thread \\ \hline
Release & $\rel{\lock}$  & Releases lock $\lock \in \plocks$, provided the executing thread holds $\lock$   \\ \hline
\end{tabular}
\end{center}
\caption{The set of program commands $\cmd_{\vars, \plocks}$ over variables
  $\vars$ and locks $\plocks$}
\label{tab:progCmd}

\end{table}

Formally, we represent a multi-threaded program as a tuple $P =
(\vars, \plocks,\threads)$ where
\begin{itemize}
\item $\vars$ is a finite set of program \emph{variables}
\item $\plocks$ is a finite set of \emph{locks}
\item $\threads$ is a finite set of \emph{thread identifiers}. Each
  thread $t \in
  \threads$ has an associated control-flow graph of the
  form $\cfg_t = (\locs_t, \ent_t, \instr_t)$ where
  \begin{itemize}
  \item $\locs_t$ is a finite set of \emph{locations} of thread $t$
  \item $\ent_t \in \locs_t$ is the \emph{entry} location of thread $t$
  \item $\instr_t \subseteq \locs_t \times \cmd_{\vars,\plocks} \times \locs_t$ is a
    finite set of \emph{instructions} of thread $t$.
  \end{itemize}
\end{itemize}

Some definitions related to threads will be useful going forward.
We denote by $\locs_P = \bigcup_{t \in \threads}\locs_t$ the disjoint
union of the thread locations.
We denote by $\ent_P$ the set $\{\ent_t \ | \ t \in \threads\}$ of all
entry locations of $P$.
Henceforth, whenever $P$ is clear from the context we will drop the
subscript $P$ from $\locs_P$ and its decorations.
For a location $\ploc \in \locs$, we denote by $\tidof{\ploc}$ the
thread $t$ which contains location $\ploc$.
We denote the set of instructions of $P$ by $\instr_P = \bigcup_{t\in
  \threads} \instr_t$.
For an instruction $\inst \in \instr_t$, we will
also write $\tidof{\inst}$ to mean the thread $t$ containing $\inst$.
For an instruction $\inst = \abracks{\ploc_s,c,\ploc_t}$, we call $\ploc_s$
the \emph{source}
location, and $\ploc_t$ the \emph{target} location of $\inst$.
%We allow two instructions in a thread to share either their source or
%target locations, but not both.
% \marginpar{Why is this needed?}
We expect instructions
pertaining to % assignments,
$\acq{}$ and $\rel{}$ commands to have
unique source and target locations.
Let $\rellocs[t]$ be the set of
program locations in thread $t$ which are the target of a $\rel{}$
instruction.
We refer to $\rellocs[t]$ as $t$'s \emph{post-release points}
and denote the set of \emph{release points} in the program by $\rellocs
= \bigcup_{t \in \threads} \rellocs[t] $.
Similarly, we define $t$'s
\emph{pre-acquire points}, denoted  $\acqlocs[t]$, and denote a
program's \emph{acquire points} by $\acqlocs =  \bigcup_{t \in
  \threads} \acqlocs[t] $. We denote the sets of post-release and
pre-acquire points pertaining to operations on lock $\lock$ by
$\rellocs[\lock]$ and $\acqlocs[\lock]$, respectively.

We denote the set of commands appearing in program $P$ by
$\cmd(P)$. We consider an assignment $\assgn{x}{e}$ to be a
\emph{write-access} to $x$, and as a \emph{read-access} to every
variable that appears in the expression $e$.
Similarly, an $\assume{b}$ statement is considered a read-access to
every variable that occurs in the boolean condition $b$.
%% \marginpar{Why?}
%% Without loss of
%% generality, we assume variables appearing in conditions of $\assume{}$
%% commands in instructions of some thread $t$
%% do not appear in any instruction of any other thread $t'\neq t$.

We illustrate these definitions for the
example program from Fig. \ref{fi:motExProg}. Here $\vars =
\{\mathtt{x},\mathtt{y},\mathtt{z}\}$, $\plocks = \{\mathtt{m}\}$, and
$\threads = \{t_1, t_2\}$.
Some example instructions in
this program are $\abracks{2, x := y, 3}$ and $\abracks{10, \mathtt{acquire(m)},
    11}$. The set $\locs_{t_1}$ of program
locations in thread $t_1$, is $\{1,2,3,4,5,6,7\}$, while $\tidof{8} =
t_2$. In this program, the set $\rellocs[t_2]$ of post-release points
in $t_2$, is  $\{13\}$.
The set of post-release points of
the whole program $\rellocs$ is $\{7, 13\}$. The set of pre-acquire
points of the whole program $\acqlocs$ is $\{1,10\}$. Since this program
has a single lock, $\lock$, $\rellocs[\lock] = \{7, 13\}$ and
$\acqlocs[\lock] = \{1, 10\}$.

Many other standard commands can be expressed using the basic commands
in our language.
A \texttt{goto} instruction from program location $l$ to $l'$ can be
simulated by the instruction $\abracks{l, \mathtt{assume(true)}, l'}$.
Constructs like \texttt{if} and \texttt{while} can be simulated using
\texttt{assume} statements in a standard way.

\subsection{Interleaving Semantics}
\label{sec:intSemantics}

We now define the standard interleaving semantics of a multi-threaded program.
We first introduce some notation that will be useful in the sequel.
Given a program $P = (\vars, \plocks, \threads)$, an \emph{environment}
for $P$ is a valuation $\env : \vars \rightarrow \vals$, which assigns
values in $\vals$ to the variables of $P$.
We denote by $\Env_P$ the set % $\vars \rightarrow \vals$
of all environments for $P$.
A \emph{lock map} for $P$ is a partial map $\lmm: \plocks \partialto
\threads$ which assigns to each lock the thread that holds it (if such
a thread exists).
We denote by $\LM_P$ the set of lock maps for $P$.
Finally, a \emph{program counter} for $P$ is a map $\pc : \threads
\rightarrow \locs_P$ which assigns a location to each thread in $P$,
such that for each $t \in \threads$, $\pc(t) \in \locs_t$.
We denote by $\PC_P$ the set of program counters of $P$.
As usual, whenever $P$ is clear from the context we will drop the
subscript $P$ from these symbols.
Fig. \ref{Fi:SemDom}
summarizes the semantic domains, and the meta-variables ranging
over them\label{Fn:SemDom}, that we will make use of in this section
and subsequently.

\begin{figure}[!htb]
%\centering
\frame{
$
\begin{array}{c}
\begin{array}{lcl}
\begin{array}{rclcll}
x,y & \in & \vars & & & \text{Variable identifiers}  \\
\lock & \in & \plocks & & & \text{Lock identifiers}  \\
t & \in & \threads & & & \text{Thread identifiers} \\
n & \in & \locs & & & \text{Program locations}  \\
v & \in & \vals & & & \text{Values}  \\
r & \in & \regions & & & \text{Region identifiers}  \\
%\end{array}
%& \quad &
%\begin{array}{rclcll}
\env & \in & \Env & \equiv & \vars \to \vals & \text{Environments} \\
\lmm & \in & \LM & \equiv &  \plocks \partialto \threads   & \text{Lock map} \\
pc & \in & \PC & \equiv & \threads \to \locs & \text{Program counters} \\
\rv & \in & \RV & \equiv & \vars \to \nat & \text{Variable versions} \\
\ve & \in & \VE & \equiv & \Env \times \RV & \text{Versioned environments}
\\
\\
\end{array}
\end{array}\\[3pt]
\begin{array}{rclcll}
s = \langle \pc, \lmm, \env  \rangle & \in & \mathcal{S} & \equiv & \PC \times \LM \times \Env     & \text{Standard States}    \\
\cs = \langle \pc, \lmm, \tview, \lview \rangle & \in & \cS & \equiv & \PC \times \LM \times (\threads \to \VE) \times (\rellocs \to \VE) & \text{Thread-Local States}    \\
\end{array}
\end{array}
$
}
\caption{Some of the semantic domains associated with a program $P =
  (\vars, \plocks, \threads)$.}
\label{Fi:SemDom}
\end{figure}

Let us fix a program $P = (\vars, \plocks, \threads)$.
We define the interleaving semantics of $P$
% standard interleaving semantics \cite{lamport1997make} of $P$
using an LTS $\ltsi_P = (\stdS, \threads, \stds_\ent, \TR^\std_P)$
whose components are defined below.
The set of states $\stdS$ is $\PC \times \LM \times \Env$.
Thus each state is of the form $\abracks{\pc,\lmm,\env}$, where $\pc$ is a
program counter, $\lmm$ is a lock map, and $\env$ is an environment for
$P$.
The transition labels come from the set $\threads$ of thread
identifiers of $P$.
The initial state $\stds_\ent$ is
$\abracks{\lambda t.\,\ent_t,\lambda \lock.\bot,\lambda x.\,0}$.
Thus, in $\stds_\ent$,
every thread is at its entry program location, no thread holds
a lock,
% \footnote{We do this for simplicity. To handle fork and
% join, the initialization of locks will be defined as explained in
% Sec. \ref{Se:PL} and Fig. \ref{fi:locks-with-acqrel}}
and all the variables are initialized to zero.

\paragraph{Transition Relation.}
\label{Se:stdTR}
The transition relation $\TR^\std_P$ is the union of the transition
relations $\TR^\std_i$ induced by each instruction $\inst$ in
$\instr_P$. We elaborate on this below.

The transition relation for each instruction depends on the command
associated with it.
Intuitively, the semantics of the program commands are as follows. An
assignment $\assgn{x}{e}$ command updates the value of the variable $x$
according to the (possibly non-deterministic) expression $e$. An
$\assume{b}$ command generates transitions only from states in which
the (deterministic) boolean interpretation of the
condition $b$ is $\mtrue$. An $\acq{\lock}$ command executed by thread
$t$ sets $\lmm(\lock) = t$, provided the lock $\lock$ is not held by
any other thread. A $\rel{\lock}$ command executed by thread $t$ sets
$\lmm(\lock) = \bot$ provided $t$ holds $\lock$.
%A thread attempting to release a lock that it does not own gets blocked.

It will be convenient
to first define a notation for the evaluation of expressions.
The evaluation of an expression $e$, in  an environment $\env$, is a
value in $\vals$.
We denote this value by $\BB{e}\env$.
The interpretation of a boolean condition $b$, in an
environment $\env$, is a boolean value  $\mtrue$ or $\mfalse$, and
we denote this value by $\BB{b}\env$.

For an instruction $\inst = \abracks{n,c,n'}$ in $\instr_P$, with
$\tidof{\inst} = t$, we define $\TR^\std_\inst$ as the set of all transitions
  $\langle \abracks{\pc,\lmm,\env}, t, \abracks{\pc',\lmm',\env'}\rangle$ such that
  $\pc(t) = n$, $\pc' = \pc[t\mapsto n']$ and the following
additional conditions are satisfied:
\begin{itemize}
\item If $c$ is a command of the form $x := e$ then $\lmm' = \lmm$,
  and $\env' = \env[x \mapsto \BB{e}\env]$.
\item If $c$ is a command of the form $\assume b$ then $\lmm' = \lmm$,
  $\BB{b}\env = \mtrue$, and $\env' = \env$.
\item If $c$ is a command of the form $\acq{\lock}$ then $\lmm(\lock)
  = \bot$, $\lmm' = \lmm[\lock \mapsto t]$, and $\env' = \env$.
\item If $c$ is a command of the form $\rel{\lock}$ then $\lmm(\lock)
  = t$, $\lmm' = \lmm[\lock \mapsto \bot]$, and $\env' = \env$.
\end{itemize}

For a transition $\tau$ caused by an instruction $\inst = \abracks{n,c,n'}$
in $\instr_t$, we denote by $\tidof{\tau}$ the
thread $t$, by $\instrof{\tau}$ the instruction $\inst$,
 and by $\cmdof{\tau}$ the command $c$.

The transition relation $\TR^\std_P$ can now be defined as:
 $$
  \TR^\std_P = \bigcup_{\inst \in \instr_P} \TR^\std_\inst.
 $$

\paragraph{Executions.}
An \emph{execution} of the program $P$ in the interleaving semantics
is simply an execution of the LTS $\ltsi_P$.
When dealing with executions in the interleaving semantics,
we will denote the transition relation $\TR^\std_P$ by
$\ltrtos{}$.
We denote by $\reach^\std(P)$ the set $\reach(\ltsi_P)$, namely the
set of reachable states in the standard interleaving semantics of $P$.

\begin{figure}[!htb]
\centering
\includegraphics[scale=0.5]{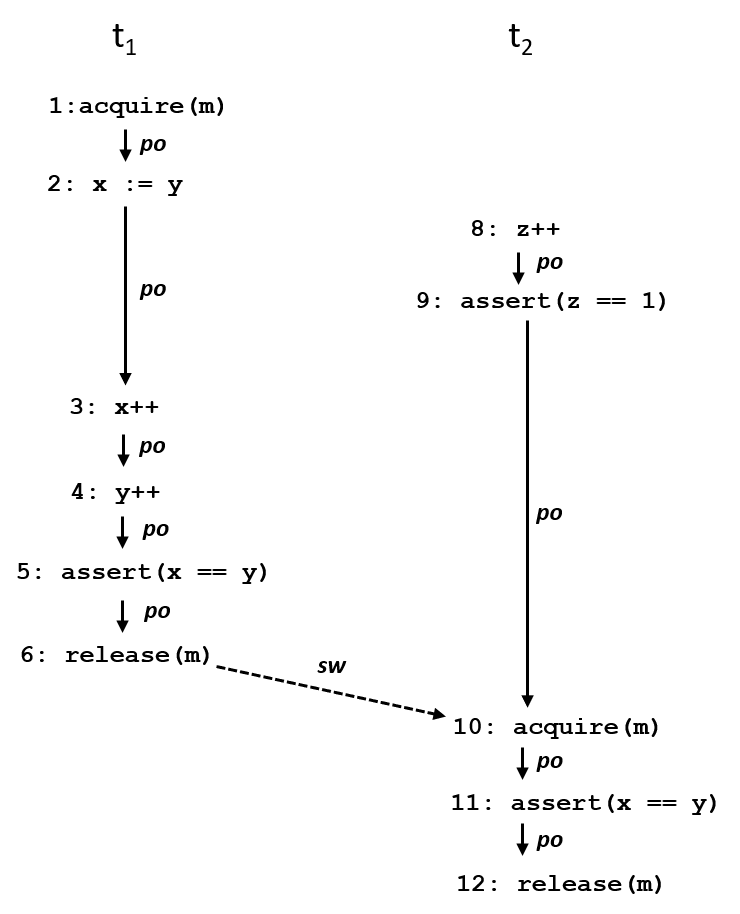}
\caption{\label{fi:standardExec} A typical execution of the program in
  Fig. \ref{fi:motExProg} with two threads, according to the standard
  interleaving semantics. Time flows from the top to the
  bottom. Instructions ordered by program-order are annotated as
  \emph{po}. The $\mathtt{release}$ executed by $t_1$ and the
  $\mathtt{acquire}$ executed by $t_2$ are related by
  synchronizes-with, and is annotated as \emph{sw}. The write of
  $\mathtt{x}$ in thread $t_1$, and its subsequent read in thread
  $t_2$, are connected by a happens-before path, comprising \emph{po}
  and \emph{sw} annotated edges. }
\end{figure}

Fig. \ref{fi:standardExec} depicts an execution of the program in
Fig. \ref{fi:motExProg} in the interleaving semantics.
To keep it simple we show only the sequence of program instructions
(from top to bottom), and the thread they belong to (column $t_1$ or
$t_2$). The states along the execution can be inferred by the standard
semantics of the commands. The other annotations in the figure will be
explained in Sec.~\ref{sec:dataraces}.

\subsection{Data Races and the Happens-Before Relation}
\label{sec:dataraces}

Now that we have formally defined the standard interleaving semantics,
we are in a position to formally define what constitutes a data
race. A standard way to  formalize the notion of \emph{data race
  freedom} (DRF), is to use the \emph{happens-before} relation
\cite{AdveH93,Lamport78} induced by executions.

For a given execution of the program $P$ in the standard interleaving semantics, the happens-before relation is defined as the reflexive and transitive closure of the \emph{program-order} and \emph{synchronizes-with} relations, formalized below.

\begin{definition}[Program order]\label{De:PO}
Let $\pi$ be an execution of $P$.  Transition $\pi_i$ is related to the
transition $\pi_j$, according to the \emph{program-order} relation in $\pi$,
denoted by $\pi_i \xrightarrow{po}_\pi \pi_j$, if
\[
j = \min \left\{ k \mid i < k < \card{\pi} \wedge \tidof{\pi_k}=\tidof{\pi_i} \right\}.
\]
That is, $\pi_i$ and $\pi_j$ are successive executions, in $\pi$, of instructions by the same
thread.\footnote{Strictly speaking, the various relations we define
  are between indices $\{0,\ldots,\card{\pi}-1\}$ of an execution, and
  not transitions, so we should have written, e.g.,
  $i\xrightarrow{po}_\pi j$ instead of $\pi_i\xrightarrow{po}_\pi
  \pi_j$. We use the informal latter notation, for readability.}
\end{definition}
The transitions related by program-order in Fig.~\ref{fi:standardExec} are marked with \emph{po}.

\begin{definition}[Synchronizes-with]\label{De:SW}
Let $\pi$ be an execution of $P$. Transition $\pi_i$ is related in $\pi$, by the \emph{synchronizes-with} relation, to the transition $\pi_j$, denoted by $\pi_i \xrightarrow{sw}_\pi \pi_j$, if $\cmdof{\pi_i}=\rel{{\lock}}$ for some lock $\lock$,
and
\[
j = \min \{ k \mid i < k < \card{\pi} \wedge \cmdof{\pi_k}=\acq{\lock} \}.
\]
That is, $\pi_i$ is a release of lock $\lock$ in $\pi$, and $\pi_j$ is a subsequent acquire of $\lock$, and there are no intervening acquires of $\lock$.

%$\pi_i$ and $\pi_j$ are successive release and acquire commands of the same lock in the execution.
\end{definition}
The transitions related by synchronizes-with in Fig.~\ref{fi:standardExec} are marked with \emph{sw}.

\begin{definition}[Happens before]\label{De:HB}
The \emph{happens-before} relation pertaining to an execution $\pi$ of $P$, denoted by $\cdot
\xrightarrow{hb}_\pi \cdot$, is the reflexive and transitive closure of the union of the program-order and synchronizes-with relations induced by the execution $\pi$.
\end{definition}

Note that transitions executed by the same thread are always related by program-order, and are thus always related according to the happens-before relation.

\begin{definition}[Data Race]\label{De:DR} Let $\pi$ be an execution of $P$. Transitions $\pi_i$ and $\pi_j$, in $\pi$, constitute a racing pair, or a data-race, if the following conditions are satisfied:
\begin{enumerate}
 \item $\cmdof{\pi_i}$ and $\cmdof{\pi_j}$ are \emph{conflicting}
   accesses to a variable $x$ (i.e.\@ they both access the variable
   $x$, and at least one of them is a write-access),
 and
  \item neither $\pi_i\xrightarrow{hb}_\pi \pi_j$ nor $\pi_j\xrightarrow{hb}_\pi \pi_i$ holds.
\end{enumerate}
\end{definition}

To illustrate these definitions, consider the execution of the program
of Fig.~\ref{fi:motExProg}, shown in Fig.~\ref{fi:standardExec}.
The program-order relation between transitions is
shown using edges marked \emph{po}, while the synchronizes-with
relation is shown using edges marked \emph{sw}. 
For example, the transitions where $t_1$ executes
\texttt{x := y} and the one where $t_1$ executes \texttt{x++} are
related by program-order. The transition where $t_1$ releases the lock
$\lock$, and the subsequent transition where $t_2$ acquires $\lock$,
are related by the synchronizes-with relation. There is a
happens-before path, namely the path comprising \emph{po} and
\emph{sw} annotated edges in Fig.~\ref{fi:standardExec}, between the
write to $\mathtt{x}$ by $t_1$, and the subsequent read of
$\mathtt{x}$ by $t_2$. %This path is made up of pairs of transitions
                       %which are either related by program-order, or
                       %synchronizes-with. 
Note that even though the instruction \texttt{x := y} is executed by $t_1$ before $t_2$ executes \texttt{z++} in the execution in Fig.~\ref{fi:standardExec}, these two instructions are \emph{not} related by happens-before. Consider, for a moment, if $t_2$ did \emph{not} have the \texttt{acquire($\lock$)} instruction. Then, the transitions made by $t_1$ could never be happens-before related to the ones in $t_2$ (due to the absence of \emph{sw} edges). In particular, the write to $\mathtt{x}$ by $t_1$ would \emph{not} be happens-before ordered with the read of $\mathtt{x}$ in $t_2$, and we would have a data race in the execution.

A program in which every execution is free from data races is said to
be \emph{data race free}. The program in Fig.~\ref{fi:motExProg} is an
example of such a race free program.
% The focus of this work is on the class of race free programs.

We say an instruction $\inst$ in $P$ is \emph{racy} if there is an execution
$\pi$ of $P$ in which two transitions $\pi_i$ and $\pi_j$ are involved
in a race and $\instrof{\pi_i} = \inst$.

We can now define the notion of the set of variables ``owned'' by a
thread at one of its locations.
We say variable $x$ is \emph{owned} by a thread $t$ at a location $n
\in \locs_t$, in program $P$, if the introduction of a read of $x$ at
location $n$ is not racy. In other words, if we introduce the
instruction $\inst$ with command $\assume{x == x}$ at point $n$ in
$t$, to get the
program $P'$, then instruction $\inst$ is not racy in $P'$.
For example, in the program of Fig.~\ref{fi:motExProg}, at location~3,
thread $t_1$ owns the variables $x$ and $y$.
However it does \emph{not} own the variable $z$ at location~3, since a
read of $z$ introduced at this point would be racy (it would race with the
write to $z$ at line~8 in $t_2$).

%%% Local Variables:
%%% mode: latex
%%% TeX-master: "fmsd.tex"
%%% End:

%%% Local Variables:
%%% mode: latex
%%% TeX-master: "fmsd.tex"
%%% End:

\section{The Thread-Local Semantics \ldrf}
\label{sec:ldrf}

In this section, we introduce a novel semantics for the class of data
race free programs, which we refer to as the \ldrf\ semantics
\cite{sas17-ldrf}. 
The ``\local'' highlights the fact that the semantics is
thread-\emph{local} in nature, while \textit{DRF} emphasizes that we
deal exclusively with data race free programs. The \ldrf\ semantics
paves the way towards devising efficient ``thread-local" data flow
analyses for race free concurrent programs. Like the standard
interleaving semantics we saw in Sec.~\ref{sec:intSemantics}, we
present the \ldrf\ semantics of a program as a labeled transition
system. We then prove that the \ldrf\ semantics is sound and complete
with respect to the standard semantics, in the sense that for each
execution of the program in the standard semantics, there is an
``equivalent" execution in the \ldrf\ semantics, and vice versa.

\subsection{The \ldrf\ Semantics}
\label{subsec:ldrf}

Our thread-local semantics, like the standard one defined in Sec.~\ref{sec:intSemantics}, is based on the interleaving of transitions made by different threads, and the use of a lock map to coordinate the use of locks. However, \emph{unlike} the standard semantics, where  the threads share access to a single \emph{global} environment, in the \ldrf\ semantics, every thread has its own \emph{local} environment which it uses to evaluate conditions and perform assignments.

Threads exchange information through \emph{release buffers}: every post-release point $\ploc \in \rellocs[t]$ \footnote{Recall that $\rellocs[t]$ is the set of all post-release points in the thread $t$.} of a thread $t$ is associated with a \emph{buffer} $\lview(\ploc)$ which records a snapshot of $t$'s local environment the last time $t$ ended up at the program point $\ploc$. Recall that this  happens right after $t$ executes the instruction $\langle n', \mathtt{release(\lock)}, n \rangle \in \instr_P$. When a thread $t'$ subsequently acquires the lock $\lock$, it updates its local environment using the snapshots stored in all the buffers pertaining to the release of $\lock$.

To ensure that $t$ updates its environment such that the value of
every variable is up-to-date, every thread maintains its own
\emph{version map} $\rv:\vars \totalto \nat$, which associates a count
to each variable. A thread increments $\rv(x)$ whenever it writes to
$x$. Along any execution, the version $\rv(x)$, for $x \in \vars$, in
the version map $\rv$ of thread $t$, associates a unique prior write
with this particular valuation of $x$. It also reflects the total
number of write accesses made (across threads) to $x$ to obtain the
value of $x$ stored in the map. A thread stores both its local
environment and version map in the buffer after releasing a lock
$\lock$. When a thread subsequently acquires lock $\lock$, it copies
from the release buffers at $\rellocs[\lock]$ \footnote{Recall that
  $\rellocs[\lock]$ is the set of all post-release points in the
  program associated with the release of lock $\lock$.} the most
up-to-date value (according to the version numbers) of every
variable. We prove that for data race free programs, there can be only
one such value. If the version of $x$ is the local state of $t$ is
higher than the versions of $x$ in the associated release buffers,
then the value of $x$ in the local state persists.

Let us fix a concurrent race free program $P = (\vars,
\plocks, \threads)$. As in Sec.~\ref{sec:intSemantics}, we define the
\ldrf\ semantics of $P$ in terms of a labeled transition system $\ltsl_P
= (\cS, \threads, \cs_\ent, \TR^\local_P)$ whose components we define
below. 
%% Note that all the semantic domains are implicitly parameterized with
%% the program $P$ under consideration. For example, once we fix $P$, the
%% set of environments $\Env$ maps the program variables in $P$ to
%% values, the function $\threads$ maps the threads in $P$ to their entry
%% program location, $\cs_\ent$ maps the program variables of $P$ to
%% their initial values, and so forth. We omit explicitly stating the
%% program $P$ for each domain for clarity.

\paragraph{States.} A \emph{state} $\cs \in \cS$ in the
\ldrf\ semantics of $P$ 
is a tuple $\langle \pc, \lmm, \tview,\lview \rangle$, where $\pc$ and
$\lmm$ are the program counter and lock map, as in the standard
interleaving semantics (Sec.~\ref{sec:intSemantics}). A
\emph{versioned environment} is a pair $\abracks{\env, \rv}$, where $\env
\in \Env$ is an environment and $\rv\,:\,\vars \totalto \nat$ is a
version map, which assigns a version count to each variable.
We denote by $\VE_P$ (or just $\VE$ when $P$ is clear from the
context) the set of versioned environments of program $P$.
%\[
%\ve = \abracks{\env,\rv}\in\VE =\Env \times (\vars \totalto \nat)
%\]
%
%is a pair comprising an environment $\env$ and a version map $\rv$. For a given versioned environment $\ve$, the environment $\ve\cdot\env$ is a valuation of the program variables. The version map $\ve \cdot \rv$ assigns a ``version count" to each variable. The version count of a variable $x$ is incremented by a thread $t$ whenever it writes to $x$. The version counts, as we explain in more detail later, are used by a thread to ensure it has the most up-to-date value and version of the variables in its local state.
The local environment map $\tview:\threads \totalto \VE $ maps every
thread to a local versioned environment, and the release buffer map
$\lview:\rellocs \totalto 
\VE $ records the snapshots of versioned environments stored in
buffers associated with post-release points.

\paragraph{Initial State.} The initial state $\cs_\ent$ is defined to be

\[\cs_\ent=\abracks{\lambda t.\,\ent_t,\lambda \lock.\,\bot,\lambda t.\,\ve_\ent,\lambda l \in \rellocs.\, \ve_\ent}\]

where $\ve_\ent = \abracks{\lambda x.0,\lambda x.0}$.
Thus, in $\cs_\ent$, every thread is at its entry program
location, no thread holds a lock, and
all the thread-local versioned environments have all the variables
and versions initialized to $0$. The release buffers are also
initialized to the versioned environment where all variable values
and versions are $0$.

\paragraph{Transition Relation.} The transition relation $TR^\local_P
\subseteq \cS \times \threads \times \cS$ captures the
interleaving nature of the \ldrf\ semantics of $P$. Like the
interleaving semantics in Sec.~\ref{sec:intSemantics}, $TR^\local_P$ is
the union of the transition relations $TR^\local_\inst$ induced by each
instruction $\inst \in \instr_P$.
%% A transition $\cs\ltrto{t}\cs'$ says that thread $t$ can execute
%% an instruction which transforms state $\cs \in \cS$ to state
%% $\cs'\in\cS$. As is the case in the standard semantics, in these
%% transitions, one thread executes a command and changes its program
%% counter accordingly, while all other threads remain stationary.

For an instruction $\inst = \abracks{n,c,n'}$ in $\instr_P$, with
$\tidof{\inst} = t$, we define $\TR^\local_\inst$ as the set of all transitions
  $\langle \abracks{\pc,\lmm,\tview, \lview}, t, \abracks{\pc',\lmm',\tview',\lview'}\rangle$ such that
  $\pc(t) = n$, $\pc' = \pc[t\mapsto n']$ and the following
additional conditions are satisfied:
\begin{itemize}
\item \textit{Assignment.}
  If $c$ is a command of the form $x := e$ then $\lmm' = \lmm$,
  and $\tview' = \tview[t \mapsto \abracks{\env',\rv'}]$,
  where $\env'$ and $\rv'$ are given as follows.
  Let $\tview(t) = \abracks{\env,\rv}$. Then
  $\env' = \env[x \mapsto \BB{e}\env]$, and
  $\rv' = \rv[x \mapsto \rv(x)+1]$.
  For subsequent use, we define the interpretation of an
  assignment statement $x:=e$ on a versioned environment
  $\abracks{\env,\rv}$, denoted
  $\BB[\local]{x:=e}(\abracks{\env,\rv})$,
  to be $\abracks{\env',\rv'}$,
  where $\env' = \env[x \mapsto \BB{e}\env]$ and
        $\rv' = \rv[x \mapsto \rv(x)+1]$.

\item \textit{Assume.}
  If $c$ is an assume statement of the form $\assume{b}$, then $\lmm' = \lmm$,
  and $\tview' = \tview$, $\lview' = \lview$, and
  $\BB[\local]{b}(\tview(t))$ is true.
  Here by $\BB[\local]{b}\abracks{\env,\rv}$ we simply mean $\BB[]{b}\env$.

  We note that for instructions which execute either assignment or
assume commands, the executing thread accesses and modifies only its
\emph{own} local versioned environment.

\item \textit{Acquire.} An $\acq{\lock}$ command, executed by a
  thread $t$, has the same effect on the lock map component as in
  the standard semantics (see Sec.~\ref{sec:intSemantics}). In
  addition, it updates the versioned environment $\tview(t)$ based
  on the contents of the \emph{relevant} release buffers. The
  release buffers relevant to a thread when it acquires $\lock$
  are the ones at $\rellocs[\lock]$.
  %DD: Removing this as it introduces unnecessary complication at
  % this point.
  %% Conversely, for any
  %% post-release point $n \in \rellocs[\lock]$, we use the symbol
  %% $\rfGsymb{}(n)$ to denote the set of pre-acquire points which
  %% can observe the buffer $\lview(n)$. In other words,
  %% $\rfGsymb{}(n)$ are the set of pre-acquire points for which
  %% $\lview(n)$ is relevant. For our purposes, for any $n \in
  %% \rellocs[\lock]$, $\rfGsymb{}(n) = \acqlocs[\lock]$ \footnote{In
  %%   Remark \ref{re:rfGsymb}, we show that it is possible to refine
  %%   the set $\rfGsymb{n}$ such that the resulting \ldrf\ semantics
  %%   is still equivalent to the interleaving semantics, but the
  %%   abstract analyses derived from such an \ldrf\ semantics has
  %%   increased precision}.

  We define an auxiliary function
  $\mathit{updEnv}$ to update the value of each $x \in
  \vars$ (along with its version) in $\tview(t)$, by taking its
  value from a snapshot stored at a relevant buffer which has the
  highest version of $x$, if the latter version is higher than
  $(\tview(t).2)(x)$. If the version of $x$ is highest in
  $(\tview(t).2)(x)$, then $t$ simply retains this
  value. Finding the most up-to-date (value, version) pairs for a
  variable $x$ from a set of versioned environments
  is the job of the auxiliary function $\take{x}$.
  %% It takes as input
  %% $\tview(t)$, as well as the versioned environments in the
  %% relevant release buffers, and returns the versioned environments
  %% for which the version associated with $x$ is the highest. 
  We will separately prove (in Lemma~\ref{lem:admissibleStates}) that
  all \emph{reachable}
  \ldrf\ states are \emph{admissible} in that in any two component
  versioned environments (i.e.\@ the thread local versioned
  environments or release buffers of the state), if the versions
  for a variable coincide, then so must their values.
  Thus if $\abracks{\env,\rv}$ and $\abracks{\env',\rv'}$ are two
  versioned environments in the components of a reachable state,
  then for each variable $x$, $\rv(x) = \rv'(x) \implies \env(x) =
  \env'(x)$.

  \medskip
  Given a set of versioned environments $Y$, we define
  $\take{x}(Y)$ to be the set of (value,version) pairs $\abracks{v,m}$
  such that there exists a versioned environment
  $\abracks{\env,\rv}$ in $Y$ with $\env(x) = v$ and $\rv(x) = m$,
  and $m$ is the \emph{highest} version of $x$ among the versioned
  environements in $Y$ (i.e.\@ $\rv(x) \geq \rv'(x)$ for each
  $\abracks{\env',\rv'}$ in $Y$).
  
  Given a versioned environment $\ve$ and a set of
  versioned environments $X$, we define
  $\mathit{updEnv}(\ve,X)$ to be the set of versioned environments
  $\abracks{\env',\rv'}$ such that for each variable $x \in
  \vars$, $\abracks{\env'(x),\rv'(x)} \in
  \take{x}(\{\ve \} \cup X)$.

  \medskip
  We can now define the transition induced by an acquire command.
  If $c$ is an aquire statement of the form $\acq{\lock}$,
  then $\lmm[\lock] = \bot$,
  $\lmm' = \lmm[\lock \mapsto t]$, 
  $\tview' = \tview[t \mapsto \ve']$, and
  $\lview' = \lview$,
  where $\ve' = \mathit{updEnv}(\tview(t),\lview^\lock)$
  and $\lview^\lock = \{\lview(n'') \ | \ n'' \in \rellocs[\lock] \}$
  is the set of versioned environments relevant to $\lock$.
  
\medskip
  As an example, consider again the execution of the program of Fig.~\ref{fi:motExProg}, as shown in Fig.~\ref{fi:standardExec}. When thread $t_2$ executes the $\mathtt{acquire(\lock)}$ instruction, the condition of the relevant buffers and the thread local state of $t_2$ is shown in Fig.~\ref{fi:takeUpdEnv}. The figure also outlines the operation of the functions $\take{x}$, $\take{y}$ and $\take{z}$, and finally the operation of the function $\mathit{updEnv}$.

\begin{figure}[!htb]
\centering
\includegraphics[width=0.7\textwidth]{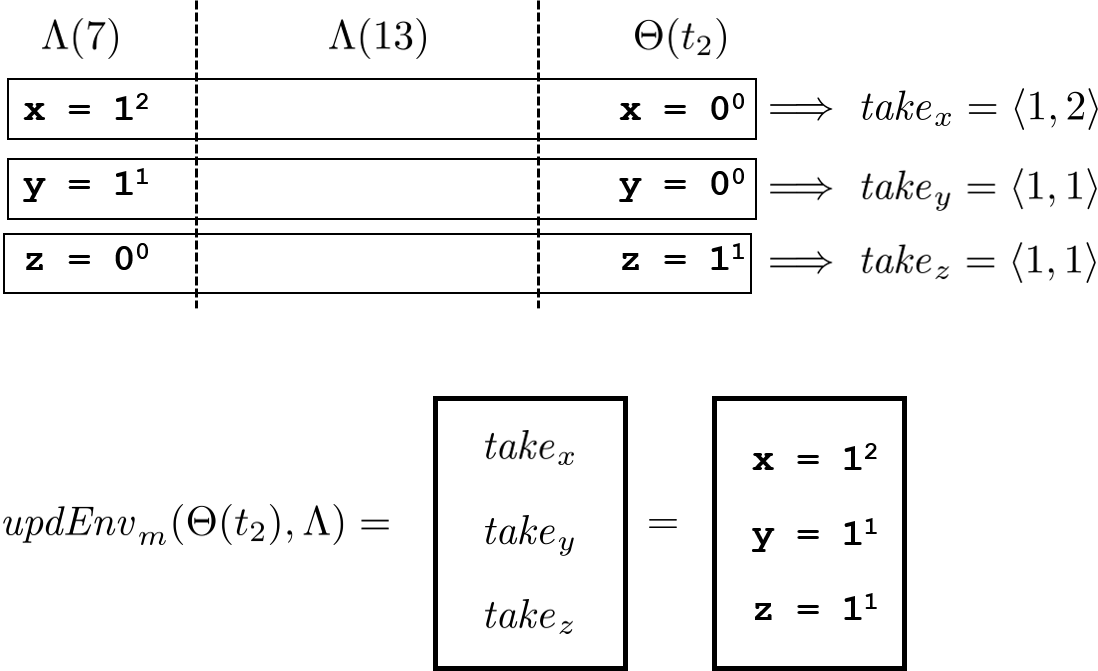}
\caption{\label{fi:takeUpdEnv} Operation of the functions $\take{x}$, $\take{y}$, $\take{z}$, and $\mathit{updEnv}$ when $t_2$ acquires $\lock$ in the execution of the program of Fig.~\ref{fi:motExProg}, as shown in Fig.~\ref{fi:standardExec}. The superscripts indicate the versions.}
\end{figure}

\item \textit{Release.}
If $c$ is a release statement of the form $\rel{\lock}$,
then $\lmm[\lock] = t$,
     $\lmm' = \lmm[\lock \mapsto \bot]$, 
  $\tview' = \tview$, and
  $\lview' = \lview[n' \mapsto \tview(t)$.

Thus an instruction $\inst$ pertaining to a $\rel{\lock}$ command
has the same effect on the lock map component of the state in the
\ldrf\ semantics that it has in the standard semantics (See
Sec.~\ref{sec:intSemantics}). In addition, it stores the local
versioned environment of thread $t$ $(=\tidof{\inst})$,
$\tview(t)$, in the buffer associated with the post-release point
of the executed $\rel{\lock}$ instruction.

\end{itemize}

\noindent The transition relation $TR^\local_P$ of program $P$ according to
the \ldrf\ semantics, is the union of the set of all possible
transitions generated by its instructions. Formally,
 \[
 \TR^\local_P = \bigcup_{\inst \in \instr_P} \TR^\local_\inst.
 \]

\medskip
This completes the description of the labelled transition system
$\ltsl_P$ capturing the \ldrf\ semantics.
An execution of program $P$ in the \ldrf\ semantics is simply an
execution of the transition system $\ltsl_P$.
When dealing with executions in the \ldrf\ semantics, we will denote
the transition relation $\TR^\local_P$ by $\ltrtol{}$.
We denote by $\reach^{\local}(P)$ the set of reachable states in this
semantics, namely $\reach(\ltsl_P)$.
%\paragraph{\textbf{Executions.}} An execution $\rfpi$
%$\pi\in\overline{\TR^\std_P}$
%of the concurrent program $P$ in the \ldrf\ semantics is a finite sequence of transitions coming from
%its transition relation $TR_P$, such that
%$\cs_\ent$ is the source of transition $\rfpi_0$, and the source state of every transition $\rfpi_i$,
%for  $0 < i < |\rfpi|$, is the target state of transition $\rfpi_{i-1}$.
 % holds that $\trgof{\pi_{i-1}}=\srcof{\pi_i}$.
%Where convenient, we also write executions as sequences of states interleaved with thread identifiers:
%\[
%\rfpi = \cs_0 \ltrto{t_1} \cs_1 \ltrto{t_2} \ldots \ltrto{t_n} \cs_n\ .
%\]
%where $\cs_i = \abracks{\pc_i, \lmm_i, \tview_i, \lview_i}$.

\subsection{Soundness and Completeness of \ldrf}
\label{Se:drfSemProof}

In this section, we show that for the class of data race free
programs, the thread local semantics \ldrf\ is sound and complete with
respect to the standard interleaving semantics. Intuitively, the
\ldrf\ and the standard semantics are ``equivalent" in the sense that
for each
execution of a program $P$ in the standard semantics, one can find a
corresponding execution in the \ldrf\ semantics which coincides with
the values read from the variables. Likewise, every execution of
program $P$ in the \ldrf\ semantics has a corresponding execution in
the standard semantics. 

Let us fix a race free program $P = (\vars, \plocks, \threads)$.
To formalize the above claim, we first define a function which extracts a
state in the interleaving semantics from a state in the
\ldrf\ semantics.

\begin{definition}[Extraction Function $\f$]\label{De:ExtractionFunction}
The extraction function $\f:\cS \partialto \stdS$ is defined for
\emph{admissible} states (see Sec.~\ref{subsec:ldrf}) in $\cS$ as follows:
\[
\f(\abracks{\pc, \lmm, \tview, \lview}) = \abracks{\pc, \lmm,
  \env},
\]
where $\env$ is defined as follows.
For each $x \in \vars$, $\env(x) = v$, provided there exists a
version value $m$, with $\abracks{v,m} \in \take{x}(\bigcup_{t\in
  \threads}\{\tview(t)\})$.
The function $\f$ thus preserves the values of the program
counters and the lock map, while it takes the value of a
variable $x$ from the thread which has the maximal version count
for $x$ in its local environment.
The map $\f$ is clearly well-defined for \emph{admissible} states.
%% where if $\tview(t)\cdot\rv(x) = \tview(t')\cdot\rv(x)$, then
%% $\tview(t)\cdot\env(x) = \tview(t')\cdot\env(x)$.
%We denote the set of admissible states by $\tilde\cS$.
%% As we prove later in Lemma~\ref{lem:admissibleStates}, the
%% \ldrf\ semantics only produces admissible states.
\end{definition}

The function $\f$ can be extended to executions in the
\ldrf\ semantics, in the following sense.
Given an execution
$\rfpi = \cs_0 \ltrtol{t_1} \ldots \ltrtol{t_n} \cs_n$
of program $P$ in the \ldrf\ semantics, and
an execution
$\pi = \stds_0 \ltrtos{t_1} \ldots \ltrtos{t_l} \stds_l$ of $P$ in
the standard semantics, we say $\pi = \f(\rfpi)$ if
$l = n$ and for each $i: 0 \leq i \leq n$, $\stds_i = \f(\cs_i)$.
%% \[
%% \f\left(\rfpi \right) \eqdef \f\left(\cs_\ent \right) \ltrto{t_1} \ldots \ltrto{t_n} \f\left(\cs_n \right)
%% \]

%% We prove in Theorem~\ref{thm:completeness} that the sequence $\f\left(\rfpi \right)$ is indeed a valid trace of $P$ in the interleaving semantics. The following theorems state our soundness and completeness results.

\begin{theorem}[Completeness]
\label{thm:completeness}
For any execution $\pi$ of $P$ in the
standard interleaving semantics, there exists an execution $\rfpi$ of
$P$ in the \ldrf\ 
semantics such that $\f(\rfpi) = \pi$.

%Moreover, for any transition $\pi_i$, if $c(\pi_i)$ involves a read of variable $x \in \vars$, then $
%\stds_{i-1}\env(x) = \cs_{i-1}\tview(t_i)\env(x)$. In other words, in $\rfpi$, the valuation of a variable $x$ in the local environment of a thread $t$ coincides with the corresponding valuation in the standard semantics only at points where $t$ reads $x$.
\end{theorem}

\begin{theorem}[Soundness]
\label{thm:soundness}
For any execution $\rfpi$ of $P$ in the \ldrf\ semantics,
there is an execution $\pi$ in the standard interleaving semantics
of $P$, with $\pi = \f(\rfpi)$.
\end{theorem}

In order to prove Theorem~\ref{thm:completeness} and Theorem~\ref{thm:soundness}, we need to establish a few intermediate results.

\begin{lemma}
\label{lm:max-version}
In any execution $\rfpi$ in the \ldrf\ semantics of $P$, the version of any variable $x \in \vars$, in any component versioned environment of any state $\cs$ in $\rfpi$, is bounded by the total number of writes to $x$ preceding it.
\end{lemma}

\begin{proof}
In $\rfpi$, the only transitions which can increment the version of variable $x$ pertain to instructions containing commands which write to $x$, of the form \texttt{x:=e}. Instructions containing other commands ($\mathtt{assume}$, $\mathtt{acquire}$ and $\mathtt{release}$) only make copies of existing version counts. If there are $n$ such transitions containing instructions writing to $x$ in $\rfpi$, and the initial version count of $x$ is $0$ in all the component versioned environments of the initial state $\cs_\ent$, the version of $x$, in any component versioned environment of any state $\cs$ in $\rfpi$ can be at most $n$. \qed
\end{proof}

\begin{lemma}
\label{lm:hb-persist-version}
Let
\begin{math}
\rfpi =
\abracks{\pc_0, \lmm_0, \tview_0, \lview_0}
\ltrtol{t_1}
\ldots
\ltrtol{t_{N}}
\abracks{\pc_{N}, \lmm_{N}, \tview_{N}, \lview_{N}}
\end{math}
be an execution in the \ldrf\ semantics of program $P$. Let
\[\tau_j = \abracks{\pc_{j-1}, \lmm_{j-1}, \tview_{j-1}, \lview_{j-1}} \ltrtol{t_{j}} \abracks{\pc_{j}, \lmm_{j}, \tview_{j}, \lview_{j}}
\]
be a transition in $\rfpi$ which contains an access (read or
write) to the variable $x$.
Suppose there is a prior write to $x$ in $\rfpi$, and let
\[
\tau_i = \abracks{\pc_{i-1}, \lmm_{i-1}, \tview_{i-1}, \lview_{i-1}} \ltrtol{t_{i}} \abracks{\pc_{i}, \lmm_{i}, \tview_{i}, \lview_{i}}
\]
be the last transition, prior to $\tau_j$, which contains an assignment to $x$. Then,
\[
(\tview_{j-1}(t_j).2)(x) \ge (\tview_i(t_i).2)(x).
\]
In other words, the version of $x$ in $\tview(t_j)$ is no less than the version of $x$ in the local state of $t_i$ post the write at $\tau_i$.
\end{lemma}

\begin{figure}[!htb]
\centering
\includegraphics[scale=0.3]{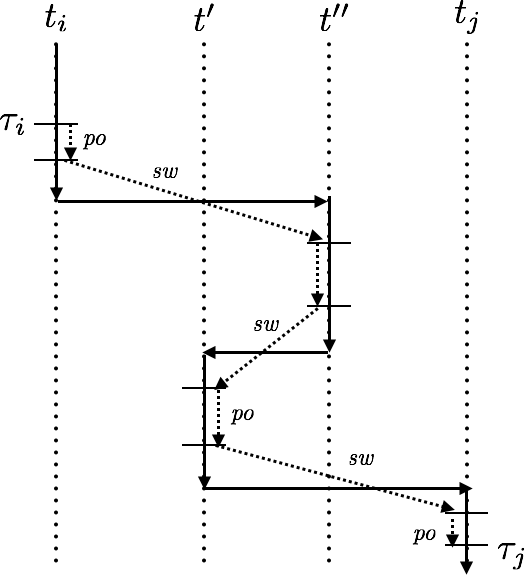}
\caption{\label{fi:exec_hb}A typical execution of a program $P$ in
  the \ldrf\ semantics. The solid arrows represent the interleaved
  execution of the instructions from different threads. The dotted
  arrows denote the happens-before path induced by this
  execution. The figure marks the sections of the happens-before
  path which are program-order related (\emph{po}), and the
  transitions related by synchronizes-with (\emph{sw}).}
\end{figure}

\begin{proof}
Fig.~\ref{fi:exec_hb} provides a pictorial description of the
situation we are considering.
We lift the notion of a happens-before path, which we defined
for the interleaving semantics, in a natural way to
\ldrf\ executions. The sequence of transitions in $\rfpi$ can also
be viewed as an standard execution, and the resulting
happens-before path in $\rfpi$ contains the same sequence of
transitions as the happens-before path in the execution in the
standard interleaving semantics.
Since $\tau_i$ and $\tau_j$ are
conflicting accesses to the variable $x$, and since the program
$P$ is assumed to be free from races, we have $\tau_i 
\xrightarrow{hb}_{\rfpi} \tau_j$ (indicated by the path comprising
dotted arrows in Fig.~\ref{fi:exec_hb}).

Let $\rho$ be such a happens-before path between $\tau_i$ and
$\tau_j$, excluding both $\tau_i$ and $\tau_j$.
If $\rho$ is of 0 length, then $\tau_j$ must immediately follow
$\tau_i$ in the same thread, and the lemma clearly holds.
Suppose $\rho$ is of length at least one, and consider a
transition $\tau_k$ in $\rho$.
By induction on the position $n$ of $\tau_k$ in $\rho$, we claim
that $(\tview_k(t_k).2)(x) \ge (\tview_i(t_i).2)(x)$.
%% \underline{Claim}: For each post-state $\cs_k$ of a transition
%% $\tau_k$, other than $\tau_j$, in some happens-before path $\rho$
%% between $\tau_i$ and $\tau_j$, we must have
%% $(\tview_k(t_k).2)(x) \ge (\tview_i(t_i).2)(x)$. We prove
%% the claim using induction on the position $n$ of a transition in
%% $\rho$. \\ 

\textit{Base Case.} If $n=1$, then $\tau_i$ and $\tau_k$ must be
related by program order, which implies $t_i =  t_k$ and $k =
i+1$.
Thus clearly $(\tview(t_k).2)(x) \geq (\tview(t_i).2)(x)$.

\textit{Inductive Case.} Assume that the hypothesis holds for all
transitions at positions less than or equal to $n$ in $\rho$,
and let us suppose $\tau_k$ occurs at position $n+1$ in $\rho$.
Let the $n$-th transition in $\rho$ be
\[
\tau_u = \abracks{\pc_{u-1}, \lmm_{u-1}, \tview_{u-1},
  \lview_{u-1}} \ltrtol{t_{u}} \abracks{\pc_{u}, \lmm_{u},
  \tview_{u}, \lview_{u}}.
\]

There are two possible cases here. Either $\tau_u
\xrightarrow{po}_{\rfpi} \tau_{k}$, and consequently $t_u =
t_k$.
In this case too, clearly $(\tview(t_k).2)(x) \geq
(\tview(t_u).2)(x)$, which, by the induction hypothesis,
is greater than or equal to $(\tview(t_i).2)(x)$.
Hence this case is taken care of.

On the other hand, if  $\tau_u \xrightarrow{sw}_{\rfpi} \tau_k$, then
$\tau_u$ must be the $\mathtt{release}$ of some lock $\lock$,
and $\tau_k$ must be the $\mathtt{acquire}$ of $\lock$. By
the \ldrf\ semantics of $\mathtt{acquire}$, thread $t_k$ will
observe the buffer associated with the $\mathtt{release}$ command
of $\tau_u$. Consequently,
$(\tview_k(t_k).2)(x) \ge (\tview_u(t_u).2).(x)$, by the semantics
of the $\mathtt{acquire}$ command and
$(\tview_{u}(t_{u}).2)(x) \ge (\tview_i(t_i).2)(x)$, by the
induction hypothesis.
Thus, the hypothesis holds in this case as well. This proves the
claim.

The lemma now follows directly from the claim.
%% Returning to the proof of the lemma, let $\tau' = \abracks{\pc', \lmm', \tview', \lview'}$ be the last transition, before $\tau_j$, in the happens-before path between $\tau_i$ and $\tau_j$. Since $\tau_j$ is not a synchronization operation, it must be the case that $\tau' \xrightarrow{po}_{\rfpi} \tau_j$, which implies $t' = t_j$. By the earlier claim, $\tview'(t_j)\cdot \rv(x) \ge \tview_i(t_i)\cdot \rv(x)$, and the absence of any intervening instructions, between $\tau'$ and $\tau_j$, which can alter the version of $x$, we infer that $\tview_{j-1}(t_j)\cdot \rv(x) \ge \tview_i(t_i)\cdot \rv(x)$.
\qed
\end{proof}

\begin{lemma}
\label{lm:max-write}
Let
\begin{math}
\rfpi = \abracks{\pc_0, \lmm_0, \tview_0, \lview_0}
\ltrtol{t_1}
\ldots
\ltrtol{t_{N}}
\abracks{\pc_{N}, \lmm_{N}, \tview_{N}, \lview_{N}}
\end{math}
be an execution in the \ldrf\ semantics of program $P$, and
let the $i$-th transition in the execution be
\[
\tau_i = \abracks{\pc_{i-1}, \lmm_{i-1}, \tview_{i-1}, \lview_{i-1}} \ltrtol{t_{i}} \abracks{\pc_{i}, \lmm_{i}, \tview_{i}, \lview_{i}}.
\]
Consider a transition $\tau_k$ with $\cmd(\tau_k)$
\footnote{By abuse of notation we use
  $\cmd(\tau)$ to denote the command of the instruction $\inst$
  causing the transition $\tau$.}
being an assignment to a variable $x$.
Then
\[
(\tview_k(t_k).2)(x) = \card{
  \{ i \ : \ i \leq k \mathrm{\ and \ } \cmd(\tau_i)
  \text{ is an assignment to } x \}
  }.
\]
That is, in the post-state of an assignment to a variable $x$ by thread
$t$, the version of $x$ in the local versioned environment of $t$
equals the total number of writes made to $x$ till that point.
\end{lemma}

\begin{proof}
We prove the lemma by induction on $k$.
  
\textit{Base Case.} If $k = 1$, then clearly $(\tview_k(t_k).2)(x)
= 1$, and we are done.

\textit{Inductive Case.} Let $k = n+1$ and assume the lemma holds
for all earlier writes to $x$ in $\rfpi$.
Let the last write to $x$, prior to $c(\tau_{n+1})$, be in the
transition $\tau_{i}$. By the induction hypothesis,

\begin{align}
(\tview_{i} (t_{i}).2)(x) = & \card{\left\{
j \ : \ j \leq i \land \cmd(\tau_j) \text{ is an assignment to } x \right\}} \nonumber \\
=& \,w \,\, \text{(say)} \nonumber
\end{align}

We now infer the following:
\begin{align}
(\tview_n(t_{n+1}).2)(x) & \ge w \quad \text{from Lemma~\ref{lm:hb-persist-version} and} \nonumber \\
(\tview_n(t_{n+1}).2)(x) & \le w \quad \text{from
    Lemma~\ref{lm:max-version}} \nonumber
\end{align}
Therefore $(\tview_n(t_{n+1}).2)(x) = w$.
Since $\tau_{n+1}$ increments the version of $x$ in
$\tview_{n+1}(t_{n+1})$, we have

\begin{align}
(\tview_{n+1} (t_{n+1}).2)(x) &= (\tview_{n} (t_{n+1}).2)(x) + 1 \nonumber \\
							 &=  w + 1 \nonumber \\
							=& \card{\left\{
							j \ :
                                                                \ j \leq i \land \cmd(\tau_j) \text{ is an assignment to } x\right\}} + 1 \nonumber \\
							=& \card{\left\{
							j \ : \ j \leq n+1 \land \cmd(\tau_j) \text{ is an assignment to } x \right\}}. \nonumber
\end{align}
This completes the proof of the lemma. \qed
\end{proof}

\begin{corollary}
\label{cor:max-version-at-access}
Let 
\begin{math}
\rfpi = \cs_0
\ltrtol{t_1}
\ldots
\ltrtol{t_{N}}
\cs_N
\end{math}
be an execution in the \ldrf\ semantics of program $P$.
Let $\cs_{i} = \abracks{\pc_i, \lmm_i, \tview_i, \lview_i}$, and
let the $i$-th transition in the execution be
\[
\tau_i = \cs_{i-1} \ltrtol{t_{i}} \cs_{i}.
\]
Suppose $\tau_k$
contains an access (read or write) to the variable $x$. Let $m$ be the
highest version count of $x$ among all component versioned
environments in $\cs_{k-1}$. Then $(\tview_{k-1}(t_k).2)(x) =
m$. In other words, whenever a thread accesses a variable $x$, the
version of $x$ is the highest in its local versioned environment.
\end{corollary}

\begin{proof}
Suppose $\tau_k$ is the first write to $x$ in $\rfpi$.
Then by Lemma~\ref{lm:max-version},
$\tview_{k-1}(t) = 0$ for each $t \in \threads$, and we are done.
Otherwise, let there be $m \geq 1$ earlier writes to $x$ before $\tau_k$, and
let $\tau_i$ be the last such write.
Then by Lemma~\ref{lm:max-version}, $(\tview_{k-1}(t).2)(x) \leq m$
for each $t \in \threads$,
and also $(\lview_{k-1}(n).2)(x) \leq m$ for each $n \in \rellocs{}$.
Further, by Lemma~\ref{lm:max-write}, $(\tview_{i}(t_i).2)(x) = m$, and by
Lemma~\ref{lm:hb-persist-version},
$(\tview_{k-i}(t_k).2)(x) \geq m$.
Hence $(\tview_{k-i}(t_k).2)(x) = m$, and we have the corollary.
\qed
\end{proof}

The next Lemma proves that the \ldrf\ semantics generates only
admissible states. 

\begin{lemma} 
\label{lem:admissibleStates} 
Let $\rfpi = \cstate_{\mathit{ent}}
\ltrtol{t_1} \ldots \ltrtol{t_N} \cstate_N$ be an execution of $P$ in
the \ldrf\ semantics. Then, for any $\cstate_k$, with two component
versioned environments (in thread local states or buffers)
$\abracks{\env_1, \rv_2}$ and $\abracks{\env_2, \rv_2}$, and any
variable $x \in \vars$,
if $\rv_1(x) = \rv_2(x)$, then
$\env_1(x) = \env_2(x)$.
\end{lemma}

\begin{proof}
We prove the lemma using induction on the position $k$ in $\rfpi$.
Let the $i$-th transition in $\rfpi$ be 
$\tau_{i} = \cs_{i-1} \ltrtol{t_{i}} \cs_{i}$, and let each
$\cstate_i$ be $\abracks{\pc_i,\lmm_i,\tview_i,\lview_i}$.

\textit{Base Case} 
When $k = 0$, we have $\cstate_k =
\cstate_\ent$. Since all versions and values are 0, the hypothesis
clearly holds.

\textit{Inductive Case.} 
Let us assume that for all $k \leq n$ the claim of the lemma holds,
and consider $k = n+1$.
We consider the different cases for $\cmd(\tau_{n+1})$.
If $\cmd(\tau_{n+1})$ is either an $\mathtt{assume}$ or a
$\mathtt{release}$ statement, then the claim clearly holds since by
assumption it holds for $\cstate_{n}$ and these
commands do not alter any versions or values in going from
$\cstate_{n}$ to $\cstate_{n+1}$.

If $\cmd(\tau_{n+1})$ is of the form $\mathtt{acquire(m)}$, 
then $t_{n+1}$ updates its local versioned environment based on its
local versioned environment and the versioned environments at relevant
buffers. By the induction hypothesis, the versioned environmments in
$\cstate_n$ satisfy the property of the lemma.
By the semantics of the $\mathtt{acquire}$ command, $t_{n+1}$ copies
over the version \emph{and} the valuation of $x$ from one such $\ve$
(including, possibly, $t_{n+1}$'s local versioned environment) in
$\cstate_n$. Thus, the hypothesis also holds for $\cstate_{n+1}$ in
this case.

If $\cmd(\tau_{n+1})$ is of the form $x := e$, then $t_{n+1}$ updates
the version and valuation of $x$ in its local versioned
environment. By Lemma~\ref{lm:max-version}, in any component versioned
environment $\abracks{\env,\rv}$ in $\cstate_n$, we must have
\[
\env(x) \le m,
\]
where $m$ is the total number of writes to $x$ preceeding
$\tau_{n+1}$. 
By Lemma \ref{lm:max-write},
\[
(\tview_{n+1}(t_{n+1}).2)(x) = m+1.
\]
This implies that for any component versioned environment
$\abracks{\env',\rv'}$ in $\cs_{n+1}$, other than the local versioned
environment of $t_{n+1}$,
\[
\rv'(x) < (\tview_{n+1}(t_{n+1}).2)(x).
\]
Since none of the other versioned environments is modified,
the claim of the lemma continues to hold for $\cstate_{n+1}$.
This completes the proof the lemma. \qed
\end{proof}

%DD: Not needed now.
%% \begin{corollary}[$\f\left(\rfpi \right)$ is well-defined]\label{cor:fwelldefined} For any execution $\rfpi$ in the \ldrf\ semantics of $P$, $\f(\rfpi)$ is well-defined.
%% \end{corollary}

%% \begin{proof}
%% The function $\f$ is only defined for admissible states, and by Lemma~\ref{lem:admissibleStates}, the \ldrf\ semantics only produces executions containing admissible states. Thus, for any trace $\rfpi$, $\f(\rfpi)$ is well-defined. \qed
%% \end{proof}

We now proceed to prove the completeness and soundness results
(Theorem~\ref{thm:completeness} and Theorem~\ref{thm:soundness}).

\begin{figure}[!htb]
\centering
\includegraphics[width=0.5\textwidth]{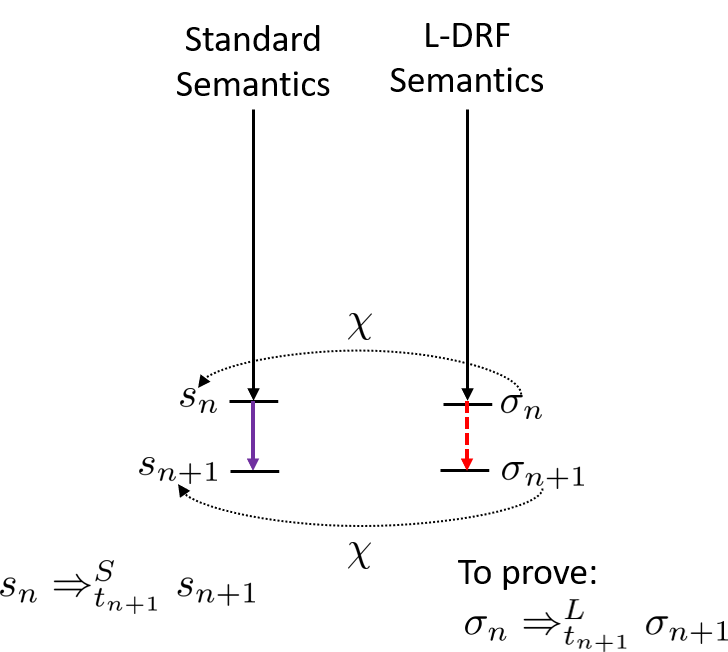}
\caption{\label{fi:completeness}The inductive proof obligation for Completeness. If we hypothesize that every $n$ length trace $\pi$ of program $P$ in the standard semantics has an equivalent trace $\rfpi$ in \ldrf\ semantics, and if we can extend the trace $\pi$ by a single step to reach state $\stds_{n+1}$, then there exists a state $\cstate_{n+1}$, with $\f(\cstate_{n+1}) = \stds_{n+1}$, by which we can extend the $\rfpi$ trace by a single step as well.}
\end{figure}

\begin{proof}[Completeness, Theorem~\ref{thm:completeness}]
We first outline the idea behind the proof, using
Fig.~\ref{fi:completeness}. For any trace $\pi$ of $P$ in the
interleaving semantics, we obtain a corresponding trace $\rfpi$ in the
\ldrf\ semantics by taking the same interleaving of instructions from
the threads. Our inductive hypothesis is that every $N$ length
standard interleaving execution has a corresponding $N$ length
\ldrf\ execution. We 
now consider a $N+1$ length execution $\pi$ in the standard
interleaving semantics, and we show that there exists a state
$\cs_{n+1}$, using which we can extend the $N$ length \ldrf\ trace to
create a $N+1$ length trace which is $\f$-equivalent to $\pi$.

We prove the result using induction on the length of the execution. 
Let $\mathbf{P}(N)$ denote the following hypothesis. For any trace
\[
\pi =
\abracks{\pc_0, \env_0, \lmm_0}
\ltrtos{t_1}
\ldots
\ltrtos{t_{N}}
\abracks{\pc_{N}, \env_{N}, \lmm_{N}}
\]
of program $P$ in the standard semantics, there exists a trace
\[
\rfpi =
\abracks{\pc_0, \lmm_0, \tview_0, \lview_0}
\ltrtol{t_1}
\ldots
\ltrtol{t_{N}}
\abracks{\pc_{N}, \lmm_{N}, \tview_{N}, \lview_{N}}
\]
in the \ldrf\ semantics such that $\f\left(\rfpi\right) = \pi$. 

We outline the inductive arguments.

\textit{Base Case.} For $N=0$, the execution $\pi$ contains the single
state $\stds_\ent$. The length 
$0$ \ldrf\ execution contains the single state $\cstate_\ent$. Since $\f(\cstate_\ent) = \stds_\ent$, $\mathbf{P}(0)$ holds.

\textit{Inductive Case.} Assume that $\mathbf{P}(k)$ holds for all
executions of length $k$, where $0 \le k \le n$. We prove that
$\mathbf{P}(n+1)$ holds. 
Consider a $n+1$ length execution
\[
\pi =
\abracks{\pc_0, \env_0, \lmm_0}
\ltrtos{t_1}
\ldots
\ltrtos{t_{n+1}}
\abracks{\pc_{n+1}, \env_{n+1}, \lmm_{n+1}}
\]
of program $P$ in the interleaving semantics. 
Let the instruction corresponding to the last transition in $\pi$ be
$\abracks{l,c,l}$.
We denote by $\pi[1
  \dots n]$ the $n$-length prefix of $\pi$. By the induction
hypothesis, there exists a trace
\[
\rfpi' =
\abracks{\pc_0, \lmm_0, \tview_0, \lview_0}
\ltrtos{t_1}
\ldots
\ltrtos{t_{n}}
\abracks{\pc_{n}, \lmm_{n}, \tview_{n}, \lview_{n}}
\]
of length $n$ in the \ldrf\ semantics, such that $\pi[1 \dots n] = \f\left(\rfpi'\right)$. Note that this implies that
\begin{equation}
\label{eq:soundness-hyp}
\f\left(\cs_n\right) = \stds_n
\end{equation}
where $\cs_n = \abracks{\pc_{n}, \lmm_{n}, \tview_{n},
  \lview_{n}}$ and $\stds_n = \abracks{\pc_{n}, \lmm_{n},\env_n}$. 

We show that there exists a state
$\cs_{n+1} = \abracks{\pc_{n+1}, \lmm_{n+1}, \tview_{n+1},
  \lview_{n+1}}$ in the \ldrf\ semantics, such that
$\f(\cs_{n+1}) = \stds_{n+1}$ and
$\cs_n\ltrtol{t_{n+1}}\cs_{n+1}$ via the same instruction
$\abracks{l,c,l'}$ used in the
transition $\stds_n \ltrtos{t_n+1} \stds_{n+1}$.
Let $\rfpi$ be the resulting \ldrf\ execution 
$\cstate_0 \ltrtol{t_1} \cdots \ltrtol{t_n-1} \cstate_{n} \ltrtol{t_n}
\cstate_{n+1}$.
Then this would prove that $\rfpi$ satisfies the property 
$\f(\rfpi) = \pi$. 
We show this proof obligation diagrammatically in
Fig.~\ref{fi:completeness}. 

% Note that, by Corollary~\ref{cor:fwelldefined}, the function $\f$ is
% well-defined for all the \ldrf\ traces we deal with. 
% Let the last instruction in $\pi_{n+1}$ be $\abracks{l, c, l'}$, where
%We let case split on $\langle l, c, l' \rangle \in \instr_{t_{n+1}}$,
%where
We note that since $\abracks{l,c,l'}$ is the last instruction in $\pi$, we
must have:
\begin{align}
\pc_n(t_{n+1}) &= l   \nonumber \\
% c &= \cmd(\pi_{n+1}) \\
\pc_{n+1}(t_{n+1}) &= l'.  \nonumber
\end{align}
Note that, by construction, the $\pc$ and $\lmm$ components of
$\cs_{n+1}$ and $\stds_{n+1}$ are made equal. 
Thus the components $\pc_{n+1}$ and $\lmm_{n+1}$ of $\cstate_{n+1}$
are already fixed, and it remains to define $\tview_{n+1}$ and
$\lview_{n+1}$ appropriately.
We now case split on the
command $c$.

\begin{itemize}
\item $c = \mathtt{acquire(\lock)}$: We define the components of 
  state $\cs_{n+1}$ as follows:
\begin{align}
\tview_{n+1}  &= \tview_n[t_{n+1} \mapsto \mathit{updEnv}(\tview_n(t_{n+1}), \lview_n^\lock)] \nonumber \\
\lview_{n+1}  &= \lview_n. \nonumber
\end{align}
Since the lock maps in both $\cs_n$ and $\stds_n$ are the same, the
lock acquisition succeeds from $\cs_n$ as well. 
By the \ldrf\ semantics of $\mathtt{acquire}$,
$\cs_n\ltrtol{t_{n+1}}\cs_{n+1}$.
Since the $\mathtt{acquire}$ does not change the maximum version, and
the corresponding value, of each $x \in \vars$ between $\cs_n$ and
$\cs_{n+1}$, we have $\f(\cs_{n+1}) = \stds_{n+1}$. Thus
$\mathbf{P}(n+1)$ holds in this case.

\item $c = \mathtt{release(\lock)}$: We define the components of 
  state $\cs_{n+1}$ as follows:
\begin{align}
\tview_{n+1} &= \tview_{n} \nonumber \\
\lview_{n+1} &= \lview_{n}[l' \mapsto \tview_n(t_{n+1})]. \nonumber
\end{align}
Once again the lock release must succeed from $\cs_n$ as well. 
By the \ldrf\ semantics of $\mathtt{release}$,
$\cs_n\ltrtol{t_{n+1}}\cs_{n+1}$. Since the $\mathtt{release}$ does
not change the maximum version, and the corresponding value, of each
$x \in \vars$ between $\cs_n$ and $\cs_{n+1}$, we have
$\f\left(\cs_{n+1}\right) = \stds_{n+1}$. Thus $\mathbf{P}(n+1)$
holds in this case as well.

\item $c = \mathtt{assume(b)}$: 
We define the components of 
state $\cs_{n+1}$ as follows:
\begin{align}
\tview_{n+1}  &= \tview_{n} \nonumber \\
\lview_{n+1}  &= \lview_{n} \nonumber.
\end{align}

Consider an arbitrary variable $x$ that is read in the condition
$\mathtt{b}$. By Corollary~\ref{cor:max-version-at-access}, in
$\cs_n$, the version of $x$ is highest in $\tview_n(t_{n+1})$. 
Given that by the induction hypothesis 
$\f(\cs_n) = \stds_n$, this
implies that for any such variable $x$,
$\env_n(x) = (\tview_n(t_{n+1}).2)(x)$.
Hence, it follows that
$\BB{b}\env_n = \BB[\local]{b}(\tview_n(t_{n+1}))$.

Since, by assumption, $\stds_n\ltrtos{t_{n+1}}\stds_{n+1}$, it follows that 
$\cs_n\ltrtol{t_{n+1}}\cs_{n+1}$. 
Since the $\mathtt{assume}$ does not alter the maximum version, and
the corresponding value, of each $x \in \vars$ between $\cs_n$ and
$\cs_{n+1}$, 
we have $\f(\cs_{n+1}) = \stds_{n+1}$. Thus $\mathbf{P}(n+1)$ holds in
this case as well.

\item $c = \mathtt{x := e}$: 
We define the components of 
state $\cs_{n+1}$ as follows:
\begin{align}
\tview_{n+1} &= \tview_n[t_{n+1} \mapsto \abracks{\env',\rv'}] \nonumber \\
\lview_{n+1} &= \lview_n \nonumber
\end{align}

where $\env'$ and $\rv'$ are defined as follows.
Let $\tview_n(t_{n+1})$ be $\abracks{\env,\rv}$.
Then
\begin{align}
\env' &= \env[x \mapsto \env_{n+1}(x)] \nonumber \\
\rv' &=  \rv[x \mapsto \rv(x) + 1]. \nonumber
\end{align}
Consider an arbitrary variable $y$ that is read in the expression
$e$. By Corollary~\ref{cor:max-version-at-access}, in $\cs_n$, the
version of $y$ is highest in $\tview_n(t_{n+1})$. This
implies that for any such variable $y \in \vars$,
\begin{align}
\label{eq:soundness-3}
\env_n(y) &= \env(y) \nonumber \\
\implies \BB{e}\env_n &= \BB[\local]{e}(\tview_n(t_{n+1})) \nonumber \\
\implies \env_{n+1}(x) & = (\tview_{n+1}(t_{n+1}).1)(x) \nonumber
\end{align}

Coupled with the definition of $\rv'$, this proves that
\begin{equation}
\abracks{\env',\rv'} = \BB[\local]{x:=e}\tview_n(t_{n+1}) \nonumber
\end{equation}
which allows us to conclude that $\cs_n\ltrtol{t}\cs_{n+1}$. 
By Lemma~\ref{lm:max-write} and the construction of $\cs_{n+1}$, the
version of $x$ is highest in $\tview_{n+1}(t_{n+1})$, among all other
component versioned environments of $\cs_{n+1}$. This, coupled with
the fact that no other versions are modified, lets us conclude that
$\f(\cs_{n+1}) = \stds_{n+1}$. Consequently,
$\mathbf{P}(n+1)$ holds here as well.
\end{itemize}
This completes the induction argument, and hence the lemma.
\qed
\end{proof}

\begin{figure}[!htb]
\centering
\includegraphics[width=0.5\textwidth]{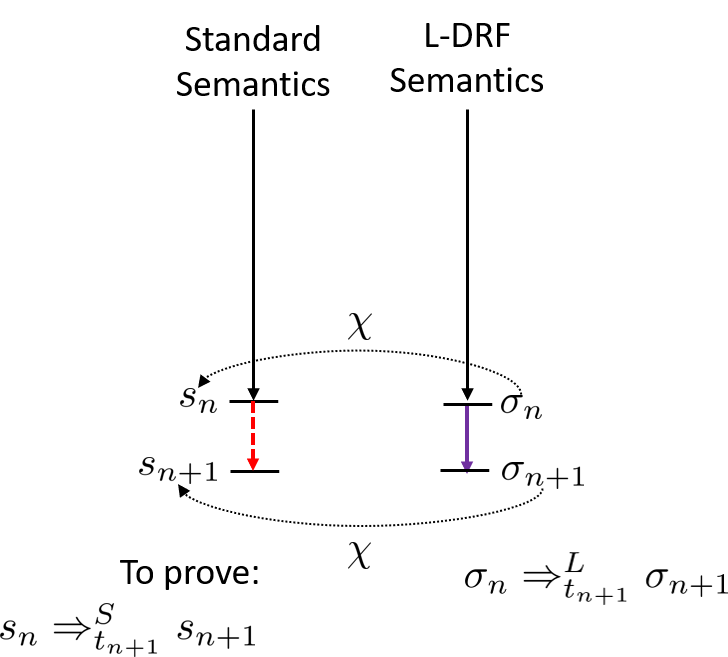}
\caption{\label{fi:soundness}The inductive proof obligation for
  Soundness. If we hypothesize that every $n$ length execution $\rfpi$ of
  program $P$ in the \ldrf\ semantics has an equivalent execution $\pi$ in
  the standard semantics, and if we can extend the execution $\rfpi$ by a
  single step to reach state $\cstate_{n+1}$, then there exists a
  state $\stds_{n+1}$, with $\f(\cstate_{n+1}) = \stds_{n+1}$, by
  which we can extend the execution $\pi$ by a single step as well.}
\end{figure}

\begin{proof}[Soundness, Theorem~\ref{thm:soundness}]
We outline the proof idea using Fig.~\ref{fi:soundness}. Here the
situation is the inverse of that in Fig.~\ref{fi:completeness}. Given
any execution $\rfpi$ in the \ldrf\ semantics of $P$, we show
that the sequence of states induced by the $\f$-map, is a valid execution of $P$
in the interleaving semantics. Our induction hypothesis is on the
length $n$ of the \ldrf\ execution. When we consider a $n+1$ length
\ldrf\ execution $\rfpi'$, we know there exists an execution $\pi$ in
the interleaving semantics corresponding to the $n$ length prefix of
$\rfpi'$. We show that we can \emph{extend} $\pi$ by using
$\f(\cs_{n+1})$ in order to obtain an $n+1$ length
execution in the interleaving semantics, which is $\f$ related to
$\rfpi'$.

Consider an execution
\[
\rfpi = \cs_\ent \ltrtol{t_1} \ldots \ltrtol{t_{N}} \cs_{N}
\]
in the \ldrf\ semantics of program $P$.
We define a sequence of states of $P$ in the standard semantics
\[
\pi = \stds_\ent \ltrtos{t_1} \ldots \ltrtos{t_{N}} \stds_{N},
\]
where for each $i: 0 \leq i \leq N$, $\stds_i = \f(\cs_i)$, and claim this to be a valid
execution of $P$ in the standard semantics.
For each $i$, let $\cs_i = \abracks{\pc_i, \lmm_i, \tview_i,
  \lview_i}$ and $\stds_i = \abracks{\pc_i, \lmm_i, \env_i}$.
We prove the claim by induction on the length $N$ of the execution
$\rfpi$.

\medskip
\noindent
\textit{Base Case.} 
If $N=0$, the execution $\rfpi$ contains the single state $\cs_0 = \cs_\ent$. 
Since $\f(\cs_0) = \stds_0 = \stds_\ent$, we have that $\pi$ is a
valid length $0$ execution of $P$ in the standard semantics.

\medskip
\noindent
\textit{Inductive Case.} 
Assume that the claim holds for all \ldrf\ executions of length $n$. 
Let $N = n+1$.
If $\rfpi[1 \ldots n]$ denotes the $n$ length prefix of the execution
$\rfpi$, then by the induction hypothesis,
\[
\stds_0 \ltrtos{t_1} \ldots \ltrtos{t_n} \stds_n
\]
is a valid execution of $P$ in the interleaving semantics. 
We show that $\stds_n \ltrtos{t_{n+1}} \stds_{n+1}$,
where $\stds_{n+1} = \f(\cs_{n+1})$, using the same instruction in the
corresponding transition of $\rfpi$. 
We show the proof obligation diagrammatically in
Fig.~\ref{fi:completeness}. 

We case split on $\cmd(\tau_{n+1})$, where $\tau_{n+1}$ is the last
transition in $\rfpi$.

If $\cmd(\tau_{n+1})$ is either an $\mathtt{acquire}$ or a
$\mathtt{release}$, then since the location maps and lock maps are
identical in both
$\stds_n$ and $\cs_n$, the lock acquisition (or release) is enabled from
$\stds_n$. Moreover, since neither of the commands alter the versions
between $\cs_n$ and $\cs_{n+1}$, we have $\env_n =
\env_{n+1}$. Thus, $\stds_n \ltrtos{t_{n+1}} \stds_{n+1}$, and the
claim holds in this case.

If $\cmd(\tau_{n+1})$ is $\mathtt{assume(b)}$, then, by
Corollary~\ref{cor:max-version-at-access}, the version of any variable
$x$ read in the condition $\mathtt{b}$ is highest in
$\tview_n(t_{n+1})$. Moreover, since $\f(\cs_n) = \stds_n$,
for any variable $x$ accessed in the
condition $\mathtt{b}$, we must have 
\[
\env_n(x) = (\tview_n(t_{n+1}).1)(x).
\]
This implies that $\BB{b}\env_n = \BB[\local]{b}(\tview_n(t_{n+1}))$. Thus,
$\stds_n \ltrtos{t_{n+1}} \stds_{n+1}$ and the claim holds in this
case too.

Finally, we consider the case when $\cmd(\tau_{n+1})$ is an assignment
statement of the form $x := e$.  In a manner analogous to the case of
the $\mathtt{assume}$ earlier, we can prove that $\BB{e}\env_n =
\BB[\local]{e}(\tview_n(t_{n+1}))$. 
By, Lemma~\ref{lm:max-write}, the
version of $x$ in $\cs_{n+1}$ is highest in
$\tview_{n+1}(t_{n+1})$. Thus,
\[
\env_{n+1}(x) = (\tview_{n+1}(t_{n+1}).1)(x).
\]
Since the assignment command is always enabled, and 
the above facts hold, we obtain 
that $\stds_n \ltrtos{t_{n+1}} \stds_{n+1}$ is a valid transition, and
we are done.

This completes the proof of the claim, and hence the theorem follows.
\qed
\end{proof}

An important corollary of the proofs of these theorems is that the
\ldrf\ semantics is both sound and precise (vis-a-vis the standard
semantics) in a \emph{relational}
sense, provided we restrict our attention to variables \emph{owned} by a
thread at a program point.
For environments $\env$ and $\env'$ and a subset of variables $V$ of
$\vars$, we use the notation $\env =_V \env'$ to mean that $\env$ and
$\env'$ agree on the values of variables in $V$; i.e.\@ for all $x
\in V$ we have $\env(x) = \env'(x)$.

\begin{corollary}
\label{cor:owned}
Let $P$ be a race-free program as above.
Consider a thread $t \in \threads$ and a point $n \in \locs_t$.
Let $V \subseteq \vars$ be the set of variables owned by $t$ at $n$.
Then
\begin{enumerate}
\item
If $\abracks{\pc,\lmm,\env}$ is a reachable state in the standard
interleaving semantics of $P$, with $\pc(t)=n$,
then there exists a reachable state in the \ldrf\ semantics of the form
$\abracks{\pc, \lmm, \tview, \lview}$, with $\tview(t).1 =_V \env$.
\item
Conversely, if 
$\abracks{\pc, \lmm, \tview, \lview}$ is a reachable state in the
\ldrf\ semantics of $P$, with $\pc(t)=n$,
then there exists a reachable state in the standard semantics of the form
$\abracks{\pc,\lmm,\env}$, with $\tview(t).1 =_V \env$.
\end{enumerate}
\end{corollary}

\begin{proof}
We prove the two parts separately.
\begin{enumerate}
\item
Since $\stds = \abracks{\pc,\lmm,\env'}$ is a reachable state in the
interleaving semantics, there is an execution $\pi$ in the
standard semantics that ends at $\stds$.
By the completeness proof, there exists an execution $\rfpi$ of the
\ldrf\ semantics ending in a state $\cstate$, with $\pi = \f(\rfpi)$.
It follows that $\cstate$ must be of the form $\abracks{\pc, \lmm, \tview,
  \lview}$ with $\stds = \f(\cstate)$.
Further, it follows from Corollary~\ref{cor:max-version-at-access},
that for each variable $x \in V$, the version of $x$ must be highest in $t$.
It now follows that $\env' =_V \tview(t).\phi$.

\item
If $\cs = \abracks{\pc, \lmm, \tview, \lview}$ is a state in
$\reach^{\local}(P)$, then there must exist an execution 
$\rfpi = \cs_\ent \ltrtol{t_1} \ldots \ltrtol{t} \cs$ of $P$ in the
\ldrf\ semantics. By Theorem~\ref{thm:soundness},
there exists an execution  
$\stds_\ent \ltrtos{t_1} \ldots \ltrtos{t} \stds$ 
of $P$ in the standard semantics, with 
$\f(\cs) = \stds$.
Thus $\stds$ is of the form $(\pc, \lmm, \env)$ for some environment
$\env$.
Once again, it follows from Corollary~\ref{cor:max-version-at-access},
that the version of each $x \in
V$ is highest in $\tview(t)$, among all component versioned
environments in $\cs$. 
By the construction of the function $\f$, it follows that
for each variable $x \in V$,
$\env(x) = (\tview(t).1)(x)$.
\qed
\end{enumerate}
\end{proof}

\begin{remark}
\label{re:rfGsymb}
Until now we assumed that buffers associated with every post-release
point in $\rellocs[\lock]$ are ``relevant'' to each pre-acquire point in
$\acqlocs[\lock]$.
That is, for a post-release point $n$, if we take $\rfGsymb{}(n)$ to
be the set of pre-aquire points for which $n$ is relevant,
then so far we have assumed that
% That is, $\forall n \in \rellocs[\lock]\,:\,
$\rfGsymb{}(n) = \acqlocs[\lock]$.
However, if no (standard) execution of the program $P$ contains a
transition $\tau_i$ (with the target location being $n$) which
synchronizes-with a transition $\tau_j$ (with source location $n' \in
\acqlocs[\lock]$),
then %\Cref{thm:sound-complete-ldrf}
Theorem~\ref{thm:completeness} (as well as Theorem~\ref{thm:soundness}) holds  even if we remove $n'$ from $\Gsymb{}(n)$.
This is true because in race-free programs, conflicting accesses are  ordered by the happens-before relation.
Thus, if the most up-to-date value of a variable accessed by $t$ was written by another thread $t'$, then in between these accesses there must be a (sequence of) synchronization operations starting at a lock released by $t'$ and ending at a lock acquired by $t$. This refinement of the set $\rfGsymb{}$ based on the above observation can be used to improve the precision of the analyses derived from \ldrf, as it reduces the set of possible release points an acquire can observe.
\end{remark}

\section{Abstract Analyses based on \ldrf}
\label{sec:abstractions}

In this section we introduce and illustrate a few static program
analyses which are based on the \scfg\ representation of a program and
are, in turn, derived from the \ldrf\ semantics. We also reason about
the correctness of such analyses using the notion of consistent
abstractions.
We begin by adapting the standard notion of abstract
interpretation \cite{cousot1977abstract} to our setting, and
recalling the theory of consistent abstractions.

\subsection{Abstract Interpretation of programs}
\label{se:consAbs}

Let us fix a program $P = (\vars, \plocks, \threads)$ for the rest
of this section.

An \emph{abstract interpretation} (or \emph{data-flow analysis})
of $P$ is a structure of the form \mbox{$\aia = (D, \leq, d_o, F)$} where
\begin{itemize}
\item $D$ is the set of \emph{abstract states} and $\leq$ represents a partial ordering over $D$.
\item $(D, \leq)$ forms a complete lattice. We denote the join
  (least upper bound) in this lattice by $\join_{\leq}$, or simply
  $\join$ when the ordering is clear from the context.
\item $d_0 \in D$ is the initial abstract state.
\item $F : \instr_P \rightarrow (D \rightarrow D)$ associates a
  \emph{transfer funcion} $F(\inst)$ with each instruction $\inst$
  of $P$. In what follows, we will write $F_\inst$ instead of
  $F(\inst)$ for ease of presentation.
  We require each transfer function $F_{\inst}$ to be
  \emph{monotonic}, in that
  whenever $d \leq d'$ we have $F_{\inst}(d) \leq F_\inst(d')$.
\end{itemize}

An abstract interpretation $\aia = (D, \leq, d_0, F)$ of $P$
induces a ``global'' transfer
function $\mathcal{F} : D \rightarrow D$, given by
\[
\mathcal{F}(d) = d_0 \join \bigjoin_{\inst \in \instr_P} F_\inst(d).
\]
This transfer function can also be seen to be monotonic.
By the Knaster-Tarski theorem \cite{tarski1955lattice},
$\mathcal{F}$ has a least fixed point (\lfp) in $D$, and we define
this to be the ``semantics'' or ``meaning'' associated to $P$ by the
interpretation $\aia$, and denote it as $\bb{P}_{\aia}$.
Formally,
\[
\bb{P}_{\aia} \eqdef \lfp(\mathcal{F}).
\]

Given two analyses $\aic = (D, \leq, d_0, F)$ and
$\aia = (D', \leq', d_0',F')$ for $P$,
we say $\aia$ is a \emph{consistent abstraction} of
$\aic$ if there exists functions $\alpha:\, D \rightarrow D'$
(called the \emph{abstraction function}), and $\gamma:\,D' \rightarrow D$
(called the \emph{concretization function}), such that:
\begin{enumerate}
\item $\alpha$ and $\gamma$ form a Galois connection, which entails
  the following:
  \begin{enumerate}
    \item $\alpha$ and $\gamma$ are monotonic
    \item $\alpha$ and $\gamma$ satisfy the following conditions
      \begin{itemize}
        \item $\forall d \in D: \, \gamma(\alpha(d)) \ge d$
        \item $\forall d' \in D': \, \alpha(\gamma(d')) = d'$
      \end{itemize}
    \end{enumerate}
\item $\alpha(\bb{P}_{\aic}) \leq' \bb{P}_{\aia}$ (or, equivalently,
  $\bb{P}_{\aic} \leq \gamma(\bb{P}_{\aia})$).
\end{enumerate}

A sufficient condition for consistent abstraction, that can be checked
``locally'' for each instruction, was proposed in
\cite{cousot1977abstract}:

\begin{theorem}[\cite{cousot1977abstract}]
\label{thm:cons-abs}
Let $\aic = (D, \leq, d_0, F)$ and $\aia = (D', \leq',
d_0', F')$ be analyses for $P$.
A sufficient condition for $\aia$ to be a consistent abstraction of
$\aic$ is that there exist maps $\alpha:\, D \rightarrow D'$, and
$\gamma:\,D' \rightarrow D$, which satisfy:
\begin{enumerate}
\item $\alpha$ and $\gamma$ form a Galois connection,
\item for each $\inst \in \instr_P$,
$F_{\inst}'$ safely approximates $F_{\inst}$, in that
\[
\forall d \in D: \, \alpha(F_{\inst}(d)) \leq' F_{\inst}'(\alpha(d)),
\]
\item and $\alpha(d_0) \leq' d_0'$.
\qed
\end{enumerate}
\end{theorem}

\subsection{Collecting Analyses}
\label{subsec:collecting-analyses}

The interleaving semantics of Sec.~\ref{sec:intSemantics} induces a
``collecting'' analysis of $P$,
\[
\aia^{\std} = (\pset{\stdS}, \subseteq, \{\stds_\ent\}, F^{\std}),
\]
where, for any instruction $\inst \in \instr_P$, with $\tidof{\inst} =
t$ say, and for any subset $X \subseteq \stdS$, 
$F_{\inst}^{\std}(X) = \{ \stds' \ | \ \exists \stds \in X \mathrm{\
  with\ } \stds \ltrtos{t} \stds'\}$. 
It turns out that the LFP of this analysis is exactly the
reachable set of 
states in the transition system $\ltsi_{P}$:
\[
 \bb{P}_{\aia^{\std}} = \reach(\ltsi_P).
\]

In a similar way, the \ldrf\ semantics of Sec.~\ref{sec:ldrf} induces
a collecting analysis $\ldrfa$ given by
\[
\ldrfa = (\pset{\cS}, \subseteq, \{\cs_\ent\}, F^{\local}),
\]
where, for any instruction $\inst \in \instr_P$, with $\tidof{\inst} =
t$ say, and for any subset $X \subseteq \cS$, 
$F^{\local}_{\inst}(X) = \{ \cs \ | \ \cs \in X \mathrm{ \ with\ } \cs
\ltrtol{t} \cs'\}$
Once again, the LFP of this analysis can be seen to coincide with the
reachable set of states in the transition system $\ltsl_{P}$ of
Sec.~\ref{sec:ldrf} for the \ldrf\ semantics:
\[
 \bb{P}_{\ldrfa} = \reach(\ltsl_P).
\]

\subsection{Sync-CFG based analyses}
\label{subsec:scfgBasedAnalyses}
We now introduce the class of \scfg\ based analyses, so called because they
analyze concurrent programs using their ``\scfg''.
The \scfg\ representation of a
concurrent program $P$ comprises the control flow graphs of each
static thread code, augmented with \emph{synchronizes-with} edges
between synchronization operations (like releases and acquires of the
same lock).
Each thread operates on local copies of the data
states, and communication between the threads is limited to
synchronization points alone. 
Such an analysis was first introduced in \cite{de2011dataflow}, while
analyses similar in spirit have been proposed in the literature (for
example the thread-modular
shape analysis of \cite{gotsman2007thread}).

A \scfg\ differs from the standard ``product-graph" representation of
concurrent programs in two important ways:

\begin{enumerate}
\item The \scfg\ contains nodes corresponding to each control location in the concurrent program $P$. In contrast, the product graph
  contains nodes corresponding to every possible \emph{combination} of
  control locations in $P$.

\item Each execution of $P$ corresponds to some path in its product
  graph representation. A \scfg\ does \emph{not} maintain such a
  property in general. On the other hand, a key property maintained by
  the \scfg\ is that for each execution of $P$, every
  \emph{happens-before} path induced by the execution corresponds to
  some path in the \scfg.
\end{enumerate}

\begin{figure}[!htb]
\centering
\begin{minipage}{0.4\textwidth}
\includegraphics[scale=0.4]{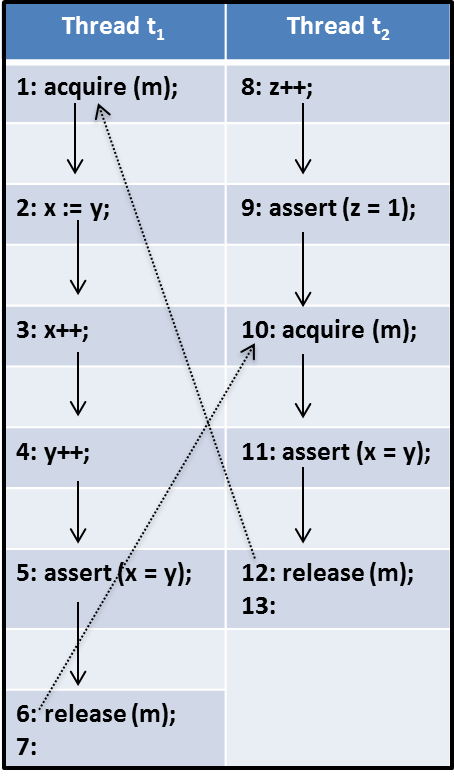}
\end{minipage}%
\begin{minipage}{0.6\textwidth}
\includegraphics[scale=0.3]{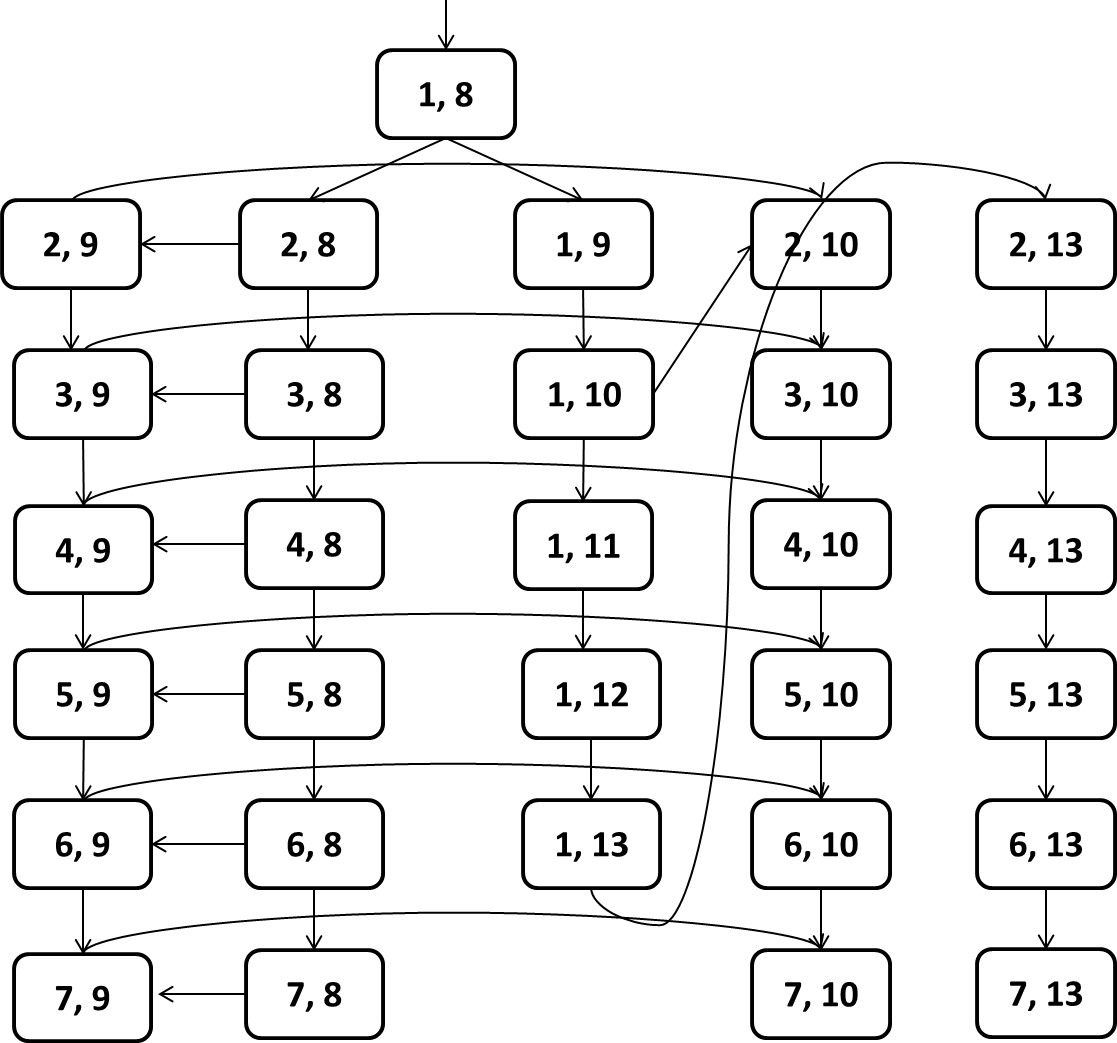}
\end{minipage}
\caption{\label{fi:abssem-graphsCompare}The sync-CFG representation of
  the program of Fig.~\ref{fi:motExProg} is presented on the left. On
  the right is an excerpt of the standard product graph representation
  of the same program.}
\end{figure}

As an example, consider again the program in
Fig.~\ref{fi:motExProg}. The \scfg\ representation of the program is
given on the left in Fig.~\ref{fi:abssem-graphsCompare} (also shown in the center of Fig.~\ref{fi:motEx}). On the other
hand, an excerpt of the far larger product-graph of this program is shown on the
right of the same figure. As one may expect, any analysis based on the
product graph would be intractable for large programs.

More precisely, we say an abstract interpretation $\aia$ of a program
$P$ is a \scfg\ based analysis if:
\begin{enumerate}
\item The domain of abstract states of
  $\aia$ is of the form $\locs_P \rightarrow D'$. Thus the domain
  associates an abstract fact from $D'$ with each location in $P$.
\item The transfer function for each instruction $\inst = (n, c, n')$
  depends only on the
  abstract fact at $n$ for commands other than $\acq{}$, while for
  $\acq{}$ commands the transfer function depends on the abstract
  facts at $n$ \emph{and} associated $\rel{}$ points.
\end{enumerate}

The soundness of the facts computed by a \scfg\ based analysis needs
to be qualified.
The abstract fact computed by the analysis at each program point may \emph{not} be an over-approximation of the
set of concrete (interleaving) states arising at that point. However,
the facts are sound as long as they are interpreted in the window of
variables \emph{owned} by the thread at that point 
(cf.\@ Sec.~\ref{sec:dataraces}).
This property of soundness of \scfg\ analyses was hitherto proved by a
direct and
somewhat involved argument that the \lfp\ of the analysis will
over-approximate the owned portion of the concrete state along an
execution \cite{de2011dataflow,gotsman2007thread}.
In particular, it appears difficult to argue soundness by showing that
the analysis 
is a consistent abstraction of the standard interleaving semantics.
% (defined in Sec.~\ref{sec:intSemantics}).

Instead, we give a way of arguing soundness of \scfg-based analyses by
showing them to be consistent abstractions of the \ldrf\ semantics.
In this sense, the \ldrf\ semantics is a kind of canonical or
reference analysis for \scfg\ based analyses.
We elaborate on this in Sec.~\ref{sec:scfg-soundness}. Before that,
however, we outline several \scfg\ based analyses, as examples, which
can be derived from the \ldrf\ semantics.

\subsection{Some Sync-CFG based analyses induced by \ldrf}
\label{subsec:someScfgAnalyses}

We introduce and illustrate some \scfg\ analyses that are derived from
the \ldrf\ semantics.
We call these analyses (in decreasing order of precision) $\vrel$ (for
``Versioned Relational''),
$\rela$ (for ``Relational'') and $\vset$ (for ``Value
Set''\cite{de2011dataflow}).
We will use the race free program in
Fig.~\ref{fig:program-vrel-code} as an example to illustrate these analyses.

\begin{figure}
\begin{center}
  \begin{minipage}{0.5\textwidth}
\begin{lstlisting}
    Thread t$_1$() {
1:    acquire(l);
2:    x := y;
3:    x++;
4:    y++;
5:    release(l);
6:  }
\end{lstlisting}
\end{minipage}%
\begin{minipage}{0.5\textwidth}
\begin{lstlisting}
     Thread t$_2$() {
 7:    acquire(l);
 8:    x++;
 9:    y++;
10:    release(l);
11:  }
\end{lstlisting}
\end{minipage}
\end{center}
\caption{A simple race-free program on which we illustrate the analyses $\vrel$,
$\rela$ and $\vset$. All the variables are shared.}
\label{fig:program-vrel-code}
\end{figure}

\subsubsection{The $\vrel$ analysis}

The $\vrel$ analysis keeps track of sets of versioned environments at
each program point.
The abstract states are functions mapping
program locations to sets of environments, ordered by point-wise
inclusion.
We call these states \emph{cartesian}, since they now lose the
correlation between thread locations in the program counter.

We define $\vrel = (\locs \rightarrow \pset{\VE}, \preceq, d_0^\vrel,
F^{\vrel})$, where
\begin{itemize}
\item $f \preceq g$ iff for each $n \in \locs$ we have $f(n) \subseteq
  g(n)$.
\item The initial abstract state is
\[
  d_0^\vrel = \lambda n. \left \{ \begin{array}{ll}
                                \{\ve_{\ent}\} & \mbox{ if } n \in \ent_P \\
                                \emptyset & \mbox{ otherwise}.
                                \end{array}
                        \right.
\]
Here $\ve_{\ent}$ is the versioned environment 
$\abracks{\lambda x.0, \lambda x.0}$.
\item The transfer function $F^{\vrel}_{\inst}$, for an instruction
  $\instr = (n,c,n')$ of $P$ is given by
\[
F^{\vrel}_\inst = \lambda f . (f \join_{\preceq} f')
\]
where $f'$ is defined based on the command $c$ as follows.
If $c$ is an assignment command $x := e$,
\[
f'(l) = \left \{ \begin{array}{ll}
                 \BB[\local]{x:=e}(f(n)) & \mbox{ if } l = n'\\
                 \emptyset    & \mbox{ otherwise}.
                 \end{array}
        \right .
\]
By $\BB[\local]{c}(f(n))$ we mean the application of the semantics of the
command $c$, $\BB[\local]{c}$, pointwise on the set of versioned
environments $f(n)$. The case when $c$ is an $\assume{b}$ command is
handled similarly.

When $c$ is an $\acq{m}$ command, we define
\[
f'(l) = \left \{ \begin{array}{ll}
                 \bigcup_{\ve \in f(n)} \mathit{UpdEnv}(\ve,X) & \mbox{ if } l = n'\\
                 \emptyset    & \mbox{ otherwise},
                 \end{array}
        \right .
\]
where $X = \bigcup_{\bar{n} \in \rellocs[\lock]} f(\bar{n})$.

Interestingly, the effect of release commands in the cartesian
semantics is the same as $\mathtt{skip}$:
This is because the abstraction neither tracks ownership of locks nor
explicitly manipulates the contents of buffers.
Thus when $c$ is a release command, we define
\[
f'(l) = \left \{ \begin{array}{ll}
                 f(n) & \mbox{ if } l = n'\\
                 \emptyset    & \mbox{ otherwise},
                 \end{array}
        \right .
\]
%% In other words,
%% \(
%% F^{\vrel}_\inst = \lambda f . f[n' \mapsto (f(n') \union f(n)].
%% \)
\end{itemize}

\remark{We note here that we have chosen to define the transfer
  function in the form of $F_{\inst} = \lambda d. (d \join d')$
  instead of simply $F_{\inst} = \lambda d. d'$. This is because (a)
  it is easy to see that the LFP of the analyses coincide in both
  forms, and (b) the latter form will be convenient for showing the
  sufficient conditions for consistent abstraction in
  Sec.~\ref{sec:adrf-cons-abs}.}

Fig.~\ref{fig:exec-vrel} shows a sequence of instructions from the
program in Fig.~\ref{fig:program-vrel-code}, along with the abstract
states obtained by running the $\vrel$ analysis along this path. This
is shown in the column marked $\vrel$.
We show only the state at the relevant locations of the active thread
along the execution.
The leftmost column shows the $\ldrf$ states along the execution.
Each $\ldrf$ state shown has four rows corresponding to the location
counter, the local state of the thread $t_1$, the local state of
thread $t_2$, and finally the contents of the release buffers. We
ignore the lock maps here.
It is instructive to see how the $\vrel$ analysis over-approximates
the $\ldrf$ analysis at each step along the execution path.
The abstraction map here maps a set of \ldrf\ states $X$ to a set of
versioned environments $Y_n$ at point $n$ in a thread $t$, which
contains the thread-local versioned environments of $t$ in the states
of $X$ where thread $t$ is a point $n$.
Finally, Fig.~\ref{fig:program-vrel} shows the fixed point solutions
of the three analyses we consider here,
for the program of Fig.~\ref{fig:program-vrel-code}.
The leftmost columns on the two sides of the program show the values
for the $\vrel$ analysis, with version tags abstracted away.

\subsubsection{The $\adrf$ Analysis}
\label{subsec:adrf}

We now define the $\adrf$ analysis, which abstracts the
$\vrel$ analysis by abstracting away the version numbers.
This is a more practicable analysis, and is one of the analyses
we focus on subsequently in our experiments.

We define $\adrf = (\AbsState[\Cart], \AbsLeq[\Cart], \absstate[\Cart]^\ent,
F^{\Cart})$, where
\begin{itemize}
\item The set of abstracts states is $\locs \rightarrow \pset{\Env}$,
  which we call $\AbsState[\Cart]$, and we range over it using the meta-variable
  $\absstate[\Cart]$.

\item We have $\absstate[\Cart] \AbsLeq[\Cart]\absstate[\Cart]'$
iff
$\forall \ploc\in\locs$ we have $\absstate[\Cart](\ploc)\subseteq
\absstate[\Cart]'(\ploc)$.

\item The initial abstract state is
\[
  \absstate[\Cart]^\ent = \lambda n. \left \{ \begin{array}{ll}
                                \{\lambda x . 0\} & \mbox{ if } n \in \ent_P \\
                                \emptyset & \mbox{ otherwise}.
                                \end{array}
                        \right.
\]
The initial state thus maps the entry location of every thread to the
set containing the single environment, where all the variables are
initialized to $0$. Every other program location is mapped to the
empty set.

\item The transfer function $F^{\Cart}_{\inst}$, for an instruction
  $\instr = (n,c,n')$ of $P$ is given as follows. We define
\[
F^{\Cart}_\inst = \lambda \absstate[\Cart] . (\absstate[\Cart] \AbsJoin[\Cart] \absstate[\Cart]'),
\]
where $\absstate[\Cart]'$ is defined as follows.

When $c$ is an assignment command $x:=e$, we define
\[
\absstate[\Cart]'(l) = \left \{ \begin{array}{ll}
                 \BB[\std]{x:=e}(\absstate[\Cart](n)) & \mbox{ if } l = n'\\
                 \emptyset    & \mbox{ otherwise}.
                 \end{array}
        \right .
\]
Here $\BB[\std]{c}$ is the interpretation of the command
$c$ according to the standard semantics, assumed to apply pointwise on
a set of environments. The case of an assume command is defined
similarly.

When $c$ is a release command, we have
\[
\absstate[\Cart]'(l) = \left \{ \begin{array}{ll}
                 \absstate[\Cart](n) & \mbox{ if } l = n'\\
                 \emptyset    & \mbox{ otherwise},
                 \end{array}
        \right .
\]
More directly,
\(
F^{\Cart}_\inst = \lambda \absstate[\Cart] . \absstate[\Cart][n'
  \mapsto (\absstate[\Cart](n') \cup \absstate[\Cart](n))].
\)

When $c$ is an $\acq{m}$ command, we define
\[
\absstate[\Cart]'(l) = \left \{ \begin{array}{ll}
                 E_{\mathit{mix}} & \mbox{ if } l = n'\\
                 \emptyset    & \mbox{ otherwise},
                 \end{array}
        \right .
\]
where
\[
\begin{array}{l}
   \qquad E_{\mathit{mix}} =
   \mathit{mix}(\absstate[\Cart](\ploc')\cup \bigcup \{\absstate[\Cart](\bar\ploc)
            \mid
            \bar{\ploc} \in \rellocs[\lock] \land \ploc\in \Gsymb(\bar{\ploc})\} )

   \text{, and  } \\[3pt]
   \qquad     \mixfunc:\pset{\Env} \to\pset{\Env} \equiv
   \lambda B_\Cart .
   \{ \env' \mid
   \forall x \in \vars, \exists \env \in B_\Cart:\env'(x) =        \env(x)
   \}.
\end{array}
\]
\end{itemize}

\begin{figure}[!htb]
\centering
\includegraphics[scale=0.3]{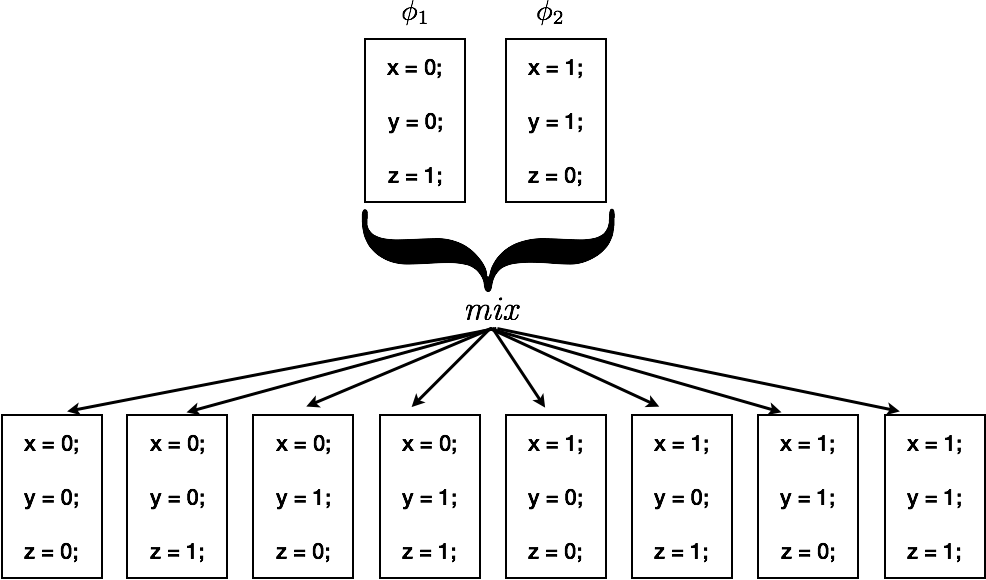}
\caption{\label{fi:abs-mix}Illustrating the $\mixfunc$ on a set of containing two environments $\phi_1$ and $\phi_2$. Observe that the invariant $x = y$ holds in the input environments. However, since this $\mixfunc$ operates at the granularity of single variables, the correlation is lost in the output states.}
\end{figure}

\begin{figure}
\begin{center}
    \begin{picture}(0,0)%
\includegraphics{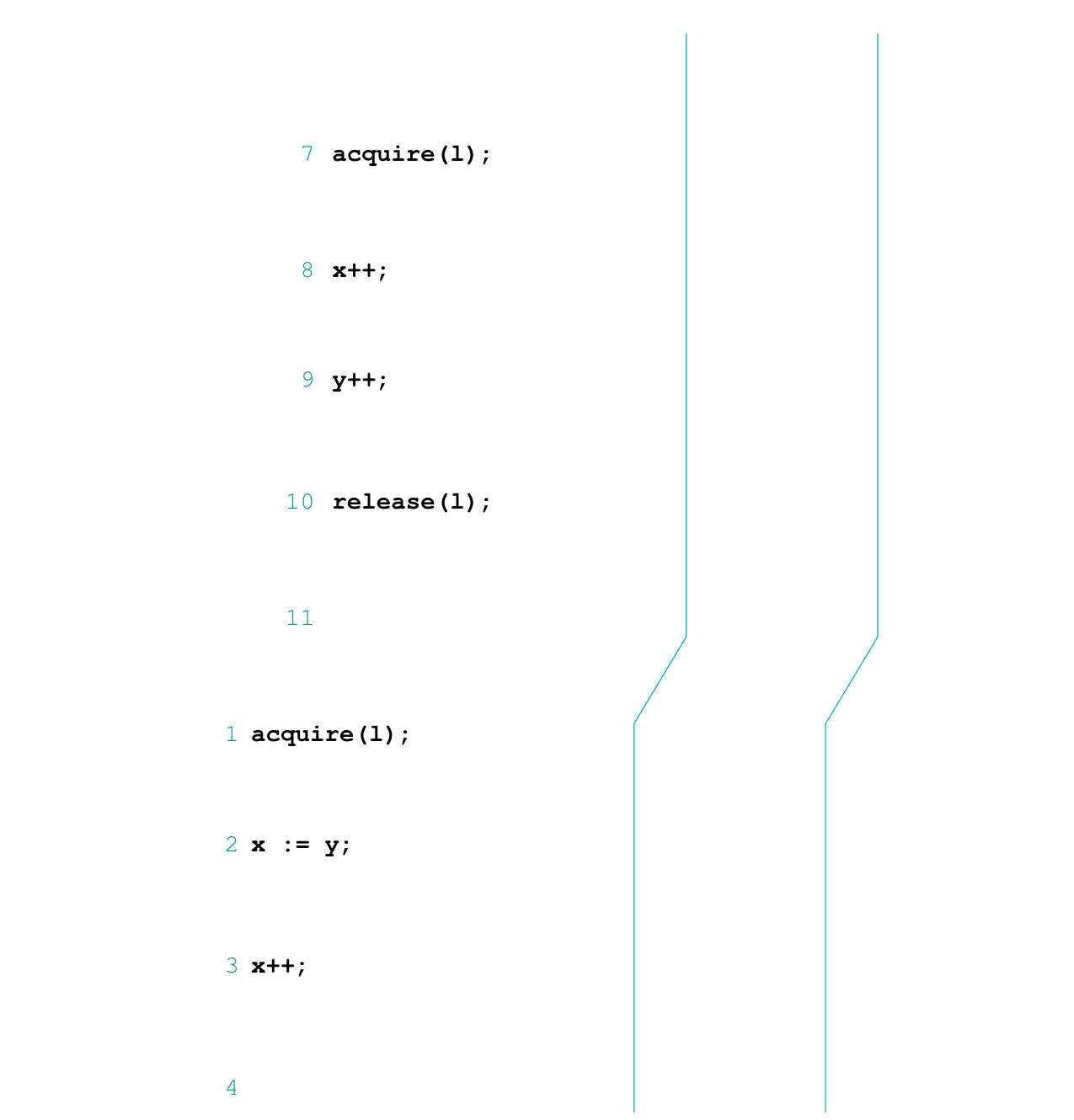}%
\end{picture}%
\setlength{\unitlength}{2901sp}%
\begingroup\makeatletter\ifx\SetFigFont\undefined%
\gdef\SetFigFont#1#2#3#4#5{%
  \reset@font\fontsize{#1}{#2pt}%
  \fontfamily{#3}\fontseries{#4}\fontshape{#5}%
  \selectfont}%
\fi\endgroup%
\begin{picture}(8304,8664)(3049,-7228)
\put(3556,254){\makebox(0,0)[lb]{\smash{{\SetFigFont{7}{8.4}{\familydefault}{\mddefault}{\updefault}{\color[rgb]{0,0,0}$\ldrfsevend$}%
}}}}
\put(3556,434){\makebox(0,0)[lb]{\smash{{\SetFigFont{7}{8.4}{\familydefault}{\mddefault}{\updefault}{\color[rgb]{0,0,0}$\ldrfsevenc$}%
}}}}
\put(8416,254){\makebox(0,0)[lb]{\smash{{\SetFigFont{7}{8.4}{\familydefault}{\mddefault}{\updefault}{\color[rgb]{0,0,0}$\relseven$}%
}}}}
\put(3556,614){\makebox(0,0)[lb]{\smash{{\SetFigFont{7}{8.4}{\familydefault}{\mddefault}{\updefault}{\color[rgb]{0,0,0}$\ldrfsevenb$}%
}}}}
\put(3556,794){\makebox(0,0)[lb]{\smash{{\SetFigFont{7}{8.4}{\familydefault}{\mddefault}{\updefault}{\color[rgb]{0,0,0}$\ldrfsevena$}%
}}}}
\put(6886,254){\makebox(0,0)[lb]{\smash{{\SetFigFont{7}{8.4}{\familydefault}{\mddefault}{\updefault}{\color[rgb]{0,0,0}$\vrelseven$}%
}}}}
\put(3556,-1501){\makebox(0,0)[lb]{\smash{{\SetFigFont{7}{8.4}{\familydefault}{\mddefault}{\updefault}{\color[rgb]{0,0,0}$\ldrfnined$}%
}}}}
\put(8416,-1501){\makebox(0,0)[lb]{\smash{{\SetFigFont{7}{8.4}{\familydefault}{\mddefault}{\updefault}{\color[rgb]{0,0,0}$\relnine$}%
}}}}
\put(3556,-1321){\makebox(0,0)[lb]{\smash{{\SetFigFont{7}{8.4}{\familydefault}{\mddefault}{\updefault}{\color[rgb]{0,0,0}$\ldrfninec$}%
}}}}
\put(6886,-1501){\makebox(0,0)[lb]{\smash{{\SetFigFont{7}{8.4}{\familydefault}{\mddefault}{\updefault}{\color[rgb]{0,0,0}$\vrelnine$}%
}}}}
\put(3556,-2446){\makebox(0,0)[lb]{\smash{{\SetFigFont{7}{8.4}{\familydefault}{\mddefault}{\updefault}{\color[rgb]{0,0,0}$\ldrftend$}%
}}}}
\put(8416,-2446){\makebox(0,0)[lb]{\smash{{\SetFigFont{7}{8.4}{\familydefault}{\mddefault}{\updefault}{\color[rgb]{0,0,0}$\relten$}%
}}}}
\put(3556,-2266){\makebox(0,0)[lb]{\smash{{\SetFigFont{7}{8.4}{\familydefault}{\mddefault}{\updefault}{\color[rgb]{0,0,0}$\ldrftenc$}%
}}}}
\put(6886,-2446){\makebox(0,0)[lb]{\smash{{\SetFigFont{7}{8.4}{\familydefault}{\mddefault}{\updefault}{\color[rgb]{0,0,0}$\vrelten$}%
}}}}
\put(3556,-3346){\makebox(0,0)[lb]{\smash{{\SetFigFont{7}{8.4}{\familydefault}{\mddefault}{\updefault}{\color[rgb]{0,0,0}$\ldrfelevend$}%
}}}}
\put(8416,-3346){\makebox(0,0)[lb]{\smash{{\SetFigFont{7}{8.4}{\familydefault}{\mddefault}{\updefault}{\color[rgb]{0,0,0}$\releleven$}%
}}}}
\put(3556,-3166){\makebox(0,0)[lb]{\smash{{\SetFigFont{7}{8.4}{\familydefault}{\mddefault}{\updefault}{\color[rgb]{0,0,0}$\ldrfelevenc$}%
}}}}
\put(6886,-3346){\makebox(0,0)[lb]{\smash{{\SetFigFont{7}{8.4}{\familydefault}{\mddefault}{\updefault}{\color[rgb]{0,0,0}$\vreleleven$}%
}}}}
\put(3151,-4246){\makebox(0,0)[lb]{\smash{{\SetFigFont{7}{8.4}{\familydefault}{\mddefault}{\updefault}{\color[rgb]{0,0,0}$\ldrfoned$}%
}}}}
\put(8011,-4246){\makebox(0,0)[lb]{\smash{{\SetFigFont{7}{8.4}{\familydefault}{\mddefault}{\updefault}{\color[rgb]{0,0,0}$\relone$}%
}}}}
\put(3151,-4066){\makebox(0,0)[lb]{\smash{{\SetFigFont{7}{8.4}{\familydefault}{\mddefault}{\updefault}{\color[rgb]{0,0,0}$\ldrfonec$}%
}}}}
\put(6481,-4246){\makebox(0,0)[lb]{\smash{{\SetFigFont{7}{8.4}{\familydefault}{\mddefault}{\updefault}{\color[rgb]{0,0,0}$\vrelone$}%
}}}}
\put(3151,-5101){\makebox(0,0)[lb]{\smash{{\SetFigFont{7}{8.4}{\familydefault}{\mddefault}{\updefault}{\color[rgb]{0,0,0}$\ldrftwod$}%
}}}}
\put(8011,-5101){\makebox(0,0)[lb]{\smash{{\SetFigFont{7}{8.4}{\familydefault}{\mddefault}{\updefault}{\color[rgb]{0,0,0}$\reltwoa$}%
}}}}
\put(3151,-4921){\makebox(0,0)[lb]{\smash{{\SetFigFont{7}{8.4}{\familydefault}{\mddefault}{\updefault}{\color[rgb]{0,0,0}$\ldrftwoc$}%
}}}}
\put(6481,-5101){\makebox(0,0)[lb]{\smash{{\SetFigFont{7}{8.4}{\familydefault}{\mddefault}{\updefault}{\color[rgb]{0,0,0}$\vreltwo$}%
}}}}
\put(3151,-6046){\makebox(0,0)[lb]{\smash{{\SetFigFont{7}{8.4}{\familydefault}{\mddefault}{\updefault}{\color[rgb]{0,0,0}$\ldrfthreed$}%
}}}}
\put(8011,-6046){\makebox(0,0)[lb]{\smash{{\SetFigFont{7}{8.4}{\familydefault}{\mddefault}{\updefault}{\color[rgb]{0,0,0}$\relthreea$}%
}}}}
\put(3151,-5866){\makebox(0,0)[lb]{\smash{{\SetFigFont{7}{8.4}{\familydefault}{\mddefault}{\updefault}{\color[rgb]{0,0,0}$\ldrfthreec$}%
}}}}
\put(6481,-6046){\makebox(0,0)[lb]{\smash{{\SetFigFont{7}{8.4}{\familydefault}{\mddefault}{\updefault}{\color[rgb]{0,0,0}$\vrelthree$}%
}}}}
\put(3151,-6991){\makebox(0,0)[lb]{\smash{{\SetFigFont{7}{8.4}{\familydefault}{\mddefault}{\updefault}{\color[rgb]{0,0,0}$\ldrffourd$}%
}}}}
\put(8011,-6991){\makebox(0,0)[lb]{\smash{{\SetFigFont{7}{8.4}{\familydefault}{\mddefault}{\updefault}{\color[rgb]{0,0,0}$\relfoura$}%
}}}}
\put(3151,-6811){\makebox(0,0)[lb]{\smash{{\SetFigFont{7}{8.4}{\familydefault}{\mddefault}{\updefault}{\color[rgb]{0,0,0}$\ldrffourc$}%
}}}}
\put(6481,-6991){\makebox(0,0)[lb]{\smash{{\SetFigFont{7}{8.4}{\familydefault}{\mddefault}{\updefault}{\color[rgb]{0,0,0}$\vrelfour$}%
}}}}
\put(3556,-1141){\makebox(0,0)[lb]{\smash{{\SetFigFont{7}{8.4}{\familydefault}{\mddefault}{\updefault}{\color[rgb]{0,0,0}$\ldrfnineb$}%
}}}}
\put(3556,-961){\makebox(0,0)[lb]{\smash{{\SetFigFont{7}{8.4}{\familydefault}{\mddefault}{\updefault}{\color[rgb]{0,0,0}$\ldrfninea$}%
}}}}
\put(3556,-2086){\makebox(0,0)[lb]{\smash{{\SetFigFont{7}{8.4}{\familydefault}{\mddefault}{\updefault}{\color[rgb]{0,0,0}$\ldrftenb$}%
}}}}
\put(3556,-1906){\makebox(0,0)[lb]{\smash{{\SetFigFont{7}{8.4}{\familydefault}{\mddefault}{\updefault}{\color[rgb]{0,0,0}$\ldrftena$}%
}}}}
\put(3556,-2986){\makebox(0,0)[lb]{\smash{{\SetFigFont{7}{8.4}{\familydefault}{\mddefault}{\updefault}{\color[rgb]{0,0,0}$\ldrfelevenb$}%
}}}}
\put(3556,-2806){\makebox(0,0)[lb]{\smash{{\SetFigFont{7}{8.4}{\familydefault}{\mddefault}{\updefault}{\color[rgb]{0,0,0}$\ldrfelevena$}%
}}}}
\put(3151,-3886){\makebox(0,0)[lb]{\smash{{\SetFigFont{7}{8.4}{\familydefault}{\mddefault}{\updefault}{\color[rgb]{0,0,0}$\ldrfoneb$}%
}}}}
\put(3151,-3706){\makebox(0,0)[lb]{\smash{{\SetFigFont{7}{8.4}{\familydefault}{\mddefault}{\updefault}{\color[rgb]{0,0,0}$\ldrfonea$}%
}}}}
\put(3151,-4741){\makebox(0,0)[lb]{\smash{{\SetFigFont{7}{8.4}{\familydefault}{\mddefault}{\updefault}{\color[rgb]{0,0,0}$\ldrftwob$}%
}}}}
\put(3151,-4561){\makebox(0,0)[lb]{\smash{{\SetFigFont{7}{8.4}{\familydefault}{\mddefault}{\updefault}{\color[rgb]{0,0,0}$\ldrftwoa$}%
}}}}
\put(3151,-5686){\makebox(0,0)[lb]{\smash{{\SetFigFont{7}{8.4}{\familydefault}{\mddefault}{\updefault}{\color[rgb]{0,0,0}$\ldrfthreeb$}%
}}}}
\put(3151,-5506){\makebox(0,0)[lb]{\smash{{\SetFigFont{7}{8.4}{\familydefault}{\mddefault}{\updefault}{\color[rgb]{0,0,0}$\ldrfthreea$}%
}}}}
\put(3151,-6631){\makebox(0,0)[lb]{\smash{{\SetFigFont{7}{8.4}{\familydefault}{\mddefault}{\updefault}{\color[rgb]{0,0,0}$\ldrffourb$}%
}}}}
\put(3151,-6451){\makebox(0,0)[lb]{\smash{{\SetFigFont{7}{8.4}{\familydefault}{\mddefault}{\updefault}{\color[rgb]{0,0,0}$\ldrffoura$}%
}}}}
\put(8011,-5281){\makebox(0,0)[lb]{\smash{{\SetFigFont{7}{8.4}{\familydefault}{\mddefault}{\updefault}{\color[rgb]{0,0,0}$\reltwob$}%
}}}}
\put(8011,-5461){\makebox(0,0)[lb]{\smash{{\SetFigFont{7}{8.4}{\familydefault}{\mddefault}{\updefault}{\color[rgb]{0,0,0}$\reltwoc$}%
}}}}
\put(8011,-5641){\makebox(0,0)[lb]{\smash{{\SetFigFont{7}{8.4}{\familydefault}{\mddefault}{\updefault}{\color[rgb]{0,0,0}$\reltwod$}%
}}}}
\put(8011,-6226){\makebox(0,0)[lb]{\smash{{\SetFigFont{7}{8.4}{\familydefault}{\mddefault}{\updefault}{\color[rgb]{0,0,0}$\relthreeb$}%
}}}}
\put(8011,-7171){\makebox(0,0)[lb]{\smash{{\SetFigFont{7}{8.4}{\familydefault}{\mddefault}{\updefault}{\color[rgb]{0,0,0}$\relfourb$}%
}}}}
\put(8461,1154){\makebox(0,0)[lb]{\smash{{\SetFigFont{8}{9.6}{\rmdefault}{\mddefault}{\itdefault}{\color[rgb]{0,0,0}\rela}%
}}}}
\put(3556,-646){\makebox(0,0)[lb]{\smash{{\SetFigFont{7}{8.4}{\familydefault}{\mddefault}{\updefault}{\color[rgb]{0,0,0}$\ldrfeightd$}%
}}}}
\put(8416,-646){\makebox(0,0)[lb]{\smash{{\SetFigFont{7}{8.4}{\familydefault}{\mddefault}{\updefault}{\color[rgb]{0,0,0}$\releight$}%
}}}}
\put(3556,-466){\makebox(0,0)[lb]{\smash{{\SetFigFont{7}{8.4}{\familydefault}{\mddefault}{\updefault}{\color[rgb]{0,0,0}$\ldrfeightc$}%
}}}}
\put(6886,-646){\makebox(0,0)[lb]{\smash{{\SetFigFont{7}{8.4}{\familydefault}{\mddefault}{\updefault}{\color[rgb]{0,0,0}$\vreleight$}%
}}}}
\put(3556,-286){\makebox(0,0)[lb]{\smash{{\SetFigFont{7}{8.4}{\familydefault}{\mddefault}{\updefault}{\color[rgb]{0,0,0}$\ldrfeightb$}%
}}}}
\put(3556,-106){\makebox(0,0)[lb]{\smash{{\SetFigFont{7}{8.4}{\familydefault}{\mddefault}{\updefault}{\color[rgb]{0,0,0}$\ldrfeighta$}%
}}}}
\put(9946,-646){\makebox(0,0)[lb]{\smash{{\SetFigFont{7}{8.4}{\familydefault}{\mddefault}{\updefault}{\color[rgb]{0,0,0}$\vseight$}%
}}}}
\put(9901,254){\makebox(0,0)[lb]{\smash{{\SetFigFont{7}{8.4}{\familydefault}{\mddefault}{\updefault}{\color[rgb]{0,0,0}$\vsseven$}%
}}}}
\put(9901,-1501){\makebox(0,0)[lb]{\smash{{\SetFigFont{7}{8.4}{\familydefault}{\mddefault}{\updefault}{\color[rgb]{0,0,0}$\vsnine$}%
}}}}
\put(9901,-2446){\makebox(0,0)[lb]{\smash{{\SetFigFont{7}{8.4}{\familydefault}{\mddefault}{\updefault}{\color[rgb]{0,0,0}$\vsten$}%
}}}}
\put(9901,-3346){\makebox(0,0)[lb]{\smash{{\SetFigFont{7}{8.4}{\familydefault}{\mddefault}{\updefault}{\color[rgb]{0,0,0}$\vseleven$}%
}}}}
\put(9496,-4246){\makebox(0,0)[lb]{\smash{{\SetFigFont{7}{8.4}{\familydefault}{\mddefault}{\updefault}{\color[rgb]{0,0,0}$\vsone$}%
}}}}
\put(9496,-5101){\makebox(0,0)[lb]{\smash{{\SetFigFont{7}{8.4}{\familydefault}{\mddefault}{\updefault}{\color[rgb]{0,0,0}$\vstwo$}%
}}}}
\put(9496,-6046){\makebox(0,0)[lb]{\smash{{\SetFigFont{7}{8.4}{\familydefault}{\mddefault}{\updefault}{\color[rgb]{0,0,0}$\vsthree$}%
}}}}
\put(9496,-6991){\makebox(0,0)[lb]{\smash{{\SetFigFont{7}{8.4}{\familydefault}{\mddefault}{\updefault}{\color[rgb]{0,0,0}$\vsfour$}%
}}}}
\put(9901,1154){\makebox(0,0)[lb]{\smash{{\SetFigFont{8}{9.6}{\rmdefault}{\mddefault}{\itdefault}{\color[rgb]{0,0,0}\vset}%
}}}}
\put(6886,1154){\makebox(0,0)[lb]{\smash{{\SetFigFont{8}{9.6}{\rmdefault}{\mddefault}{\itdefault}{\color[rgb]{0,0,0}\vrel}%
}}}}
\put(3556,1154){\makebox(0,0)[lb]{\smash{{\SetFigFont{8}{9.6}{\rmdefault}{\mddefault}{\itdefault}{\color[rgb]{0,0,0}\ldrf}%
}}}}
\put(4996,1154){\makebox(0,0)[lb]{\smash{{\SetFigFont{8}{9.6}{\familydefault}{\mddefault}{\updefault}{\color[rgb]{0,0,0}$t_1$}%
}}}}
\put(5626,1154){\makebox(0,0)[lb]{\smash{{\SetFigFont{8}{9.6}{\familydefault}{\mddefault}{\updefault}{\color[rgb]{0,0,0}$t_2$}%
}}}}
\end{picture}%

\end{center}
\caption{The interpretation of $\vrel$, $\adrf$, and
  $\vset$ along an execution of the program of
  Fig.~\ref{fig:program-vrel-code}.}
\label{fig:exec-vrel}
\end{figure}

\begin{figure}
\begin{center}
  \begin{picture}(0,0)%
\includegraphics{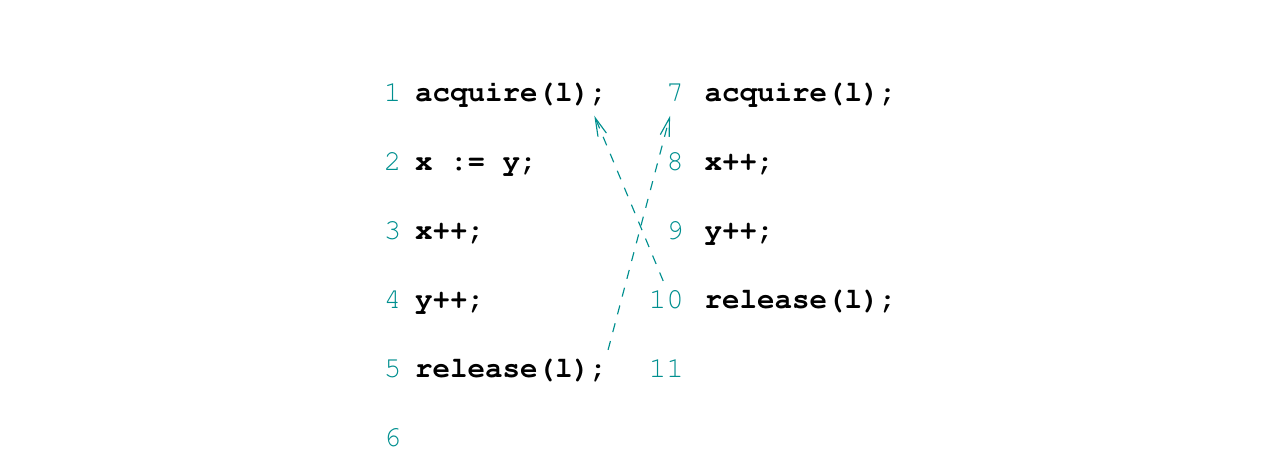}%
\end{picture}%
\setlength{\unitlength}{2901sp}%
\begingroup\makeatletter\ifx\SetFigFont\undefined%
\gdef\SetFigFont#1#2#3#4#5{%
  \reset@font\fontsize{#1}{#2pt}%
  \fontfamily{#3}\fontseries{#4}\fontshape{#5}%
  \selectfont}%
\fi\endgroup%
\begin{picture}(8259,3039)(1024,-2863)
\put(1081,-61){\makebox(0,0)[lb]{\smash{{\SetFigFont{8}{9.6}{\rmdefault}{\mddefault}{\itdefault}{\color[rgb]{0,0,0}\vrel}%
}}}}
\put(1081,-421){\makebox(0,0)[lb]{\smash{{\SetFigFont{7}{8.4}{\familydefault}{\mddefault}{\updefault}{\color[rgb]{0,0,0}$x \!=\! y \!=\! 0$}%
}}}}
\put(1081,-871){\makebox(0,0)[lb]{\smash{{\SetFigFont{7}{8.4}{\familydefault}{\mddefault}{\updefault}{\color[rgb]{0,0,0}$0 \!\leq\! x \!=\! y$}%
}}}}
\put(1081,-1321){\makebox(0,0)[lb]{\smash{{\SetFigFont{7}{8.4}{\familydefault}{\mddefault}{\updefault}{\color[rgb]{0,0,0}$0 \!\leq\! x \!=\! y$}%
}}}}
\put(1081,-2221){\makebox(0,0)[lb]{\smash{{\SetFigFont{7}{8.4}{\familydefault}{\mddefault}{\updefault}{\color[rgb]{0,0,0}$0 \!<\! x \!=\! y$}%
}}}}
\put(1081,-2671){\makebox(0,0)[lb]{\smash{{\SetFigFont{7}{8.4}{\familydefault}{\mddefault}{\updefault}{\color[rgb]{0,0,0}$0 \!<\! x \!=\! y$}%
}}}}
\put(1981,-61){\makebox(0,0)[lb]{\smash{{\SetFigFont{8}{9.6}{\rmdefault}{\mddefault}{\itdefault}{\color[rgb]{0,0,0}\rela}%
}}}}
\put(1981,-421){\makebox(0,0)[lb]{\smash{{\SetFigFont{7}{8.4}{\familydefault}{\mddefault}{\updefault}{\color[rgb]{0,0,0}$x \!=\! y \!=\! 0$}%
}}}}
\put(1981,-1321){\makebox(0,0)[lb]{\smash{{\SetFigFont{7}{8.4}{\familydefault}{\mddefault}{\updefault}{\color[rgb]{0,0,0}$0 \!\leq\! x\!=\!y$}%
}}}}
\put(1981,-2221){\makebox(0,0)[lb]{\smash{{\SetFigFont{7}{8.4}{\familydefault}{\mddefault}{\updefault}{\color[rgb]{0,0,0}$0 \!<\! x \!=\! y$}%
}}}}
\put(1981,-2671){\makebox(0,0)[lb]{\smash{{\SetFigFont{7}{8.4}{\familydefault}{\mddefault}{\updefault}{\color[rgb]{0,0,0}$0 \!<\! x \!=\! y$}%
}}}}
\put(2836,-61){\makebox(0,0)[lb]{\smash{{\SetFigFont{8}{9.6}{\rmdefault}{\mddefault}{\itdefault}{\color[rgb]{0,0,0}\vset}%
}}}}
\put(2836,-421){\makebox(0,0)[lb]{\smash{{\SetFigFont{7}{8.4}{\familydefault}{\mddefault}{\updefault}{\color[rgb]{0,0,0}$x \!=\! y \!=\! 0$}%
}}}}
\put(6886,-871){\makebox(0,0)[lb]{\smash{{\SetFigFont{7}{8.4}{\familydefault}{\mddefault}{\updefault}{\color[rgb]{0,0,0}$0\!\leq\! x\!=\!y$}%
}}}}
\put(6886,-61){\makebox(0,0)[lb]{\smash{{\SetFigFont{8}{9.6}{\rmdefault}{\mddefault}{\itdefault}{\color[rgb]{0,0,0}\vrel}%
}}}}
\put(6886,-421){\makebox(0,0)[lb]{\smash{{\SetFigFont{7}{8.4}{\familydefault}{\mddefault}{\updefault}{\color[rgb]{0,0,0}$x \!=\! y \!=\! 0$}%
}}}}
\put(6886,-1771){\makebox(0,0)[lb]{\smash{{\SetFigFont{7}{8.4}{\familydefault}{\mddefault}{\updefault}{\color[rgb]{0,0,0}$0\!<\!x\!=\!y$}%
}}}}
\put(6886,-2221){\makebox(0,0)[lb]{\smash{{\SetFigFont{7}{8.4}{\familydefault}{\mddefault}{\updefault}{\color[rgb]{0,0,0}$0\!<\!x\!=\!y$}%
}}}}
\put(7741,-61){\makebox(0,0)[lb]{\smash{{\SetFigFont{8}{9.6}{\rmdefault}{\mddefault}{\itdefault}{\color[rgb]{0,0,0}\rela}%
}}}}
\put(7741,-421){\makebox(0,0)[lb]{\smash{{\SetFigFont{7}{8.4}{\familydefault}{\mddefault}{\updefault}{\color[rgb]{0,0,0}$x \!=\! y \!=\! 0$}%
}}}}
\put(8461,-421){\makebox(0,0)[lb]{\smash{{\SetFigFont{7}{8.4}{\familydefault}{\mddefault}{\updefault}{\color[rgb]{0,0,0}$x \!=\! y \!=\! 0$}%
}}}}
\put(8461,-61){\makebox(0,0)[lb]{\smash{{\SetFigFont{8}{9.6}{\rmdefault}{\mddefault}{\itdefault}{\color[rgb]{0,0,0}\vset}%
}}}}
\put(1981,-736){\makebox(0,0)[lb]{\smash{{\SetFigFont{7}{8.4}{\familydefault}{\mddefault}{\updefault}{\color[rgb]{0,0,0}$0 \leq x$}%
}}}}
\put(1981,-871){\makebox(0,0)[lb]{\smash{{\SetFigFont{7}{8.4}{\familydefault}{\mddefault}{\updefault}{\color[rgb]{0,0,0}$0 \leq y$}%
}}}}
\put(2836,-736){\makebox(0,0)[lb]{\smash{{\SetFigFont{7}{8.4}{\familydefault}{\mddefault}{\updefault}{\color[rgb]{0,0,0}$0 \leq x$}%
}}}}
\put(2836,-871){\makebox(0,0)[lb]{\smash{{\SetFigFont{7}{8.4}{\familydefault}{\mddefault}{\updefault}{\color[rgb]{0,0,0}$0 \leq y$}%
}}}}
\put(2836,-1186){\makebox(0,0)[lb]{\smash{{\SetFigFont{7}{8.4}{\familydefault}{\mddefault}{\updefault}{\color[rgb]{0,0,0}$0 \leq x$}%
}}}}
\put(2836,-1321){\makebox(0,0)[lb]{\smash{{\SetFigFont{7}{8.4}{\familydefault}{\mddefault}{\updefault}{\color[rgb]{0,0,0}$0 \leq y$}%
}}}}
\put(2836,-1636){\makebox(0,0)[lb]{\smash{{\SetFigFont{7}{8.4}{\familydefault}{\mddefault}{\updefault}{\color[rgb]{0,0,0}$0 < x$}%
}}}}
\put(2836,-1771){\makebox(0,0)[lb]{\smash{{\SetFigFont{7}{8.4}{\familydefault}{\mddefault}{\updefault}{\color[rgb]{0,0,0}$0 \leq y$}%
}}}}
\put(1981,-1636){\makebox(0,0)[lb]{\smash{{\SetFigFont{7}{8.4}{\familydefault}{\mddefault}{\updefault}{\color[rgb]{0,0,0}$x = y+1$}%
}}}}
\put(1981,-1771){\makebox(0,0)[lb]{\smash{{\SetFigFont{7}{8.4}{\familydefault}{\mddefault}{\updefault}{\color[rgb]{0,0,0}$0 \leq y$}%
}}}}
\put(2836,-2086){\makebox(0,0)[lb]{\smash{{\SetFigFont{7}{8.4}{\familydefault}{\mddefault}{\updefault}{\color[rgb]{0,0,0}$0 < x$}%
}}}}
\put(2836,-2221){\makebox(0,0)[lb]{\smash{{\SetFigFont{7}{8.4}{\familydefault}{\mddefault}{\updefault}{\color[rgb]{0,0,0}$0 < y$}%
}}}}
\put(2836,-2536){\makebox(0,0)[lb]{\smash{{\SetFigFont{7}{8.4}{\familydefault}{\mddefault}{\updefault}{\color[rgb]{0,0,0}$0 < x$}%
}}}}
\put(2836,-2671){\makebox(0,0)[lb]{\smash{{\SetFigFont{7}{8.4}{\familydefault}{\mddefault}{\updefault}{\color[rgb]{0,0,0}$0 < y$}%
}}}}
\put(1081,-1636){\makebox(0,0)[lb]{\smash{{\SetFigFont{7}{8.4}{\familydefault}{\mddefault}{\updefault}{\color[rgb]{0,0,0}$x = y+1$}%
}}}}
\put(1081,-1771){\makebox(0,0)[lb]{\smash{{\SetFigFont{7}{8.4}{\familydefault}{\mddefault}{\updefault}{\color[rgb]{0,0,0}$0 \leq y$}%
}}}}
\put(6886,-1186){\makebox(0,0)[lb]{\smash{{\SetFigFont{7}{8.4}{\familydefault}{\mddefault}{\updefault}{\color[rgb]{0,0,0}$x=y+1$}%
}}}}
\put(6886,-1321){\makebox(0,0)[lb]{\smash{{\SetFigFont{7}{8.4}{\familydefault}{\mddefault}{\updefault}{\color[rgb]{0,0,0}$0\leq y$}%
}}}}
\put(7741,-736){\makebox(0,0)[lb]{\smash{{\SetFigFont{7}{8.4}{\familydefault}{\mddefault}{\updefault}{\color[rgb]{0,0,0}$0 \leq x$}%
}}}}
\put(7741,-871){\makebox(0,0)[lb]{\smash{{\SetFigFont{7}{8.4}{\familydefault}{\mddefault}{\updefault}{\color[rgb]{0,0,0}$0\leq y$}%
}}}}
\put(8461,-736){\makebox(0,0)[lb]{\smash{{\SetFigFont{7}{8.4}{\familydefault}{\mddefault}{\updefault}{\color[rgb]{0,0,0}$0 \leq x$}%
}}}}
\put(8461,-871){\makebox(0,0)[lb]{\smash{{\SetFigFont{7}{8.4}{\familydefault}{\mddefault}{\updefault}{\color[rgb]{0,0,0}$0\leq y$}%
}}}}
\put(7741,-1186){\makebox(0,0)[lb]{\smash{{\SetFigFont{7}{8.4}{\familydefault}{\mddefault}{\updefault}{\color[rgb]{0,0,0}$0 < x$}%
}}}}
\put(7741,-1321){\makebox(0,0)[lb]{\smash{{\SetFigFont{7}{8.4}{\familydefault}{\mddefault}{\updefault}{\color[rgb]{0,0,0}$0\leq y$}%
}}}}
\put(8461,-1186){\makebox(0,0)[lb]{\smash{{\SetFigFont{7}{8.4}{\familydefault}{\mddefault}{\updefault}{\color[rgb]{0,0,0}$0 < x$}%
}}}}
\put(8461,-1321){\makebox(0,0)[lb]{\smash{{\SetFigFont{7}{8.4}{\familydefault}{\mddefault}{\updefault}{\color[rgb]{0,0,0}$0\leq y$}%
}}}}
\put(7741,-1771){\makebox(0,0)[lb]{\smash{{\SetFigFont{7}{8.4}{\familydefault}{\mddefault}{\updefault}{\color[rgb]{0,0,0}$0 < y$}%
}}}}
\put(7741,-1636){\makebox(0,0)[lb]{\smash{{\SetFigFont{7}{8.4}{\familydefault}{\mddefault}{\updefault}{\color[rgb]{0,0,0}$0 < x$}%
}}}}
\put(8461,-1636){\makebox(0,0)[lb]{\smash{{\SetFigFont{7}{8.4}{\familydefault}{\mddefault}{\updefault}{\color[rgb]{0,0,0}$0 < x$}%
}}}}
\put(8461,-1771){\makebox(0,0)[lb]{\smash{{\SetFigFont{7}{8.4}{\familydefault}{\mddefault}{\updefault}{\color[rgb]{0,0,0}$0 < y$}%
}}}}
\put(7741,-2086){\makebox(0,0)[lb]{\smash{{\SetFigFont{7}{8.4}{\familydefault}{\mddefault}{\updefault}{\color[rgb]{0,0,0}$0 < x$}%
}}}}
\put(7741,-2221){\makebox(0,0)[lb]{\smash{{\SetFigFont{7}{8.4}{\familydefault}{\mddefault}{\updefault}{\color[rgb]{0,0,0}$0 < y$}%
}}}}
\put(8461,-2086){\makebox(0,0)[lb]{\smash{{\SetFigFont{7}{8.4}{\familydefault}{\mddefault}{\updefault}{\color[rgb]{0,0,0}$0 < x$}%
}}}}
\put(8461,-2221){\makebox(0,0)[lb]{\smash{{\SetFigFont{7}{8.4}{\familydefault}{\mddefault}{\updefault}{\color[rgb]{0,0,0}$0 < y$}%
}}}}
\put(3736,-61){\makebox(0,0)[lb]{\smash{{\SetFigFont{8}{9.6}{\familydefault}{\mddefault}{\updefault}{\color[rgb]{0,0,0}Thread $t_1$}%
}}}}
\put(5626,-61){\makebox(0,0)[lb]{\smash{{\SetFigFont{8}{9.6}{\familydefault}{\mddefault}{\updefault}{\color[rgb]{0,0,0}Thread $t_2$}%
}}}}
\end{picture}%

\end{center}
\caption{The fixed point results of the $\vrel$, $\rela$, and
  $\vset$ analyses on the program of
  Fig.~\ref{fig:program-vrel-code}. The set of variables owned at
  location 1, 6, 7 and 11 is $\emptyset$, while at other points it is
  $\{x,y\}$. The facts are sound (even in a relational sense) when
  restricted to the variables owned at each point.} 
\label{fig:program-vrel}
\end{figure}

In other words, the $\mixfunc$ returns a cartesian product of the input
states. Note that as a result of abstracting away the version numbers,
a thread cannot determine %does not
%necessarily take
the most up-to-date value of a variable, % , i.e., the one which was written last to the variable,
and thus conservatively picks any possible value found either in its own local environment or in a relevant release buffer. Fig.~\ref{fi:abs-mix} illustrates the operation of the $\mixfunc$ function on two arbitrary input environments.

We denote the LFP of the \adrf\ analysis for program $P$ by
$\BB[\Cart]{P}$.

\subsubsection{The $\vset$ Analysis}
\label{subsec:vset}

The $\vset$ analysis of \cite{de2011dataflow} can be obtained as an
abstraction of the $\adrf$ analysis.
The abstract domain of the $\vset$ analysis is of the form $\locs
\rightarrow \mathit{VS}$, where $\mathit{VS}$ is the
``value-set'' domain which which maps each program variable to a
\emph{set} of values, that is, $\mathit{VS}:\,\vars \rightarrow
\pset{\vals}$.

We define $\vset = (\locs \rightarrow \mathit{VS}, \sqsubseteq,
s_0^\vset, F^\vset)$
where
\begin{itemize}
\item $s \sqsubseteq s'$ iff $\forall n \in locs$ we have $s(n)(x)
  \subseteq s'(n)(x)$.
\item The initial abstract state is
\[
  s_0^\vset = \lambda n. \left \{ \begin{array}{ll}
                                \lambda x . \{0\} & \mbox{ if } n \in \ent_P \\
                                \lambda x . \emptyset & \mbox{ otherwise}.
                                \end{array}
                        \right.
\]

\item The transfer function $F^{\vset}$ can be defined via the
  transfer function $F^{\Cart}$ of the $\adrf$ analysis.
Let us define the value-set abstraction function $\alpha_\mathit{vs}:
\AbsState[\Cart] \rightarrow (\locs \rightarrow \mathit{VS})$ as
\[
\alpha_\mathit{vs}(\absstate[\Cart]) = \lambda \ploc.\left(\lambda x . \{v \mid \exists \env \in \absstate[\Cart](n):\,
\env(x) = v \} \right),
\]
and the value-set concretization function $\gamma_{\mathit{VS}}: (\locs
\rightarrow \mathit{VS}) \rightarrow \AbsState[\Cart]$ as
\[
\gamma_{\mathit{VS}}(s) = \lambda n . \{ \env \ | \ \forall x \in
\vars:\ \env(x) \in s(n)(x) \}.
\]

The transfer function of the $\vset$ analysis for an instruction
$\inst$ can now be defined as
$F^{\vset}_{\inst}(s) =
\alpha_{\mathit{VS}}(F^{\Cart}_{\inst}(\gamma_{\mathit{VS}}(s)))$.

In the $\vset$ analysis, the abstract $\mixfunc$ operator reduces to
the standard value-set join operation (which takes a component wise
union of the value-sets).
\end{itemize}

The abstract state of the $\vset$ analysis along the example execution
is shown in the third column of Fig.~\ref{fig:exec-vrel}, and the
fixed point solution in the third column of Fig.~\ref{fig:program-vrel}.

As one can see from Fig.~\ref{fig:program-vrel}, the analysis $\vrel$
computes the most precise facts -- it is able to establish the equality
between $x$ and $y$ prior to the $\rel{}$ command in both the
threads. The $\rela$ analysis loses this correlation after the
$\acq{}$ command in thread $t_2$. Lastly, the $\vset$ analysis fails
to establish any useful relation between $x$ and $y$.

\subsection{Other abstractions of \ldrf}
\label{se:ldrf-more-abs}

We can improve upon \adrf\ in a practicable way by not forgetting the
versions entirely. We augment $\AbsState[\Cart]$ with % a set of S
``recency" information based on the versions as follows.
For a set $C$ of states of the \ldrf\ semantics, define $\recent{C}$
to be the set of threads $t \in \threads$ such that there exists
a state $\abracks{\pc,\lmm,\tview,\lview} \in C$,
and $x \in \vars$, such that $(\tview(t).2)(x) \geq
(\tview(t').2)(x)$ for each $t' \in \threads$.
%% \[
%% \{\bar{t} \in \threads \mid \exists \cs \in \csset, x \in \vars : \left(\argmax_{t \in \threads}\cs\tview(t)\rv(x)\right) = \bar{t} \}.
%% \]
In other words, $\recent{C}$ is the set of threads which
contain the most up-to-date value of some variable $x$. This
additional information can now be used to improve the precision of
$\mixfunc$.

\begin{figure}[!htb]
\centering
\includegraphics[width=0.7\textwidth]{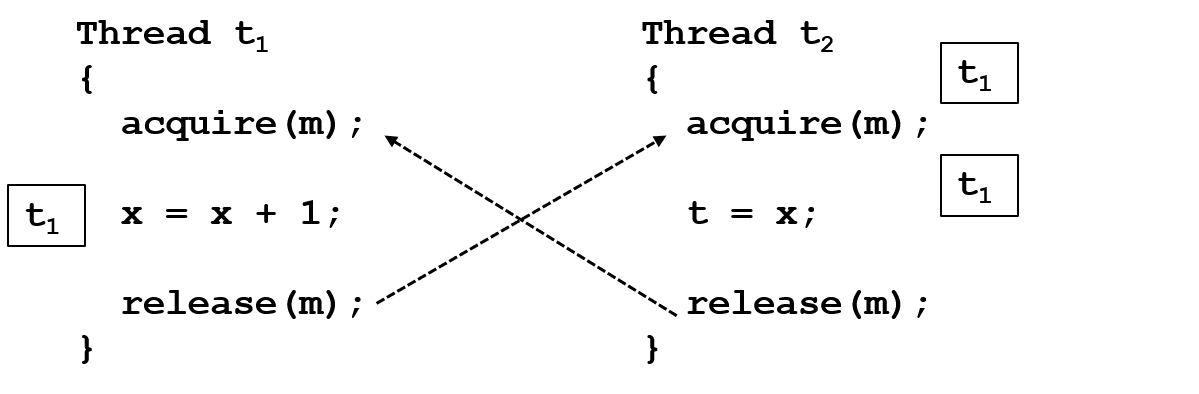}
\caption{\label{fi:tidEx}A simple race-free program to demonstrate the benefit of
  using thread-identifiers in the abstract state. In the normal
  setting, the synchronizes-with edges create a cycle in the program,
  and it is not possible to derive an upper bound on the value of
  $\mathtt{x}$. However, if we track thread-identifiers in the state,
  thread $t_1$ observes that any state it receives from
  $t_2$ is tagged with the set $\{ t_1\}$, and thus $t_1$ can safely
  drop the data flow facts.}
\end{figure}

In the program shown in Fig.~\ref{fi:tidEx}, thread $t_1$
writes to $\mathtt{x}$, while holding the lock $\lock$, whereas thread
$t_2$ reads from $\mathtt{x}$ while holding $\lock$. In the
usual \scfg\ setting, the synchronizes-with edges creates a cycle in
the program graph. Thus, the data flow facts propagate back and forth
between the threads, and the analysis, in this example, fails to
derive an upper bound for the value of $\mathtt{x}$. In the recency
based analysis, the data flow fact comprises elements from
$\AbsState[\Cart]$, as well as a set $S$ of thread-identifiers that
overapproximate the recency information. Whenever a thread writes to a
variable, it adds its identifier to $S$. Other commands do not affect
$S$. In the example, $t_1$ adds its identifier to $S$, and
this is propagated to $t_2$. However, since $t_2$ does
\emph{not} write to $\mathtt{x}$, the set $S$ is propagated back,
unaltered, to $t_1$. The thread $t_1$ now finds that
the incoming data flow fact contains a singleton $S$, with its own
thread-identifier, which indicates it is receiving a stale fact. This
allows the thread to safely drop the data flow fact along an incoming
sync-edge, thereby breaking
the cycle. An abstract analysis based on thread-identifiers can, in
fact, prove an upper bound for $\mathtt{x}$.

\subsection{Soundness of Sync-CFG analyses}
\label{sec:scfg-soundness}

Consider a \scfg\ analysis $\aia$ for program $P$. 
We can prove the ``soundness'' of
$\aia$, in the sense defined in Sec.~\ref{subsec:scfgBasedAnalyses},
with respect to the interleaving semantics, by showing $\aia$ to be a
consistent abstraction of the \ldrf\
analysis via an abstraction map $\alpha$ and concretization map $\gamma$.
Simply put, the set of environments computed by the sync-CFG analysis
$\aia$ at location $n$ in thread $t$, is guaranteed to be a safe
approximation of the actual concrete (standard) states arising whenever thread
$t$ is at location $n$, \emph{provided} we restrict our attention to
the sub-environments on the set of variables \emph{owned} by $t$ at $n$.
We state this more formally below.

\begin{theorem}
\label{thm:ldrf-abs-soundness}
Let $\aia$ be a \scfg\ analysis of a race free program $P$.
Suppose that $\aia$ has been shown to be a
consistent abstraction of the \ldrf\
analysis, via an abstraction map $\alpha$ and concretization map
$\gamma$.
Let $t \in \threads$ and $n \in \locs_t$, and let $V$ be the set of
variables owned by $t$ at location $n$.
Let $\stds = \abracks{\pc, \lmm, \env}$ be a reachable state of the
interleaving semantics, with $\pc(t) = n$.
Then there exists a state $\cstate = \abracks{\pc, \lmm, \tview, \lview}$
in $\gamma(\bb{P}_{\aia})$
with $\env =_V (\tview(t).1)$.
\end{theorem}

\begin{proof}
The proof is immediate since, by Corollary~\ref{cor:owned}, there is a
reachable state $\cstate$ of the \ldrf\ semantics which coincides
with $s$, modulo the restriction to $V$. The fact that $\aia$ is a
consistent abstraction of \ldrf\ says that the $\gamma$ image of its
LFP must contain the state $\cstate$.
\qed
\end{proof}

For example, the facts about $x$ and $y$ inferred by each of the three
analyses in Fig.~\ref{fig:program-vrel} at point 4 is sound (since
both $x$ and $y$ are owned by $t_1$ at these points). However at point
1, the inferred facts may not be sound (and in fact they are not),
since $x$ and $y$ are \emph{not} owned at point~1.

\section{Soundness of $\adrf$ analysis}
\label{sec:adrf-cons-abs}

In this section we show that the $\adrf$ analysis is a consistent
abstraction of the $\ldrfa$ analysis based on \ldrf.

\begin{claim}
\label{claim:adrf-cons-abs}
For any program $P$, the analysis $\adrf$ is a consistent abstraction
of the $\ldrfa$ analysis for $P$.
\end{claim}

\begin{proof}
Consider a program $P = (\vars, \plocks, \threads)$.
We will make use of the definitions of the analysis $\ldrfa$ from
Sec.~\ref{subsec:collecting-analyses}, and $\adrf$ from
Sec.~\ref{subsec:adrf}, and we refer the reader to them.
To show that $\adrf$ is a consistent abstraction of $\ldrfa$, it
suffices (by Theorem~\ref{thm:cons-abs}) to exhibit an abstraction map
$\alpha_\Cart$ and a concretization function $\gamma_\Cart$ satisfying
the conditions of Theorem~\ref{thm:cons-abs}.

The abstraction function $\alpha_\Cart$ maps a set of $\ldrf$
states $\csset\subseteq \cS$ to an abstract state
$\absstate[\Cart]\in\AbsState[\Cart]$.
The abstract value $\alpha_\Cart(\csset)(\ploc)$ contains the
collection of $t$'s environments (where $t =\tidof{\ploc}$) coming
from
any state $\cs\in\csset$ where $t$ is at location $\ploc$. In
addition, if $\ploc$ is a post-release point,
$\alpha_\Cart(\csset)(\ploc)$ also contains the contents of the buffer
$\lview(\ploc)$ for each state $\cs \in \csset$.
We define $\alpha_\Cart : \pset{\AdmStates} \to \AbsState[\Cart]$,
given by
\[
\begin{array}{ll}
\alpha_\Cart(\csset) =  \lambda \ploc . ( &
 \{ \env \mid
 \abracks{\pc,\lmm,\tview,\lview} \in \csset \land
 \tidof{n} = t \land
 \pc(t) = \ploc \land
 \tview(t) = \abracks{\env,\rv}\} \, \cup \\
 & \{ \env \mid
 \abracks{\pc,\lmm,\tview,\lview} \in \csset \land
 \ploc \in \rellocs \land
 \lview(\ploc) = \abracks{\env,\rv} \}).
\end{array}
\]

The concretization function $\gamma_\Cart$ maps a cartesian state
$\absstate[\Cart]$ to
a set of $\ldrf$  states $\csset$ in which the local
% (admissible) %DD definition permits in-admissible states
state  of a thread $t$, when $t$ is at program point
$\ploc\in\locs_t$, comes from $\absstate[\Cart](\ploc)$ and the
contents of the release buffer pertaining to the post-release location
$\ploc\in\rellocs$ also comes from $\absstate[\Cart](\ploc)$.
We define $\gamma_\Cart: \AbsState[\Cart] \totalto \pset{\Sigma}$
given by:
\[
\begin{array}{l}
\gamma_\Cart(\absstate[\Cart]) =
\left\{ \abracks{\pc,\lmm,\tview,\lview} \in \cS % \AdmStates
\left|
\begin{array}{l}
\forall t\in\threads:\, \tview(t) = \abracks{\env,\rv} \land \env \in \absstate[\Cart](\pc(t)) \land  \mbox{}\\
\forall \ploc\in\rellocs:\,\lview(\ploc) = \abracks{\env,\rv} \land \env \in
\absstate[\Cart](\ploc) \\
\end{array}
\right.\right\}.
\end{array}
\]

Let $X \subseteq \cS$ be a set of states of $P$ in the
\ldrf\ semantics.
Let $\inst = (n,c,n')$ be an instruction in $P$, with $\tidof{n} =
t$.
Let
\[
X' = F^{\local}_{\inst}(X) = \{ \cs' \ | \ \exists \cs \in
X,\  \cs \ltrtol{t} \cs' \}.
\]
Further, let $\absstate[\Cart] = \alpha_{\Cart}(X)$ and
$\absstate[\Cart]' = F^{\Cart}_{\inst}(\absstate[\Cart])$.
Then we need to show that
\begin{equation}
\alpha_{\Cart}(X') \AbsLeq[\Cart] \absstate[\Cart]'.
\label{eqn:cons-abs}
\end{equation}
This is depicted in Fig.~\ref{fig:adrf-cons-abs}.
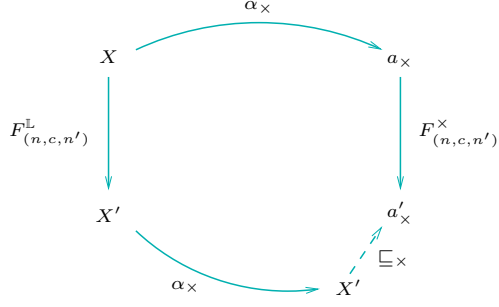
\begin{figure}
\begin{center}
\begin{picture}(0,0)%
\includegraphics{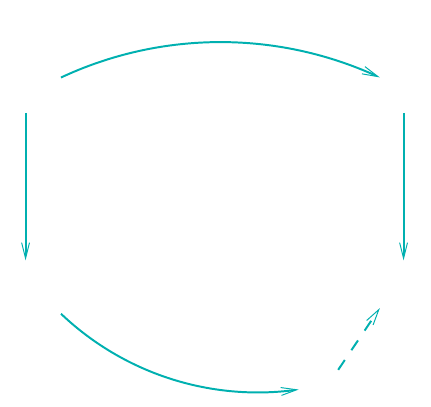}%
\end{picture}%
\setlength{\unitlength}{2486sp}%
\begingroup\makeatletter\ifx\SetFigFont\undefined%
\gdef\SetFigFont#1#2#3#4#5{%
  \reset@font\fontsize{#1}{#2pt}%
  \fontfamily{#3}\fontseries{#4}\fontshape{#5}%
  \selectfont}%
\fi\endgroup%
\begin{picture}(3270,3080)(2506,-2880)
\put(2701,-511){\makebox(0,0)[b]{\smash{{\SetFigFont{7}{8.4}{\rmdefault}{\mddefault}{\updefault}{\color[rgb]{0,0,0}$X$}%
}}}}
\put(5581,-511){\makebox(0,0)[b]{\smash{{\SetFigFont{7}{8.4}{\rmdefault}{\mddefault}{\updefault}{\color[rgb]{0,0,0}$\absstate[\Cart]$}%
}}}}
\put(2701,-2086){\makebox(0,0)[b]{\smash{{\SetFigFont{7}{8.4}{\rmdefault}{\mddefault}{\updefault}{\color[rgb]{0,0,0}$X'$}%
}}}}
\put(5581,-2041){\makebox(0,0)[b]{\smash{{\SetFigFont{7}{8.4}{\rmdefault}{\mddefault}{\updefault}{\color[rgb]{0,0,0}$\absstate[\Cart]'$}%
}}}}
\put(4186, 29){\makebox(0,0)[b]{\smash{{\SetFigFont{7}{8.4}{\rmdefault}{\mddefault}{\updefault}{\color[rgb]{0,0,0}$\alpha_{\Cart}$}%
}}}}
\put(2521,-1231){\makebox(0,0)[rb]{\smash{{\SetFigFont{7}{8.4}{\rmdefault}{\mddefault}{\updefault}{\color[rgb]{0,0,0}$F^{\local}_{(n,c,n')}$}%
}}}}
\put(5761,-1231){\makebox(0,0)[lb]{\smash{{\SetFigFont{7}{8.4}{\rmdefault}{\mddefault}{\updefault}{\color[rgb]{0,0,0}$F^{\Cart}_{(n,c,n')}$}%
}}}}
\put(5356,-2491){\makebox(0,0)[lb]{\smash{{\SetFigFont{7}{8.4}{\rmdefault}{\mddefault}{\updefault}{\color[rgb]{0,0,0}$\AbsLeq[\Cart]$}%
}}}}
\put(5071,-2816){\makebox(0,0)[b]{\smash{{\SetFigFont{7}{8.4}{\rmdefault}{\mddefault}{\updefault}{\color[rgb]{0,0,0}$X'$}%
}}}}
\put(3611,-2751){\makebox(0,0)[rb]{\smash{{\SetFigFont{7}{8.4}{\rmdefault}{\mddefault}{\updefault}{\color[rgb]{0,0,0}$\alpha_{\Cart}$}%
}}}}
\end{picture}%

\end{center}
\caption{The proof obligation to show $\adrf$ is a consistent
  abstraction of $\ldrfa$. The solid lines represent given relations,
  while the dashed line needs to be established.}
\label{fig:adrf-cons-abs}
\end{figure}

We observe that for each $\cs' = \abracks{\pc',\lmm',\tview',\lview'}$ in
$X'$ we have $\pc'(t) = n'$, and there exists a state $\cs =
\abracks{\pc,\lmm,\tview,\lview} \in X$ such that
$pc(t) = n$, $\pc' = \pc[t\mapsto n']$, and for each $t'\neq t$ we
have $\tview'(t') = \tview(t')$.
Further, every environment $\env'$ that occurs in $\tview(t')$ where
$t' \neq t$, is already present in $\absstate[\Cart]$.
This is because (a) it is present in $\sigma$ and $\alpha_{\Cart}$ ensures
that it is present in the appropriate location in $\absstate[\Cart]$;
and (b) by the definition of the transfer function
$F^{\Cart}_{\inst}$, every environment at location $l$ in $\absstate[\Cart]$
is also at location $l$ in $\absstate[\Cart]'$.
Thus to show that (\ref{eqn:cons-abs}) holds, it suffices to show for
an arbitrary $\cs' = \abracks{\pc',\lmm',\tview',\lview'}$ that the
environments in $\tview'(t)$ and $\lview'$ are present in the
appropriate locations ($n'$ and release points, respectively) in
$\absstate[\Cart]'$.

Let us fix an $\cs' = \abracks{\pc',\lmm',\tview',\lview'} \in X'$ and a $\cs =
\abracks{\pc,\lmm,\tview,\lview} \in X$ as above.
We now show this subclaim for each command $c$.

\paragraph{Assignment.} When $c$ is an assignment of the form $x := e$.
Let $\tview'(t) = \abracks{\env',\rv'}$.
Then $\env' = \BB[]{x := e}\env$, where $\tview(t) = \abracks{\env,\rv}$, for
some $\rv$.
Now $\env \in \absstate[\Cart](n)$, and by the definition of
$F^{\Cart}_{\inst}$, also in $\absstate[\Cart]'(n')$.

Further, since $\lview' = \lview$, its environments are all included
in $\absstate[\Cart]$ and hence also in $\absstate[\Cart]'$.

The case of assume commands is handled similarly.

\paragraph{Release.}
Recall that in this case $\absstate[\Cart]' =
\absstate[\Cart][n'\mapsto(\absstate[\Cart](n') \cup
                           \absstate[\Cart](n))]$.
Now $\env' = \env$ and therefore $\env' \in \absstate[\Cart](n')$.
Also, $\lview' = \lview[n' \mapsto\abracks{\env,\rv}]$.
But $\env$ already belongs to $\absstate[\Cart](n')$.

\paragraph{Acquire.}
In this case, $\tview'(t)$ chooses to take the value of a variable $x$ in the
thread-local environment of $t$, from the versioned environment $\ve$
in some relevant buffer, or the existing thread-local environment of
$t$. By the construction of $\alpha_\Cart$, if $\ve$ was chosen from
some post-release point $\bar{n}$, then this environment is guaranteed
to exist in $\absstate[\Cart](\bar{n})$. Likewise, if $\ve$ is simply
the thread-local versioned environment of $t$, then the environment
would be in $\absstate[\Cart](n)$. Since, by the semantics of the
$\mathtt{acquire}$ in the \adrf\ analysis, all the environments at all
such $\bar{n}$, and the environment at $n$, is taken into account in
the $\mixfunc$, and since this operation is performed for each
variable $x \in \vars$, we have $\tview'(t).1 \in
\absstate[\Cart](n')$.

This completes the proof of (\ref{eqn:cons-abs}) and hence of the Claim.
\qed
\end{proof}

From Theorem~\ref{thm:ldrf-abs-soundness}, it now follows that the
facts inferred by the $\adrf$ analysis about the owned set of
variables at each location in a program $P$, are indeed sound.

\section{A Region-Parameterized version of \ldrf}
\label{ch:rdrf}

In this section, we introduce a refined notion of data race freedom, based on \emph{data regions}, and derive from it a more precise abstract analysis capable of transferring \emph{some} relational information between threads at synchronization points. The objective is to modify the \ldrf\ semantics such that the abstract mix operates at a granularity higher than individual variables.

\subsection{Why do we need another semantics?}
Fig.~\ref{fi:abs-mix}, which illustrates the operation of \emph{mix}, also highlights the key issue with the \ldrf\ semantics: any abstract analysis derived from the \ldrf\ semantics must make use of an abstract $\mixfunc$ which operates at the granularity of individual variables. Thus, even though two variables may be related in the input environments to $\mixfunc$ (like $x = y$ in Fig.~\ref{fi:abs-mix}), the function must necessarily forget their correlation after the mixing. This is essential for soundness. This is the reason that prevents us from proving the assertion $x = y$ at line $11$ in the motivating example in Fig.~\ref{fi:motEx}. Even though the $\mathtt{acquire(m)}$ in $t_2$ obtains the fact $x = y$ from both its input edges, it fails to maintain this correlation post the mix.

While the $\vrel$ analysis we saw in Sec.~\ref{subsec:someScfgAnalyses} had a \emph{mix} operator which did better for the program in Fig.~\ref{fig:program-vrel-code}  -- it preserved the correlation between $x$ and $y$ after the \emph{mix} in thread $t_2$ -- the analysis is not practicable (it does not provide an abstraction of the versions, which may grow in an unbounded fashion).

Our solution is to make use of user-defined regions. Essentially, regions are  a user-defined partitioning of the set of program variables. We call each partition a \emph{region} $r$, denote the set of regions as $\regions$, and the region of a variable $x$ by $\reg(x)$.

The semantics precisely tracks correlations between variables \emph{within} regions \emph{across} inter-thread communication, while abstracting away the correlations between variables across regions. This partitioning is based on the semantics of the program: developers often write code where a group of variables forms a logical cluster. Often, some invariant holds on the variables within this cluster at specific program points. Since we make this partitioning explicit in the semantics, with suitable abstractions the tracked correlations can improve the precision of the abstract analyses for programs which conform to the notion of race freedom defined below.

\subsection{Region Race Freedom}
\label{se:regionRaces}
We present a refinement of the standard notion of data race freedom by ensuring that variables residing in the same region are manipulated atomically across threads. A \emph{region-level data race} \cite{sas17-ldrf} occurs when two concurrent threads access variables from the same region $r$ (not necessarily the same variable), with at least one access being a write, and the accesses are devoid of any ordering constraints.

A command $x := e$ constitutes a \emph{write access} to the region $\reg(x)$, and a \emph{read access} of every region $\reg(y)$, for each variable $y$ appearing in the expression $e$. Similarly, a command $\mathtt{assume(b)}$ constitutes a read access of every region $\reg(y)$, for each variable $y$ appearing in the condition $\mathtt{b}$. We are now in a position to introduce our notion of region level races.

\begin{definition}[Region-level races]\label{De:RFRaces} Let $P$ be a program and let $\regions$ be a region partitioning of $P$. An execution $\pi$ of $P$, in the standard interleaving semantics, has a \emph{region-level race} if there exists $0 \leq i < j < |\pi|$, such that $c(\pi_i)$ and
$c(\pi_j)$ both access variables in region $r \in \regions$,
at least one access is a write,
and it is not the case that $\pi_i\xrightarrow{hb}_\pi \pi_j$.
\end{definition}

The problem of checking for region races can be reduced to the problem
of checking for data races as follows. We introduce a fresh variable
$\mathtt{X}_r$ for each region $r \in \regions$. We now transform the
input program $P$ to a program $P'$ with the following additions.
We assume without loss of generality that $\assume{}$ statements in
only reference thread-local variables. For example, we replace
$\assume{x<y}$ by the statements 
``$l_x := x;\ l_y := y;\ \assume{l_x < l_y}$''.

\begin{itemize}
\item We precede every assignment statement $\mathtt{x := e}$, where $r_w$ is the region which is written to, and $r_1, \dots, r_n$ are the regions read, with a sequence of instructions $\mathtt{X}_{r_w} := \mathtt{X}_{r_1}; \, \dots \, \mathtt{X}_{r_w} := \mathtt{X}_{r_n};$.

\item Statements of the form $\assume{b}$ do not need to be changed
  because $b$ refers only to thread-private variables.

\item The $\mathtt{acquire}$ and $\mathtt{release}$ statements do not involve the access of any variable. Thus, they remain unmodified.
\end{itemize}
Note that these modifications do not alter the semantics of the
original program (for each trace of $P$ there is a corresponding trace
in $P'$, and vice versa). We now check for \emph{data races} on the
$\mathtt{X}_r$ variables.

\subsection{The \rdrf\ semantics}

The region-based version of \ldrf\ semantics, which we call here the
\rdrf\ semantics \cite{sas17-ldrf}, is obtained via a simple change to
the \ldrf\ semantics:
a write-access to a variable $x$ leads to incrementing the version of
every variable that resides in $x$'s region. In other words, the
semantics of the assignment command, $\BB{\assgn{x}{e}} \colon \VE
\totalto \VE$, is defined as follows:
\[
\BB{\assgn{x}{e}}\abracks{\env,\rv} = \abracks{\env',\rv'}\]
where $\env' = \env[x\mapsto \BB{e}\env]$, and 
$\rv'$ is given by:
\[
\begin{array}{lll}
\rv'(y) & = & \left \{
    \begin{array}{ll}
    \rv(y)+1 & \mathrm{ if\ } \reg(y) = \reg(x), \\
    \rv(y)   & \mathrm{otherwise.}
    \end{array} \right.
\end{array}
\]

It is not difficult to see that the versions of
Theorems~\ref{thm:completeness} and 
\ref{thm:soundness} hold for the completeness and soundness of the
\rdrf\ semantics vis-a-vis the standard interleaving semantics, for
programs that are region-race free.
Hence, we can analyze such programs using abstractions of \rdrf\ and
obtain sound results with respect to the standard interleaving semantics
(Sec.~\ref{sec:intSemantics}). 

\subsection{Thread-Local Abstractions of the \rdrf\ Semantics}
%\subsubsection{Thread-Local Relational Abstractions of the \RDRF\ Semantics}
\label{sec:rfdrfAbsSem}
The cartesian abstractions defined in Sec.~\ref{sec:abstractions} can
be extended to accommodate regions in a natural way.
The only difference lies in the definition of the $\mixfunc$
operation, which now operates at the granularity of \emph{regions},
rather than variables:
$$
\begin{array}{l}
\mixfunc:\pset{\Env} \to\pset{\Env} \eqdef
   \lambda B_\Cart .
   \{ \env' \mid
   \forall r \in \regions, \exists \env \in B_\Cart\mathrm{\ s.t.\ }  
        \forall x \in \vars \mathrm{\ s.t.\ } \reg(x)=r
\\
\quad\qquad\qquad\qquad
\qquad\qquad\qquad\qquad
\qquad\qquad\qquad\qquad
  \mathrm{\ we\ have\ } \env'(x) = \env(x) \}.
\end{array}
$$
Mixing  environments at the granularity of regions is permitted because
the \rdrf\ semantics ensures that all the variables in the same region have the same version.
Thus, their most up-to-date values reside in either the thread's local environment or in one of the release buffers.
As before, we can obtain an effective analysis using any sequential abstraction, provided that the abstract domain supports the (more precise) region based mix operator.

\subsection{Illustrative Example}
\begin{figure}[!htb]
\centering
\includegraphics[scale=0.3]{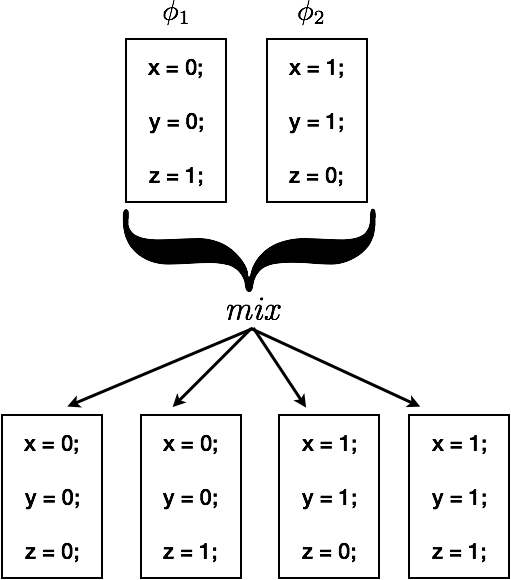}
\caption{\label{fi:region-mix}Illustrating the operation of $\mixfunc$ when it is aware of regions. In this example, with the regions being $\abracks{\{x,y\}, \{z\}}$, the function maintains the correlation between $x$ and $y$ in the output.}
\end{figure}

We illustrate the effect of the regions using some small
examples. Consider again the situation in
Fig.~\ref{fi:abs-mix}. Recall that even though the input environments
maintained $x = y$, the $\mixfunc$ was unable to preserve this
correlation because it operated at the granularity of individual
variables. However, when $\mixfunc$ is made aware of the region
definitions, it maintains the correlation between variables
\emph{within} a region. Thus, in Fig.~\ref{fi:region-mix}, the
invariant $x = y$ continues to hold in the output state. 

\begin{figure}[!htb]
\centering
\includegraphics[width=0.7\textwidth]{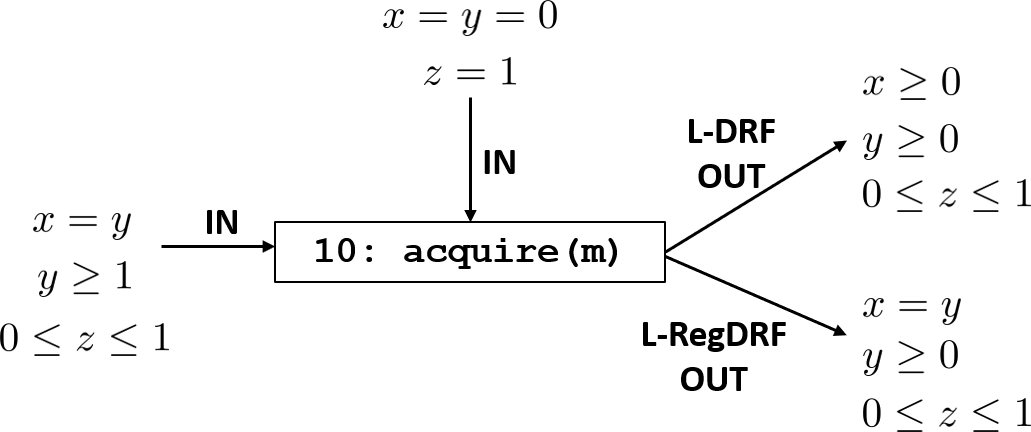}
\caption{\label{fi:r-drfex}The improved precision of the region aware $\mixfunc$ derived from the \rdrf\ semantics allows it to prove the additional assertion at line $11$ in Fig.~\ref{fi:motEx}.}
\end{figure}

Returning to the program in Fig.~\ref{fi:motEx}, consider the
situation at the $\mathtt{acquire}$ at line $10$ (illustrated in
Fig.~\ref{fi:r-drfex}). It receives the invariant $x = y$ from both
its input branches. The $\mixfunc$ in the \adrf\ abstraction of
\ldrf\ only outputs the correct bounds for the variables, and forgets
the correlation between $x$ and $y$. However, the region-aware
$\mixfunc$ preserves this invariant, which enables the region-aware
version of $\adrf$ derived from \rdrf, which we call \Reg, 
to prove the assertion at line $11$.

\section{Implementation and Experiments}
\label{ch:exp-ratcop}

\subsection{RATCOP: Relational Analysis Tool for COncurrent Programs}
In this section, we perform a thorough empirical evaluation of our
analyses using a prototype analyzer which we have developed, called
RATCOP \cite{ratcop}\footnote{The project artifacts are available at
  \texttt{https://bitbucket.org/suvam/ratcop}}, for the static
intra-procedural analysis of race-free concurrent Java
programs. RATCOP comprises around 4000 lines of Java code, and
implements a variety of relational analyses based on the theoretical
underpinnings described in earlier sections of this paper. Through
command line arguments, each analysis can be made to use any one of
the following three numerical abstract domains provided by  the Apron
library \cite{jeannet2009apron}: Convex Polyhedra (with support for
strict inequalities), Octagons and Intervals. RATCOP also makes use of
the Soot \cite{vallee1999soot} analysis framework for Java. The tool
reuses the code for fixed point computation and the graph data
structures in the implementation of \cite{de2011dataflow}.

\begin{figure}[!htb]
\centering
\includegraphics[scale=0.3]{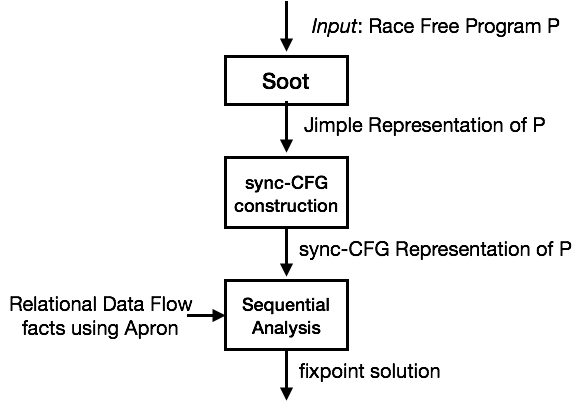}
\caption{\label{fi:ratcop-summary}Architecture of RATCOP.}
\end{figure}

The tool takes as input a Java program with assertions marked at
appropriate program points. We first checked all the programs in our
benchmarks for data races and region races using Chord
\cite{chord}. For detecting region races, we have implemented the
translation scheme outlined in Sec.~\ref{se:regionRaces}. RATCOP then
performs the necessary static analysis on the program until a fixpoint
is reached. Subsequently, the tool automatically tries to prove the
assertions using the inferred facts (which translates to checking
whether the inferred fact at a program point, projected to the variables owned at that point, implies the assertion condition): if it fails to prove an assertion, it records the corresponding inferred fact in a log file for manual inspection. Fig.~\ref{fi:ratcop-summary} summarizes the set of operations in RATCOP.

As benchmarks, we use a subset of concurrent programs from the SV-COMP 2015 suite \cite{beyer2015software}. We chose only those programs which we believe have interesting relational invariants. We ported the programs (which are originally in C) to Java and introduced locks appropriately to remove races. We also use a program from \cite{mine2013static}, which is an abstraction of a producer-consumer scenario. While these programs are not too large, they have challenging invariants to prove, and
provide a good test for the precision of the various analyses. We ran
the tool in a virtual machine with $16$GB RAM and $4$ cores. The
virtual machine, in turn, ran on a machine with $32$GB RAM and a
quad-core Intel i$7$ processor. We evaluated five analyses on the
benchmarks.
The first four are based on the \adrf\ analysis
(Sec.~\ref{subsec:adrf}),
and employ the Octagon numerical abstract domain.
The last is based on the \vset\ analysis (Sec.~\ref{subsec:vset}),
and uses the Interval domain.
These analyses are named as follows:
% with the following abstract domains:
\begin{enumerate}
\item $\ANa$: Without regions and thread identifiers \footnote{By thread-identifiers we are referring to the abstraction of the versions (recency information) outlined in Remark \ref{se:ldrf-more-abs}}.

\item$\ANb$: With regions, but with no thread identifiers.

\item $\ANc$: Without regions, but with thread identifiers.

\item $\ANd$: With regions and thread identifiers.

\item $\ANe$: The value-set analysis of \cite{de2011dataflow}.
\end{enumerate}

In terms of the precision of the abstract domains, the analyses form the following partial order: $\ANe \prec \ANa \prec \ANc \prec \ANd$ and $\ANe \prec \ANa \prec \ANb \prec \ANd$. We use $\ANe$ as the baseline.

\subsection{Evaluation}
\paragraph{Porting Sequential Analyses to Concurrent Analyses.}
\label{se:o-port}
For the sequential commands, we performed a lightweight parsing of statements and simply re-use the built-in transformers of Apron. The only operator we needed to define afresh was the abstract \emph{mix}. Since Apron exposes functions to perform each of the constituent steps, implementing the abstract \emph{mix} was straightforward as well.

\paragraph{Precision and Efficiency.}
Table. \ref{fi:results} summarizes the results of the experiments.

\begin{landscape}
\vfill
\begin{table}
\centering
%\fontsize{9}{11}\selectfont
\begin{tabular}{|c|c|c|c|c|c|c|c|c|c|c|c|c|c|}
\hline
{\color[HTML]{333333} }                                                                            & {\color[HTML]{333333} }                               & {\color[HTML]{333333} }                                   & {\color[HTML]{333333} }                                   & \multicolumn{2}{c|}{{\color[HTML]{333333} \tiny{$\ANa$}}}                                                                   & \multicolumn{2}{c|}{{\color[HTML]{333333} \tiny{$\ANb$}}}                                                                  & \multicolumn{2}{c|}{{\color[HTML]{333333} \tiny{$\ANc$}}}                                                                  & \multicolumn{2}{c|}{{\color[HTML]{333333} \tiny{$\ANd$}}}                                                                  & \multicolumn{2}{c|}{{\color[HTML]{333333} \tiny{$\ANe$}}}                                                                  \\ \cline{5-14}
\multirow{-2}{*}{{\color[HTML]{333333} \textbf{Program}}}                                          & \multirow{-2}{*}{{\color[HTML]{333333} \textbf{LOC}}} & \multirow{-2}{*}{{\color[HTML]{333333} \textbf{Threads}}} & \multirow{-2}{*}{{\color[HTML]{333333} \textbf{Asserts}}} & {\color[HTML]{333333} \textbf{\checkmark}}  & {\color[HTML]{333333} \textbf{\begin{tabular}[c]{@{}c@{}}\tiny{Time}\\ \tiny{(ms)}\end{tabular}}}  & {\color[HTML]{333333} \textbf{\checkmark}}  & {\color[HTML]{333333} \textbf{\begin{tabular}[c]{@{}c@{}}\tiny{Time}\\ \tiny{(ms)}\end{tabular}}} & {\color[HTML]{333333} \textbf{\checkmark}}  & {\color[HTML]{333333} \textbf{\begin{tabular}[c]{@{}c@{}}\tiny{Time}\\ \tiny{(ms)}\end{tabular}}} & {\color[HTML]{333333} \textbf{\checkmark}}  & {\color[HTML]{333333} \textbf{\begin{tabular}[c]{@{}c@{}}\tiny{Time}\\ \tiny{(ms)}\end{tabular}}} & {\color[HTML]{333333} \textbf{\checkmark}}  & {\color[HTML]{333333} \textbf{\begin{tabular}[c]{@{}c@{}}\tiny{Time}\\ \tiny{(ms)}\end{tabular}}} \\ \hline
{\color[HTML]{333333} \textbf{reorder\_2}}                                                         & {\color[HTML]{333333} 106}                            & {\color[HTML]{333333} 5}                                  & {\color[HTML]{333333} 2}                                  & {\color[HTML]{333333} 0(C)}        & {\color[HTML]{333333} 77}                                                            & {\color[HTML]{333333} 2(C)}        & {\color[HTML]{333333} 43}                                                           & {\color[HTML]{333333} 0(C)}        & {\color[HTML]{333333} 71}                                                           & {\color[HTML]{333333} 2(C)}        & {\color[HTML]{333333} 37}                                                           & {\color[HTML]{333333} 0}           & {\color[HTML]{333333} 25}                                                           \\ \hline
{\color[HTML]{333333} \textbf{sigma \textsuperscript{B}*}}                                                           & {\color[HTML]{333333} 118}                            & {\color[HTML]{333333} 5}                                  & {\color[HTML]{333333} 5}                                  & {\color[HTML]{333333} 0}           & {\color[HTML]{333333} 132}                                                           & {\color[HTML]{333333} 0}           & {\color[HTML]{333333} 138}                                                          & {\color[HTML]{333333} 4}           & {\color[HTML]{333333} 48}                                                           & {\color[HTML]{333333} 4}           & {\color[HTML]{333333} 50}                                                           & {\color[HTML]{333333} 0}           & {\color[HTML]{333333} 506}                                                          \\ \hline
{\color[HTML]{333333} \textbf{sssc12}}                                                             & {\color[HTML]{333333} 98}                             & {\color[HTML]{333333} 3}                                  & {\color[HTML]{333333} 4}                                  & {\color[HTML]{333333} 4}           & {\color[HTML]{333333} 76}                                                            & {\color[HTML]{333333} 4}           & {\color[HTML]{333333} 90}                                                           & {\color[HTML]{333333} 4}           & {\color[HTML]{333333} 82}                                                           & {\color[HTML]{333333} 4}           & {\color[HTML]{333333} 86}                                                           & {\color[HTML]{333333} 2}           & {\color[HTML]{333333} 28}                                                           \\ \hline
{\color[HTML]{333333} \textbf{unverif}}                                                            & {\color[HTML]{333333} 82}                             & {\color[HTML]{333333} 3}                                  & {\color[HTML]{333333} 2}                                  & {\color[HTML]{333333} 0}           & {\color[HTML]{333333} 115}                                                           & {\color[HTML]{333333} 0}           & {\color[HTML]{333333} 121}                                                          & {\color[HTML]{333333} 0}           & {\color[HTML]{333333} 84}                                                           & {\color[HTML]{333333} 0}           & {\color[HTML]{333333} 86}                                                           & {\color[HTML]{333333} 0}           & {\color[HTML]{333333} 46}                                                           \\ \hline
{\color[HTML]{333333} \textbf{spin2003}}                                                           & {\color[HTML]{333333} 65}                             & {\color[HTML]{333333} 3}                                  & {\color[HTML]{333333} 2}                                  & {\color[HTML]{333333} 2}           & {\color[HTML]{333333} 6}                                                             & {\color[HTML]{333333} 2}           & {\color[HTML]{333333} 9}                                                            & {\color[HTML]{333333} 2}           & {\color[HTML]{333333} 10}                                                           & {\color[HTML]{333333} 2}           & {\color[HTML]{333333} 10}                                                           & {\color[HTML]{333333} 2}           & {\color[HTML]{333333} 8}                                                            \\ \hline
{\color[HTML]{333333} \textbf{simpleLoop}}                                                         & {\color[HTML]{333333} 74}                             & {\color[HTML]{333333} 3}                                  & {\color[HTML]{333333} 2}                                  & {\color[HTML]{333333} 2}           & {\color[HTML]{333333} 56}                                                            & {\color[HTML]{333333} 2}           & {\color[HTML]{333333} 61}                                                           & {\color[HTML]{333333} 2}           & {\color[HTML]{333333} 57}                                                           & {\color[HTML]{333333} 2}           & {\color[HTML]{333333} 64}                                                           & {\color[HTML]{333333} 0}           & {\color[HTML]{333333} 27}                                                           \\ \hline
{\color[HTML]{333333} \textbf{simpleLoop5}}                                                        & {\color[HTML]{333333} 84}                             & {\color[HTML]{333333} 4}                                  & {\color[HTML]{333333} 1}                                  & {\color[HTML]{333333} 0}           & {\color[HTML]{333333} 40}                                                            & {\color[HTML]{333333} 0}           & {\color[HTML]{333333} 50}                                                           & {\color[HTML]{333333} 0}           & {\color[HTML]{333333} 31}                                                           & {\color[HTML]{333333} 0}           & {\color[HTML]{333333} 37}                                                           & {\color[HTML]{333333} 0}           & {\color[HTML]{333333} 20}                                                           \\ \hline
{\color[HTML]{333333} \textbf{doubleLock\_p3}}                                                     & {\color[HTML]{333333} 64}                             & {\color[HTML]{333333} 3}                                  & {\color[HTML]{333333} 1}                                  & {\color[HTML]{333333} 1}           & {\color[HTML]{333333} 11}                                                            & {\color[HTML]{333333} 1}           & {\color[HTML]{333333} 24}                                                           & {\color[HTML]{333333} 1}           & {\color[HTML]{333333} 16}                                                           & {\color[HTML]{333333} 1}           & {\color[HTML]{333333} 19}                                                           & {\color[HTML]{333333} 1}           & {\color[HTML]{333333} 9}                                                            \\ \hline
{\color[HTML]{333333} \textbf{fib\_Bench}}                                                         & {\color[HTML]{333333} 82}                             & {\color[HTML]{333333} 3}                                  & {\color[HTML]{333333} 2}                                  & {\color[HTML]{333333} 0}           & {\color[HTML]{333333} 138}                                                           & {\color[HTML]{333333} 0}           & {\color[HTML]{333333} 118}                                                          & {\color[HTML]{333333} 0}           & {\color[HTML]{333333} 129}                                                          & {\color[HTML]{333333} 0}           & {\color[HTML]{333333} 102}                                                          & {\color[HTML]{333333} 0}           & {\color[HTML]{333333} 56}                                                           \\ \hline
{\color[HTML]{333333} \textbf{\begin{tabular}[c]{@{}c@{}}fib\_Bench\_\\ Longer\end{tabular}}}      & {\color[HTML]{333333} 82}                             & {\color[HTML]{333333} 3}                                  & {\color[HTML]{333333} 2}                                  & {\color[HTML]{333333} 0}           & {\color[HTML]{333333} 95}                                                            & {\color[HTML]{333333} 0}           & {\color[HTML]{333333} 103}                                                          & {\color[HTML]{333333} 0}           & {\color[HTML]{333333} 123}                                                          & {\color[HTML]{333333} 0}           & {\color[HTML]{333333} 91}                                                           & {\color[HTML]{333333} 0}           & {\color[HTML]{333333} 35}                                                           \\ \hline
{\color[HTML]{333333} \textbf{indexer}}                                                            & {\color[HTML]{333333} 119}                            & {\color[HTML]{333333} 2}                                  & {\color[HTML]{333333} 2}                                  & {\color[HTML]{333333} 2}           & {\color[HTML]{333333} 1522}                                                          & {\color[HTML]{333333} 2}           & {\color[HTML]{333333} 1637}                                                         & {\color[HTML]{333333} 2}           & {\color[HTML]{333333} 1750}                                                         & {\color[HTML]{333333} 2}           & {\color[HTML]{333333} 1733}                                                         & {\color[HTML]{333333} 2}           & {\color[HTML]{333333} 719}                                                          \\ \hline
{\color[HTML]{333333} \textbf{twostage\_3 \textsuperscript{B}}}                                                      & {\color[HTML]{333333} 93}                             & {\color[HTML]{333333} 2}                                  & {\color[HTML]{333333} 2}                                  & {\color[HTML]{333333} 0}           & {\color[HTML]{333333} 61}                                                            & {\color[HTML]{333333} 0}           & {\color[HTML]{333333} 48}                                                           & {\color[HTML]{333333} 0}           & {\color[HTML]{333333} 57}                                                           & {\color[HTML]{333333} 0}           & {\color[HTML]{333333} 28}                                                           & {\color[HTML]{333333} 0}           & {\color[HTML]{333333} 59}                                                           \\ \hline
{\color[HTML]{333333} \textbf{\begin{tabular}[c]{@{}c@{}}singleton\_\\ with\_uninit\end{tabular}}} & {\color[HTML]{333333} 59}                             & {\color[HTML]{333333} 2}                                  & {\color[HTML]{333333} 1}                                  & {\color[HTML]{333333} 1}           & {\color[HTML]{333333} 31}                                                            & {\color[HTML]{333333} 1}           & {\color[HTML]{333333} 29}                                                           & {\color[HTML]{333333} 1}           & {\color[HTML]{333333} 14}                                                           & {\color[HTML]{333333} 1}           & {\color[HTML]{333333} 10}                                                           & {\color[HTML]{333333} 1}           & {\color[HTML]{333333} 28}                                                           \\ \hline
{\color[HTML]{333333} \textbf{stack}}                                                              & {\color[HTML]{333333} 85}                             & {\color[HTML]{333333} 2}                                  & {\color[HTML]{333333} 2}                                  & {\color[HTML]{333333} 0}           & {\color[HTML]{333333} 151}                                                           & {\color[HTML]{333333} 0}           & {\color[HTML]{333333} 175}                                                          & {\color[HTML]{333333} 0}           & {\color[HTML]{333333} 127}                                                          & {\color[HTML]{333333} 0}           & {\color[HTML]{333333} 129}                                                          & {\color[HTML]{333333} 0}           & {\color[HTML]{333333} 71}                                                           \\ \hline
{\color[HTML]{333333} \textbf{stack\_longer}}                                                      & {\color[HTML]{333333} 85}                             & {\color[HTML]{333333} 1}                                  & {\color[HTML]{333333} 2}                                  & {\color[HTML]{333333} 0}           & {\color[HTML]{333333} 1163}                                                          & {\color[HTML]{333333} 0}           & {\color[HTML]{333333} 669}                                                          & {\color[HTML]{333333} 0}           & {\color[HTML]{333333} 1082}                                                         & {\color[HTML]{333333} 0}           & {\color[HTML]{333333} 1186}                                                         & {\color[HTML]{333333} 0}           & {\color[HTML]{333333} 597}                                                          \\ \hline
{\color[HTML]{333333} \textbf{stack\_longest}}                                                     & {\color[HTML]{333333} 85}                             & {\color[HTML]{333333} 2}                                  & {\color[HTML]{333333} 2}                                  & {\color[HTML]{333333} 0}           & {\color[HTML]{333333} 1732}                                                          & {\color[HTML]{333333} 0}           & {\color[HTML]{333333} 1679}                                                         & {\color[HTML]{333333} 0}           & {\color[HTML]{333333} 1873}                                                         & {\color[HTML]{333333} 0}           & {\color[HTML]{333333} 2068}                                                         & {\color[HTML]{333333} 0}           & {\color[HTML]{333333} 920}                                                          \\ \hline
{\color[HTML]{333333} \textbf{sync01 *}}                                                           & {\color[HTML]{333333} 65}                             & {\color[HTML]{333333} 2}                                  & {\color[HTML]{333333} 2}                                  & {\color[HTML]{333333} 2}           & {\color[HTML]{333333} 7}                                                             & {\color[HTML]{333333} 2}           & {\color[HTML]{333333} 25}                                                           & {\color[HTML]{333333} 2}           & {\color[HTML]{333333} 37}                                                           & {\color[HTML]{333333} 2}           & {\color[HTML]{333333} 33}                                                           & {\color[HTML]{333333} 2}           & {\color[HTML]{333333} 10}                                                           \\ \hline
{\color[HTML]{333333} \textbf{qw2004 *}}                                                           & {\color[HTML]{333333} 90}                             & {\color[HTML]{333333} 2}                                  & {\color[HTML]{333333} 4}                                  & {\color[HTML]{333333} 0}           & {\color[HTML]{333333} 1401}                                                          & {\color[HTML]{333333} 4}           & {\color[HTML]{333333} 1890}                                                         & {\color[HTML]{333333} 0}           & {\color[HTML]{333333} 1478}                                                         & {\color[HTML]{333333} 4}           & {\color[HTML]{333333} 1913}                                                         & {\color[HTML]{333333} 0}           & {\color[HTML]{333333} 698}                                                          \\ \hline
{\color[HTML]{333333} \textbf{ \cite{mine2013static} Fig. 3.11}}                                                   & {\color[HTML]{333333} 89}                             & {\color[HTML]{333333} 2}                                  & {\color[HTML]{333333} 2}                                  & {\color[HTML]{333333} 0}           & {\color[HTML]{333333} 49}                                                            & {\color[HTML]{333333} 2}           & {\color[HTML]{333333} 46}                                                           & {\color[HTML]{333333} 0}           & {\color[HTML]{333333} 54}                                                           & {\color[HTML]{333333} 2}           & {\color[HTML]{333333} 36}                                                           & {\color[HTML]{333333} 0}           & {\color[HTML]{333333} 19}                                                           \\ \hline
{\color[HTML]{333333} \textbf{Total}}                                                              & {\color[HTML]{333333} \textbf{1625}}                  & {\color[HTML]{333333} \textbf{3 (Avg)}}                   & {\color[HTML]{333333} \textbf{42}}                        & {\color[HTML]{333333} \textbf{14}} & {\color[HTML]{333333} \textbf{\begin{tabular}[c]{@{}c@{}}361 \\ (Avg)\end{tabular}}} & {\color[HTML]{333333} \textbf{22}} & {\color[HTML]{333333} \textbf{\begin{tabular}[c]{@{}c@{}}366\\ (Avg)\end{tabular}}} & {\color[HTML]{333333} \textbf{18}} & {\color[HTML]{333333} \textbf{\begin{tabular}[c]{@{}c@{}}374\\ (Avg)\end{tabular}}} & {\color[HTML]{333333} \textbf{26}} & {\color[HTML]{333333} \textbf{\begin{tabular}[c]{@{}c@{}}406\\ (Avg)\end{tabular}}} & {\color[HTML]{333333} \textbf{10}} & {\color[HTML]{333333} \textbf{\begin{tabular}[c]{@{}c@{}}204\\ (Avg)\end{tabular}}} \\ \hline
\end{tabular}
\caption{\label{fi:results}Summary of the experiments. Superscript \textsuperscript{B} indicates that the program has an actual bug. (C) indicates the use of Convex Polyhedra as abstract data domain. ``*" indicates a program where we have altered/weakened the original assertion. The $\checkmark$ column indicates the number of assertions the tool was able to prove.}
\end{table}
\end{landscape}

%\begin{figure}[!ht]
%%\vspace{-10pt}
%\centering
%\includegraphics[scale=0.36]{Results.png}
%\caption{\label{fi:results}\small{Summary of the experiments. Superscript \textsuperscript{B} indicates that the program has an actual bug. (C) indicates the use of Convex Polyhedra as abstract data domain. * indicates a program where we have altered/weakened the original assertion.}}
%%\vspace{-15pt}
%\end{figure}
While all the analyses failed to prove the assertions in $\mathtt{reorder\_2}$, $\ANb$ and $\ANd$ were able to prove them when they used convex polyhedra instead of octagons. Since none of the analyses track arrays precisely, all of them failed to prove the original assertion in $\mathtt{sigma}$ (which involves checking a property involving the sum of the array elements). However, $\ANc$ and $\ANd$ correctly detect a potential array out-of-bounds violation in the program. The improved precision is due to the fact that $\ANc$ and $\ANd$ track thread identifiers in the abstract state, which avoids spurious read-write cycles in the analysis of $\mathtt{sigma}$. The program $\mathtt{twostage\_3}$ has an actual bug, and the assertions are expected to fail. This program provides a ``sanity check" of the soundness of the analyses. Programs marked with ``*"" contain assertions which we have altered completely and/or weakened. In these cases, the original assertion was either expected to fail or was too precise (possibly requiring a disjunctive domain in order to prove it). In $\mathtt{qw2004}$, for example, our modified assertions are of the form $x = y$. $\ANb$ and $\ANd$ perform well in this case, since we can specify a region containing $x$ and $y$, which precisely tracks their correlation across threads. The imprecision in the remaining cases are mostly due to the program requiring \emph{disjunctive} domains to discharge the assertions, or the presence of spurious write-write cycles which weaken the inferred facts. Abstracting our semantics to handle such cycles is an interesting future work.

Of the total $40$ ``valid" assertions (excluding the two in $\mathtt{twostage\_3}$), $\ANd$ is the most precise, being able to prove $65\%$ of them. It is followed by $\ANb$ ($55\%$), $\ANc$ ($45\%$), $\ANa$ ($35\%$) and, lastly, $\ANe$ ($25\%$). Thus, the new analyses derived from \ldrf\ and \rdrf\ perform significantly better than the value-set analysis of \cite{de2011dataflow}. Moreover, this total order respects the partial ordering between the analyses defined earlier.

With respect to the running times, the maximum time taken, across all the programs, is around $2$ seconds, by $\ANd$. $\ANe$ turns out to be the fastest in general, due to its lightweight abstract domain. $\ANb$ and $\ANd$ are typically slower that $\ANa$ and $\ANc$ respectively. The slowdown can be attributed to the additional tracking of regions by the former analyses. Note that for the program $\texttt{sigma}$, $\ANd$ was both more precise and faster than the baseline $\ANe$.

\subsection{Comparison with a recent abstract interpretation based tool.}

We also compared the efficiency of RATCOP with that of Batman, a tool implementing the previous state-of-the-art analyses based on abstract interpretation \cite{mine2014relational, monat2017precise} (a discussion on the precision of our analyses against those in \cite{mine2014relational} is presented in Sec.~\ref{Se:related}).
The basic structure of the benchmark programs for this experiment is as follows: each program defines a set of shared variables. A $\mathtt{main}$ thread then partitions the set of shared variables, and creates threads which access and modify variables in a unique partition. Thus, the set of memory locations accessed by any two threads is disjoint. In our experiments, each thread simply performed a sequence of writes to a specific set of shared variables. In some sense, these programs represent a ``best-case" scenario for concurrent program analyses because there are no interferences between threads. Unlike RATCOP, the Batman tool, in its current form, only supports a small toy language and does not provide the means to automatically check assertions. Thus, for the purposes of this experiment, we only compare the time required to reach a fixpoint in the two tools. We compare $\ANc$ against Batman running with the Octagon domain and the BddApron library \cite{jeannet2010some} (Bm-oct).

%\begin{figure}[ht]
%\vspace{-10pt}
%\centering
%\includegraphics[scale=0.5]{rc_vs_bm_fvar_log.png}
%\vspace{-5pt}
%\caption{\label{fi:o_vs_b_fvar_log}A comparison of the running times of RATCOP and Batman on a logarithmic scale, where the number of threads is varied.}
%\vspace{-16pt}
%\end{figure}
%\vspace{-10pt}
%\begin{figure}[t] %[ht]
%  \begin{minipage}{6cm}
%    \centering
%    \begin{small}
%\begin{tabular}{|c|c|c|} \hline
%\textbf{\#Threads} &
%\textbf{$\mathbf{A3}$ Time (ms)} &
%\textbf{Bm-oct Time (ms)} \\ \hline
%2 & 61 & 7706 \\ \hline
%3 & 86 & 82545 \\ \hline
%4 & 138 & 507663 \\ \hline
%5 & 194 & 2906585 \\ \hline
%6 & 261 & 13095977 \\ \hline
%7 & 368 & 53239574 \\ \hline
%\end{tabular}
%    \end{small}
%\end{minipage}%
%\begin{minipage}{4cm}
%\centering
%\includegraphics[scale=0.5]{ratcop/figures/rc_vs_bm_fvar_log.png}
%\end{minipage}
%\medskip
%\caption{Running times of RATCOP ($\mathbf{A3}$) and Batman
%  (Bm-oct) on loosely coupled threads. The number of shared
%  variables is fixed at $6$. The graph on the right shows the
%  running times on a log scale.}
%\label{tab:o_vs_b_fvar}
%\vspace{-17pt}
%\end{figure}

\begin{table}[!htb]
\centering
\begin{tabular}{|c|r|r|} \hline
\textbf{\#Threads} &
\textbf{\scriptsize{$\ANc$} Time (ms)} &
\textbf{\scriptsize{Bm-oct Time (ms)}} \\ \hline
2 & 61 & 7706 \\ \hline
3 & 86 & 82545 \\ \hline
4 & 138 & 507663 \\ \hline
5 & 194 & 2906585 \\ \hline
6 & 261 & 13095977 \\ \hline
7 & 368 & 53239574 \\ \hline
\end{tabular}
\caption{\label{tab:o_vs_b_fvar}Running times of RATCOP ($\ANc$) and Batman
  (Bm-oct) on loosely coupled threads. The number of shared
  variables is fixed at $6$.}
\end{table}

\begin{figure}[!htb]
\centering
\includegraphics[scale=0.45]{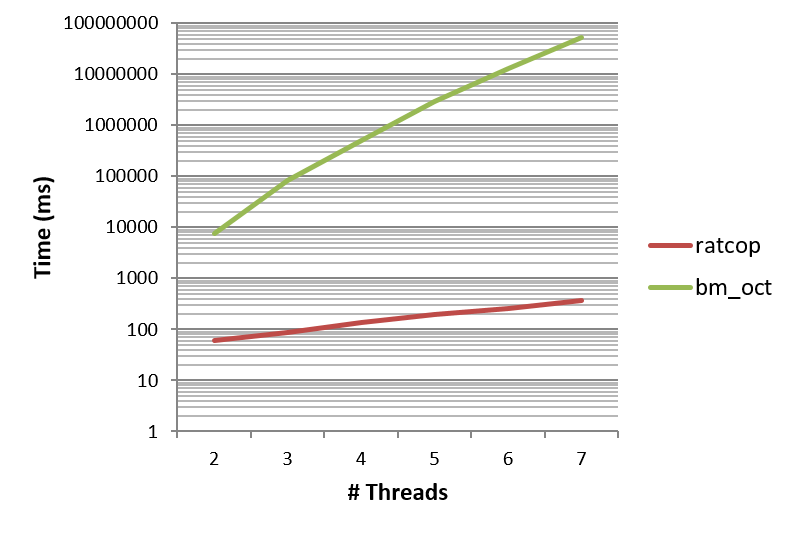}
\caption{\label{fi:o_vs_b_fvar}Graphical representation of the data in Table~\ref{tab:o_vs_b_fvar} on a logarithmic scale. RATCOP performs exponentially faster, compared to Batman, on this benchmark.
}
\end{figure}

The running times of the two analyses are given in Table
~\ref{tab:o_vs_b_fvar}. The graph in Fig.~\ref{fi:o_vs_b_fvar} plots
these running times on a logarithmic scale. In the benchmarks, with
increasing number of threads, RATCOP was upto $5$ orders of magnitude
faster than Bm-oct. The rate of increase in running time was roughly
linear for RATCOP, while it was almost exponential for Bm-oct. 
We believe the reason for this difference in running times is that the
analyses in \cite{mine2014relational, monat2017precise} compute sound
facts at \emph{every} program point. Thus, as the number of threads
increase, these analyses have to account for data flow over an
exponential number of context-switch points, which contributes to the
slowdown. RATCOP, on the other hand, does not attempt to be sound at
all program points. For these programs it performs no inter-thread
propagation, and the time increases linearly with the total number of
program points. For assertions in thread $t$ which only involve variables in the logical partition of $t$, RATCOP is at least as precise as Batman, since proving such assertions do not require inter-thread reasoning.

%!TEX root=./vmcaiReport.tex

\section{Related Work and Discussion}
\label{Se:related}

In this paper we have presented a framework for developing intra-procedural
data-flow analyses for data race free shared-memory concurrent
programs, with a statically fixed number of threads, and with
variables having primitive data types.

There is a rich literature on data flow analysis of concurrent
programs. We refer the reader to the detailed survey by Rinard
\cite{rinard2001analysis} which provides details of the main
approaches. 
In this section, we proceed to compare our work with some of the
relevant prior approaches. 

\paragraph{Degree of Inter-thread Communication.} 
Chugh et al \cite{chugh2008dataflow} automatically lift a given sequential
analysis to a sound analysis for concurrent programs, using a data
race detector. However, data-flow facts are not communicated across
threads, and this can cause a loss in precision. 
The work by Mine \cite{mine2011static} allows a greater degree of
inter-thread
communication. Here, the overall analysis can be considered to proceed
in rounds of thread-modular analyses. At the end of each round, every
thread generates a set of per-thread ``interferences" -- for each
variable $x$, a thread $t$ stores the set of values it writes to $x$
when $t$ was analyzed modularly. In the next iteration, each thread
$t' \not= t$ takes into account this interference information from
$t$, whenever it reads $x$. This, in turn, generates more
interferences for $t'$, and the process continues till fixpoint. Thus,
the inter-thread communication is flow insensitive. Unlike our
semantics, this analysis is unable to infer relational properties
between variables. 

Mine \cite{mine2014relational} presents an abstract
interpretation formulation of the rely-guarantee proof paradigm
\cite{jones1981development,xu1997rely}, and allows one to derive
analyses with varying degrees of inter-thread flow sensitivity. In
particular, the work in \cite{mine2011static} is shown to be an
abstraction of the semantics in \cite{mine2014relational}. The
semantics in \cite{mine2014relational} involves a nested fixed-point
computation, compared to our single fixed-point formulation. The
resulting analysis aims to be sound at \emph{all} program points (e.g,
in Fig. \ref{fi:motEx} the value of $y$ at line~$9$ in $t_2$), due to
which many more interferences will have to be propagated than we do,
leading to a less efficient analysis. The times clocked by Batman, in
comparison to RATCOP, is testament to this. \cite{mine2014relational}
attempts to retrieve some degree of efficiency by computing ``lock
invariants", which are essentially summaries of each critical
section. However, to make use of this, the program must be
well-synchronized -- every access of a shared variable must be protected
by a lock, which is a stronger requirement than data race freedom.
Moreover, for certain programs, our abstract analyses are more
precise. Fig.~\ref{fi:read-no-lock-example} shows a program which is
race free, even though the conflicting accesses to $x$ in lines $2$
and $12$ are not protected by a common lock. The ``lock invariants" in
\cite{mine2014relational} would consider these accesses as potentially
racy, and would allow the read at line $12$ to observe the write at
line $2$, thereby being unable to prove the assertion. However, our
analyses would ensure that the read only observes the write at line
$11$, and is able to prove the
assertion. \cite{ferreira2010parameterized} presents an operational
semantics for concurrent programs, parameterized by a relation. It
makes additional assumptions about code regions which are
unsynchronized (allowing only read-only shared variables and local
variables in such regions). Moreover, it too computes sound facts at
every point, resulting in less efficient abstractions. In this sense,
De et al \cite{de2011dataflow} strikes a sweet spot: by leveraging the
race freedom assumption, the analysis restricts data flow facts to
synchronization points alone, thereby gaining efficiency. However,
this work cannot compute relational information either, being based on
a cartesian value-set domain.

\paragraph{Control Flow Representation.} 
The methods described in \cite{dwyer1994data, grunwald1993data,
  de2011dataflow}
present concurrent data flow algorithms by building specialized
concurrent flow graphs. However, the class of analyses they address
are restricted -- \cite{dwyer1994data} handles properties expressible
as Quantified Regular Expressions, \cite{grunwald1993data} handles
reaching definitions, while \cite{de2011dataflow} only handles
value-set analyses. While our analyses also makes use of the
\scfg\ data structure of \cite{de2011dataflow}, the \ldrf\ and
\rdrf\ semantics allows us to use it in conjunction with much more
expressive abstract domains. In contrast to our approach, the
techniques in \cite{farzan2012verification, farzan2013inductive}
provide an approach to verifying properties of concurrent programs
using \emph{data flow} graphs, rather than use control flow graphs
like we do.
%%  (where the flow facts are abstractions of
%% variables mapped to sets of values). The concurrent flow graphs
%% introduced in \cite{de2011dataflow}, called \textit{sync-}CFGs, are
%% abstractions of our semantics.

%\vspace{-15pt}
%\input{read-nolock-example2}
%\vspace{-15pt}

\begin{figure}[!htb]
\centering
\begin{minipage}{0.5\textwidth}
\begin{lstlisting}
    Thread t1() {
1:    acquire(m);
2:    x := 1;
3:    y := 1;
4:    release(m);
5:  }
\end{lstlisting}
\end{minipage}%
\begin{minipage}{0.5\textwidth}
\begin{lstlisting}
     Thread t2() {
 6:    while( p != 1 ) {
 7:      acquire(m);
 8:      p := y;
 9:      release(m);
10:    }
11:    x := 2;
12:    p := x;
13:    assert(p != 1);		
14:  }
\end{lstlisting}
\end{minipage}
\caption{\label{fi:read-no-lock-example}Example demonstrating that a program can be DRF, when the accesses of a global variable (in this case, the write and read of \texttt{x} at lines~$11$ and $12$ respectively) are not directly guarded by any lock.}
\end{figure}

\paragraph{Resource Invariants vs. Regions.} 
A traditional approach to analyzing concurrent programs involves
\emph{resource invariants} associated with every lock (e.g. Gotsman et
al ~\cite{gotsman2007thread}).  This approach depends on a
\emph{locking policy} where a thread only accesses global data if it
holds a protecting lock. In contrast, our approach does not require a
particular locking policy (e.g., see
Fig. \ref{fi:read-no-lock-example}), and is based on a parameterized
notion of
data-race-freedom, which allows to encode locking policies as a
particular case. Thus, at the overhead cost of ensuring data race freedom,
our new semantics provides greater flexibility to analysis
writers. The analysis in \cite{gotsman2007thread} also works in
similar spirit as the \scfg\, a selected part of the heap protected by
a lock is made accessible to a thread only when it acquires the
lock. In contrast, the synchronization edges in a \scfg\ propagate
\emph{entire} data flow facts. The locking policy employed by
\cite{gotsman2007thread} is stronger than the notion of race freedom,
and the class of programs the analysis can handle is a subset of what
we handle in this work.
%Indeed,  it is unclear how to verify our program shown in Figure~\ref{fig:read-no-lock-example} using lock
%(resource) invariants because the read of $\mathtt{x}$ is not explicitly protected by locks.
%
%

\textbf{Region Races.} Our notion of region races is inspired by the notion of high-level data races \cite{HighLevelDataRaces}.  The concept of splitting the state space into regions was earlier used in \cite{manevich2008heap}, which used these regions to perform shape analysis for concurrent programs. However, that algorithm still performs a full interleaving analysis which results in poor scalability. The notion of variable packing \cite{blanchet2003static} is similar to our notion of data regions. However, variable packs constitute a purely \emph{syntactic} grouping of variables, while regions are semantic in nature. A syntactic block may not access all variables in a semantic region, which would result in a region partitioning more refined than what the programmer has in mind, which would result in decreased precision.

As future work, we would like to evaluate the performance of our tool when equipped with disjunctive relational domains. In this work, we do not consider dynamically allocated memory, and extending the \ldrf\ semantics to account for the heap memory is interesting future work. Abstractions of such a semantics could potentially yield efficient shape analyses for race free concurrent programs.

\begin{acknowledgements}
We would like to thank the anonymous reviewers for their insightful
and helpful comments which have greatly improved the quality of the
presentation. 
We would like to thank Mooly Sagiv for his help
and insights. We would also like to thank Antoine Min{\'e} and
Rapha{\"e}l Monat for their help with the Apron library and in setting
up Batman. This publication is part of a project that has received
funding from the European Research Council (ERC) under the European
Union's Seventh Framework Programme (FP7/2007-2013) / ERC grant
agreement n$^{\circ}$ [321174] and under the European Union's Horizon
2020 research and innovation programme (grant agreement No
[759102-SVIS]).
%This research was supported by the European Research Council under the European Union's Seventh Framework Programme (FP7/2007-2013) / ERC grant agreement n$^{\circ}$ [321174], 
This research was supported by Len Blavatnik and the Blavatnik Family foundation, and by the Blavatnik Interdisciplinary Cyber Research Center, Tel Aviv University.
\end{acknowledgements}

\pagebreak

\bibliographystyle{spmpsci}      % mathematics and physical sciences
\bibliography{theBibliography}   % name your BibTeX data base

\end{document}